\documentclass[11pt, onecolumn,superscriptaddress,nofootinbib,aps,prd]{revtex4-1}
\pdfoutput=1
\PassOptionsToPackage{usenames,dvipsnames}{xcolor} 
\usepackage{longtable}
\usepackage{graphicx,tcolorbox}
\usepackage{ulem}
\usepackage{listings}
\usepackage{makecell,booktabs,siunitx}
\usepackage{amsfonts,bm,bbm,mathrsfs}
\usepackage{mathtools}
\numberwithin{equation}{section} 
\makeatletter
\def\theequation{\arabic{section}.\arabic{equation}}
\makeatother
\usepackage{amsmath}
\usepackage{amssymb}
\usepackage{color}
\usepackage{enumitem}
\usepackage{url}
\usepackage{array}
\usepackage{tabularx}

\usepackage[usenames, dvipsnames]{xcolor}
\definecolor{lbrown}{RGB}{139,69,19}
\usepackage[%
  colorlinks = true,
  urlcolor= lbrown,
  linkcolor= lbrown,
  citecolor= lbrown
]{hyperref}
\usepackage[nolist,nohyperlinks]{acronym}
\usepackage{orcidlink}
\newcommand{\OGW}{\Omega_\mathrm{gw}}
\newcommand{\GW}{\mathrm{gw}}
\newcommand{\barell}{{\bar{\ell}}}
\newcommand{\barm}{{\bar{m}}}
\newcommand{\sinc}{{\mathrm{sinc}}}
\newcommand{\D}{{\text{d}}}
\newcommand{\ovl}{{\mathrm{ovl}}}
\newcommand{\opt}{{\mathrm{opt}}}
\newcommand\be{\begin{equation}}
\newcommand\ee{\end{equation}}

\newcommand{\realistic}{\rm{real}}
\newcommand{\Creal}{{\hat C_{\rm real}}}
\newcommand{\Copt}{{\hat C_{\rm opt}}}

\begin{acronym}
    \acro{GW}{gravitational wave}
    \acro{PSD}{power spectral density}
    \acro{FT}{Fourier transform}
    \acro{IFT}{inverse Fourier transform}
    \acro{DFT}{discrete Fourier transform}
    \acro{IDFT}{inverse discrete Fourier transform}
    \acro{GR}{general relativity}
    \acro{CBC}{compact binary coalescence}
    \acro{BH}{black hole}
    \acro{BBH}{binary black hole}
    \acro{BNS}{binary neutron star}
    \acro{NSBH}{neutron star-black hole}
    \acro{SFR}{star formation rate}
    \acro{LVK}{LIGO-Virgo-KAGRA}
    \acro{ET}{Einstein Telescope}
    \acro{CE}{Cosmic Explorer}
    \acro{PE}{parameter estimation}
    \acro{SGWB}{stochastic gravitational-wave background}
    \acro{XG}{next-generation}
    \acro{PTA}{pulsar timing array}
    \acro{MCMC}{Markov Chain Monte Carlo}
    \acro{NUTS}{No-U-Turn Sampler}
\end{acronym}

\begin{document}

\title{A thorough investigation of cross-correlation estimators for stochastic gravitational-wave background searches in ground-based detector data}
\author{Haowen Zhong \orcidlink{0000-0001-8324-5158}}
\email{zhong461@umn.edu}
\affiliation{
 School of Physics and Astronomy, University of Minnesota, Minneapolis, Minnesota 55455, USA
}

\author{Joseph D. Romano \orcidlink{0000-0003-4915-3246}}
\email{joseph.romano@utrgv.edu}
\affiliation{
Department of Physics and Astronomy, University of Texas Rio Grande Valley, One West University Boulevard, Brownsville, Texas 78520, USA}

\author{Vuk Mandic \orcidlink{0000-0001-6333-8621}}
\email{vuk@umn.edu}
\affiliation{
School of Physics and Astronomy, University of Minnesota, Minneapolis, Minnesota 55455, USA
}
\author{Shivaraj Kandhasamy
\orcidlink{0000-0002-4825-6764}}
\email{shivaraj@iucaa.in}
\affiliation{Inter-University Centre for Astronomy and Astrophysics, Pune 411007, India}

\author{Arianna I. Renzini \orcidlink{0000-0002-4589-3987}}
\email{arenzini@ethz.phys.ch}
\affiliation{Institute for Particle Physics and Astrophysics, ETH Zurich, Zurich, Switzerland}

\date{\today}

\begin{abstract}
Detecting a stochastic gravitational-wave background represents a crucial yet challenging objective within the field of gravitational-wave astronomy. 
Ground-based detectors currently rely almost exclusively on cross-correlation methods to detect stochastic gravitational-wave background signals. 
Traditionally, these methods define and optimize a broadband estimator initially constructed in the time domain. 
However, a growing number of analyses require precise narrowband estimators to accurately characterize the energy density of the underlying signal in specific frequency bins. 
Transitioning from time-domain broadband estimators to frequency-domain narrowband estimators introduces significant complexities that have not yet been fully explored in the existing literature. 
In this study, we systematically revisit and rigorously reformulate the cross-correlation method in the frequency domain, explicitly addressing and resolving issues related to non-zero covariances induced by windowing and overlapping of data in the time domain. 
We provide new expressions for the narrowband estimators and their covariances, which differ from those used in past searches. 
Fortunately, we show that the expressions that have been widely used in the field nonetheless lead to correct posterior distributions for parameter estimation and correct log-Bayes factors for model selection. 
By establishing a robust theoretical framework, our work facilitates more accurate and physically insightful interpretations of stochastic gravitational-wave background observations, laying an essential foundation for current and future research in this field. 
\end{abstract}

\maketitle

\tableofcontents
\newpage

\section{Introduction}
\label{sec: intro}

The direct detection of gravitational waves (GWs) provides an unprecedented observational window onto the universe, offering a unique source of information beyond the traditional electromagnetic observations conducted using ground-based or space-based telescopes. 
With ongoing improvements in the sensitivity of ground-based GW observatories~\cite{aLIGO, aVirgo, aVirgostatus, KAGRA:2018plz}, the detection of an astrophysical \ac{SGWB} from \ac{CBC} systems appears highly promising in the near future~\cite{KAGRA:2021duu, KAGRA:2021kbb, Renzini:2022alw, O3stoch, Bellie:2023jlq}. 
In addition to the \ac{SGWB} from \acp{CBC}, several other compelling sources of the \ac{SGWB} may exist, including first-order phase transitions~\cite{PT1, PT2, PT3, PT4, PT5, PT6, PT7}, a population of supernova explosions~\cite{Ferrari:1998ut, Buonanno:2004tp, Crocker:2015taa, Crocker:2017agi, Finkel:2021zgf}, standard inflationary models~\cite{Grishchuk:1974ny, Starobinsky:1979ty, Grishchuk:1993te}, axion inflation~\cite{Barnaby:2011qe}, and cosmic strings~\cite{Damour:2004kw, Siemens:2006yp, Olmez:2010bi, Regimbau:2011bm}. 
Detecting these subdominant \ac{SGWB} signals would provide substantial scientific benefits. Astrophysical \ac{SGWB}s carry critical information about the nature and characteristics of their underlying populations~\cite{LIGOScientific:2020kqk, KAGRA:2021duu, Bavera:2021wmw}, whereas cosmological \acp{SGWB} could offer unique insights into the earliest stages of the Universe, enabling us to probe fundamental physics at energy scales approaching the Planck scale~\cite{Grishchuk:1974ny, Starobinsky:1979ty, Grishchuk:1993te, Barnaby:2011qe, Damour:2004kw, Siemens:2006yp}.

However, due to the extremely weak nature of these astrophysical or cosmological \ac{SGWB} signals compared to the detector noise of current or even next-generation terrestrial \ac{GW} detectors (\ac{CE}~\cite{CE} and \ac{ET}~\cite{ET}), the direct detection of a \ac{SGWB} is particularly challenging, necessitating sophisticated data analysis methods. 
Several approaches have already been proposed to detect an astrophysical background signal modeled as a series of weak, sporadic, compact binary coalescences~\cite{Drasco:2002yd, Smith:2017vfk, Smith:2020lkj, TBS2, Lawrence:2023buo}.
These methods, which involve some form of marginalization over the parameters defining subthreshold mergers, demand significant computational resources, and their practical computational feasibility for real searches with GW detectors has yet to be fully demonstrated. 
So, in these circumstances, the standard \textit{cross-correlation method}~\cite{Allen:1997ad,Romano:2016dpx} still serves as the main choice for ground-based detectors to robustly measure the energy density spectrum $\Omega_\mathrm{gw}(f)$ of the \ac{SGWB} signal.
The estimated $\Omega_\mathrm{gw}(f)$ then becomes the starting point of all subsequent parameter estimation analyses~\cite{Mandic:2012pj}, which aim to connect the observational data and theoretical models. 

Due to the importance of the cross-correlation method in the field, we revisit previous studies~\cite{Allen:1997ad, Romano:2016dpx, Note_1, Note_2} that established its theoretical foundations\footnote{A related LIGO technical note~\cite{Note_3} discusses window functions in a cross-correlation search for continuous GWs, building on Ref.~\cite{Note_2}; its key ideas were later largely incorporated into the appendix of Ref.~\cite{Whelan2015_ScoX1CrossCorr}.}. 
We observe that these studies primarily focused on defining and optimizing the \textit{broadband estimator} $\hat{C}$, while neglecting an in-depth exploration of the statistical properties of the \textit{narrowband estimator} $\hat{C}(f)$\footnote{Broadband estimators are also known as {\it frequency-independent} or {\it frequency-integrated} estimators, while
narrowband estimators are frequency-dependent, i.e., they represent spectral densities.}. 
In practical analyses, one usually transforms the time-domain data into the frequency domain first and then computes desired quantities. 
Given this, window functions are necessarily applied to the raw data before taking a discrete Fourier transform to obtain frequency-domain data $\tilde{d}(f)$. 
The use of window functions introduces additional complexity to the statistical properties of the data and, consequently, the cross-correlation statistics. 
Although Refs.~\cite{Note_1, Note_2} have comprehensively discussed modifications required for optimizing broadband estimators when window functions are used, they do not address the subtle statistical properties of narrowband estimators.

We emphasize the critical need for a dedicated investigation into the statistical properties of narrowband estimators in the presence of window functions.
This is a subject that has remained unexplored for a considerable time. 
The gap in the literature is particularly pressing given the increasing number of analyses~\cite{Mandic:2012pj,Zhong:2025qno,O3stoch,Renzini:2024hiu,Cousins:2025bas,Callister:2020arv,Callister:2023tws,Turbang:2023tjk} that work within a Bayesian statistical framework, in which the correct formulation of the likelihood function strictly depends on a correct understanding of the mean, variance, and covariance of the narrowband estimators. 
The conclusions that are valid for broadband estimators are incorrectly extended to narrowband estimators, while neglecting the correlation between narrowband estimators across different frequency bins.

In this paper, we rigorously revisit and reformulate the cross-correlation method in the frequency domain, carefully addressing these complexities. 
Our work lays a vital theoretical foundation, enabling more accurate and physically insightful analyses of the \ac{SGWB}. 
As GW detector sensitivities continue to improve, this frequency-domain framework will become increasingly indispensable, guiding and informing future observational efforts and theoretical interpretations of \ac{SGWB} data. While this paper focuses exclusively on \textit{isotropic} searches of a \ac{SGWB}, similar analyses should be carried out for anisotropic searches such as radiometer searches~\cite{Ballmer2006_radiometer} and cross-correlation searches for continuous GWs~\cite{Dhurandhar2007_CrossCorr,Whelan2015_ScoX1CrossCorr}; we leave these extensions for future work.

The remainder of this paper is structured as follows: 
In Sec.~\ref{sec: data_analysis}, we present the data analysis preliminaries  necessary for the stochastic search.  
In Sec.~\ref{s:roadmap}, we provide a ``roadmap" of the calculations that follow.
In Secs.~\ref{s:single_segment}, \ref{s:covariance}, and \ref{s:multi_segment} we define and derive expressions for the narrowband and broadband estimators.
We compare in Sec.~\ref{s:comparisons} our results with those from previous searches, and explain the origin of their differences. 
In Sec.~\ref{sec: PE}, we demonstrate that, despite these differences, previous frequency-domain analyses yield posterior distributions of parameters and Bayes factors that are consistent with our newly derived expressions, since both approaches lead to identical \textit{sufficient statistics} for the likelihood function. 
In Sec.~\ref{s:bias}, we study in detail how estimating detector noise power spectral densities (PSDs) from the data introduces a bias in the inferred variance of the various estimators.
We not only recover the bias factor that has been recognized in previous studies~\cite{Matas:2020roi, pygwb}, but also identify an additional bias that further inflates the true variance of the estimator constructed in practice. 
Finally, we summarize our results and conclusions in Sec.~\ref{sec: conclusions}.

We have also included several appendices: Appendix~\ref{app: stats} collects useful and well-known definitions and results in statistics.
Appendices~\ref{s:single_seg_details}, \ref{s:covariance_details}, \ref{s:multi_seg_details}, and \ref{s:biases_details} provide technical details of derivations that we chose to omit from Secs.~\ref{s:single_segment}, \ref{s:covariance}, \ref{s:multi_segment}, and \ref{s:bias}. 
We discuss the necessity of zero-padding in App.~\ref{app: zero_padding}, extensions in App.~\ref{s:extenstions}, and provide details of our simulations in App.~\ref{app: simulation}.


\section{Data analysis preliminaries}
\label{sec: data_analysis}

We provide the data analysis preliminaries in this section as preparation for the intensive study in following sections.

\subsection{Fourier transform and two-point correlations}

Let $x(t)$ denote a continuous function of time-domain data, with $-\infty < t < \infty$. 
The \ac{FT} of $x(t)$ is defined as
\begin{equation}
    \tilde{x}(f)\equiv \int_{-\infty}^{\infty}\D t\>x(t)e^{-2\pi i ft}\,,
\end{equation}
with \ac{IFT} given by
\begin{equation}
    x(t)=\int_{-\infty}^{\infty}\D f\>\tilde{x}(f)e^{2\pi ift}\,.
\end{equation}
To go back and forth between $x(t)$ and $\tilde x(f)$, we use the relation
\begin{equation}
    \delta(f-f') =\int_{-\infty}^\infty \D t\> e^{-2\pi i (f-f')t}
\end{equation}
for the Dirac delta function.

For the actual analysis, we only have access to a sample of data having finite duration $T$, uniformly sampled with sampling interval $\Delta t$. 
Hence, there are only a finite number $N\equiv T/\Delta t$ of 
discretely-sampled data points%
\footnote{We will assume that $N$ is even. In practice, this condition is always satisfied for us because zero-padding, as described in App.~\ref{app: zero_padding}, ensures that $N$ is even.}:
\begin{equation}
    x_j\equiv x(t_j)\,,\qquad t_j=j\Delta t\,,\quad\text{where}\quad j=0,1,\cdots,N-1\,.
    \label{e:discrete_time}
\end{equation}
We define their \ac{DFT} and \ac{IDFT} as follows:%
\footnote{We include factors of $\Delta t$ and $\Delta f$ in \eqref{eq: DFT} so that $\tilde x_\ell$ and $x_j$ have the same dimensions as $\tilde x(f)$ and $x(t)$.} 
\begin{equation}
\tilde{x}_\ell\equiv \sum_{j=0}^{N-1}x_j\, e^{-2\pi i f_\ell t_j}\Delta t\,,\qquad
x_j=\sum_{\ell=0}^{N-1}\tilde{x}_\ell\, e^{2\pi i f_\ell t_j}\Delta f\,,
\label{eq: DFT}
\end{equation}
where 
\begin{equation}
f_\ell \equiv \ell\Delta f\,,\qquad 
\Delta f\equiv 1/T\equiv 1/(N\Delta t)\,,\quad\text{and}\quad \ell=0,1,\cdots,N-1\,.
\end{equation}
Throughout this paper, indices $\ell$ and $m$ label frequency bins, while indices $i$ and $j$ can label either time bins or frequency bins, depending on the context. Also $\tilde{}$ indicates a frequency domain quantity.
If the $x_j$ are real, one has $\tilde{x}_\ell=\tilde{x}^*_{N-\ell}$, which means that all of the relevant information is contained in the $N/2-1$ positive frequency bins $\ell =1,2,\cdots, N/2-1$; DC (zero frequency, corresponding to $\ell=0$); and the Nyquist frequency ($f_\mathrm{Nyquist}\equiv 1/(2\Delta t)$, corresponding to $\ell=N/2$).
Negative frequencies $\{-f_{\rm Nyquist}+\Delta f, -f_{\rm Nyquist}+2\Delta f, \cdots, -\Delta f\}$ correspond to the indices $\ell= N/2+1, N/2+2, \cdots, N-1$.
Since the GW strain data $d(t)\equiv h(t)+n(t)$ are real, the above discussion is valid for our analyses.

Let us consider a persistent, stationary, Gaussian, unpolarized, and isotropic \ac{SGWB}, and Gaussian noise that is uncorrelated across different detectors. 
For simplicity, we consider only  \textit{two} detectors for the main analyses; the extension of the formalism to multiple detectors will be discussed briefly in App.~\ref{s:ext-mult-det}. 
Given the above assumptions about the GW signal and detector noise, it follows that the two-point correlation functions for the signal and noise are~\cite{Allen:1997ad,Romano:2016dpx}
\begin{equation}
\begin{aligned}
&\langle \tilde{h}_a(f)\tilde{h}_b^*(f')\rangle =\frac{1}{2}\delta(f-f')\gamma_{ab}(f)P_\mathrm{gw}(f)\,,\\
&\langle \tilde{n}_a(f)\tilde{n}_b^*(f')\rangle =\frac{1}{2}\delta(f-f')\delta_{ab}P_{n_a}(f)\,,\\
&\langle\tilde{h}_{a}(f)\tilde{n}_{b}^*(f')\rangle =0\,,\\
\end{aligned}
\label{eq: correlation_continuous}
\end{equation}
where $\gamma_{ab}(f)$ is the overlap reduction function~\cite{Allen:1997ad} of the two detectors $a, b \in\{1,2\}$, and $P_{n_a}(f)$ denotes the one-sided noise \ac{PSD} of detector $a$. $P_\mathrm{gw}(f)$ is the one-sided \ac{PSD} of the GW strain signal\footnote{We distinguish between the \ac{PSD} of the Fourier modes of the SGWB, $P_h(f)$, and the \ac{PSD} of the strain signal, $P_\mathrm{gw}(f)$. 
For more details, see Ref.~\cite{Mingarelli:2019mvk}.}~\cite{Mingarelli:2019mvk}, which is given by
\begin{equation}
    P_\mathrm{gw}(f)=\frac{3H_0^2}{10\pi^2}f^{-3}\OGW(f)\equiv \frac{1}{5}S_0(f)\OGW(f)\,,
\end{equation}
where $S_0(f)$ is defined by
\begin{equation}
    S_0(f)\equiv\frac{3H_0^2}{2\pi^2}\frac{1}{f^3}\,,
    \label{eq: S_0_def}
\end{equation}
and the dimensionless energy density of the \ac{SGWB} is given by
\begin{equation}
    \Omega_\mathrm{gw}(f)\equiv\frac{f}{\rho_c}\frac{\D\rho_\mathrm{gw}(f)}{\D\!f}\,,
    \label{eq: omega_gw}
\end{equation}
where $\D\rho_\mathrm{gw}(f)$ is the GW energy density in the frequency band $(f\,,f+\D\!f)$, and $\rho_c\equiv 3H_0^2c^2/(8\pi G)$ is the critical energy density required to close the Universe. Physically, $S_0(f)$ defined in~\eqref{eq: S_0_def} can be interpreted as the strain power spectral density that an \ac{SGWB} would have if $\Omega_\mathrm{gw}(f)=1$.

Now, the above two-point correlations assume that we have access to the infinite duration, continuous time-series $d_1(t)$ and $d_2(t)$, meaning that we can perform the \ac{FT} assuming infinite time/frequency resolution. 
However, as mentioned above, we have access to only a finite-number of discretely sampled data, as in \eqref{e:discrete_time} and \eqref{eq: DFT}, so we need to rewrite the above correlators in terms of the  \ac{DFT} data instead:
\begin{equation}
\begin{aligned}
&\langle \tilde{h}_{a;\ell}\tilde{h}_{b;m}^*\rangle \approx\frac{T}{2}\delta_{\ell m}\gamma_{ab;\ell}P_{\mathrm{gw};\ell}\,,
\\
&\langle \tilde{n}_{a;\ell}\tilde{n}_{b;m}^*\rangle \approx\frac{T}{2}\delta_{\ell m}\delta_{ab} P_{n_a;\ell}\,,\\
&\langle\tilde{h}_{a;\ell}\tilde{n}_{b;m}^*\rangle =0\,,
\end{aligned}
\label{eq: correlation_discrete}
\end{equation}
where $\delta_{\ell m}$ denotes the Kronecker delta function, and the one-sided discrete power spectra are defined by $P_\ell\equiv (2/T)\langle|\tilde n_\ell|^2\rangle$ and thus satisfy the property that $P_\ell = P_{N-\ell}$.
Note that first two lines above have “$\approx$” signs as opposed to  the “$=$” sign in~\eqref{eq: correlation_continuous}, since the finite duration DFT data are only \textit{approximately uncorrelated} for different frequency bins.
Following the discussion in App.~D of~\cite{Romano:2016dpx}, one can show that (see~(16) in \cite{allen-romano:2025}):
\begin{equation}
    \frac{2}{T}\langle \tilde{n}_\ell\tilde{n}_m^*\rangle\equiv P_{\ell m}= T\int_{-\infty}^{\infty}\D\!f \>P(f)\,\sinc[\pi(f-f_\ell)T]\sinc[\pi(f-f_m)T]\,,
    \label{eq: Plm_integral}
\end{equation}
where $P(f)=P(|f|)$ is the one-sided \ac{PSD} of the detector noise, and ${\rm sinc}\, x\equiv \sin x/x$ is the sinc function.
Thus, we see that the right-hand side of \eqref{eq: Plm_integral}
is not equal to 
$\delta_{\ell m} P_\ell$, in general.

But suppose that $P(f)=\mathrm{const}\equiv P$.
Then the $P_{\ell m}$ integral \eqref{eq: Plm_integral} has an exact analytical expression. 
This follows by noting that the $\sinc$ function and rectangular window function $\Pi_T(t)$ (defined below) form a Fourier transform pair:
\begin{equation}
    \Pi_T(t)\equiv
    \left\{
    \begin{aligned}
        &1,\qquad|t|\leqslant T/2,\\
        &0,\qquad \mathrm{otherwise},
    \end{aligned}
    \right.
    \qquad \int_{-\infty}^{\infty}\D t \>\Pi_T(t)e^{-2\pi ift}=T\,\sinc(\pi T f)\,.
\end{equation}
Thus, we can rewrite the integral as
\begin{equation}
    P_{\ell m}={PT}\frac{1}{T^2}\int_{-\infty}^{\infty}\D t\int_{-\infty}^{\infty} \D t'\> \Pi_T(t)\Pi_T(t')e^{2\pi i (f_\ell t+f_mt')}\int_{-\infty}^{\infty}\D\!f\>e^{-2\pi if(t+t')}\,,
\end{equation}
and then evaluate the right-hand side by noting that the integral over $f$ is simply $\delta(t+t')$.
Thus,
\begin{equation}
\begin{aligned}
     P_{\ell m}&=\frac{P}{T}\int_{-\infty}^{\infty}\D t\> \Pi_T(t)\Pi_T(-t)e^{2\pi i(f_\ell-f_m)t}\\
     &=\frac{P}{T}\int_{-T/2}^{T/2}\D t\>e^{2\pi i(f_\ell-f_m)t}\\
     &=P\,\sinc[\pi T(f_\ell-f_m)]\\
     &=P\,\sinc[\pi(\ell-m)]\\
     &=P\,\delta_{\ell m}\,.
\end{aligned}
\end{equation}

\begin{figure}[!htbp]
    \centering
    \includegraphics[width=0.45\linewidth]{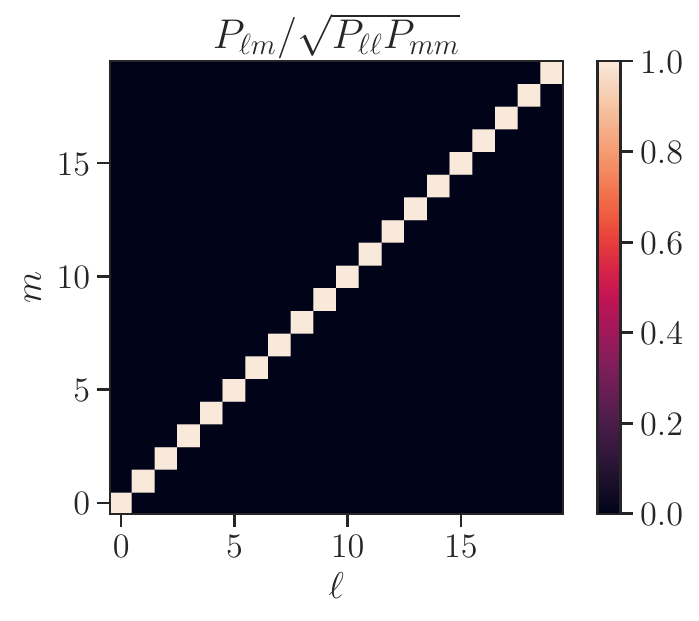}
    \includegraphics[width=0.45\linewidth]{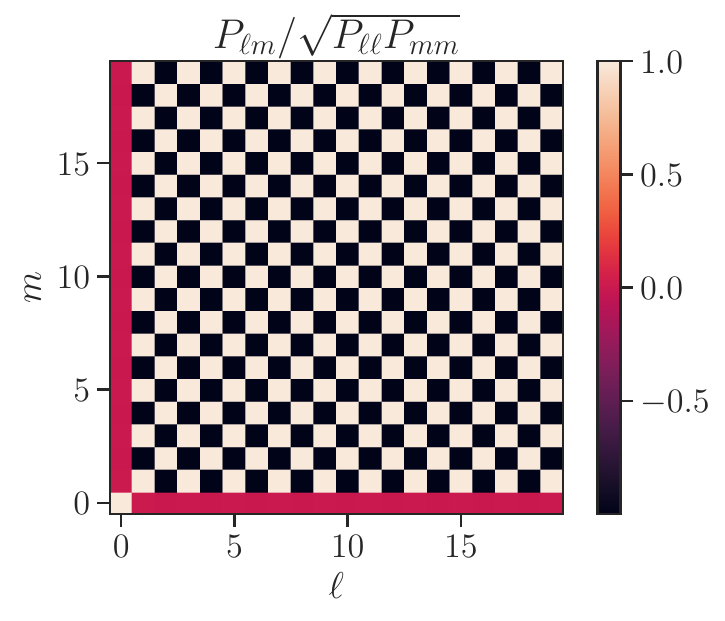}
    \caption{Normalized correlation matrix for the band-limited white noise example, where $P_{\ell m}$ is defined in~\eqref{eq: Plm_summation}. The left and right panels correspond to the limits $T\gg \tau$ and $T\ll \tau$, respectively, where $T$ is the observation time and $\tau$ is the signal correlation time. In the $T\gg\tau$ limit, the correlation matrix is strongly diagonal-dominated; in the $T\ll \tau$ limit, significant off-diagonal correlations emerge, indicating that correlations between different frequencies cannot be neglected. Since ground-based detectors operate in the $T\gg\tau$ regime, off-diagonal entries are safely negligible in practice.}
    \label{fig: P_lm_bandlimited}
\end{figure}

But when $P(f)$ is not constant, the matrix $P_{\ell m}$ is no longer strictly diagonal, and usually does not have a closed form. 
In practice, one needs to evaluate $P_{\ell m}$ numerically by replacing the integral in the above equation by a discrete sum
\begin{equation}
    P_{\ell m}\approx T\sum_i \delta\! f\>P(i\,\delta\!f)\,\sinc[\pi(i\,\delta\!f-\ell \Delta f)T]\sinc[\pi(i\,\delta\!f-m\Delta f)T]\,,
    \label{eq: Plm_summation}
\end{equation}
where $\delta\!f$ denotes the frequency grid spacing used to perform the summation. 

One must carefully choose $\delta\!f$ to ensure that the summation can accurately approximate the original integral. As implied by~\eqref{eq: Plm_integral}, two distinct time (and corresponding frequency) scales are present. The first is the observational duration $T$, which determines the frequency resolution $\Delta f\equiv 1/T$ of the \ac{DFT}. The second is the characteristic time scale associated with the \ac{PSD} $P(f)$, commonly referred to as the correlation time $\tau$. The underlying time series $n(t)$ is assumed to be correlated only for time separations $|t-t'|\lesssim \tau$. Consequently, the characteristic frequency scale over which $P(f)$ varies is given by $\Delta f_0=1/\tau$. To ensure that the summation faithfully reproduces the continuum integral, the frequency grid spacing must satisfy $\delta\!f\lesssim \mathrm{min}(\Delta f,~\Delta f_0)$. 

To further illustrate the dependence of $P_{\ell m}$ on the correlation time $\tau$ and the observational duration $T$, we  consider two specific examples: 
\begin{enumerate}[label=(\roman*)]
\item 
band-limited white noise:
\begin{equation}
    P(f)=
    \begin{cases}
    C, &|f|\leqslant f_0={1}/{\tau},\\
    0, &\mathrm{otherwise}.
    \end{cases}
\end{equation}

\item 
Gaussian-shaped \ac{PSD}:
\begin{equation}
    P(f)=e^{-\frac{1}{2}\left(\frac{f}{f_0}\right)^2}\,,\quad\text{where}\quad f_0=\frac{1}{\tau}.
\end{equation}
\end{enumerate}
For each example, we study two limiting regimes: (a) $T\gg \tau$, which is relevant for ground-based detectors, and (b) $T\ll\tau$, which is the case for \ac{PTA} analyses.
\begin{figure}[!htbp]
    \centering
    \includegraphics[width=0.45\linewidth]{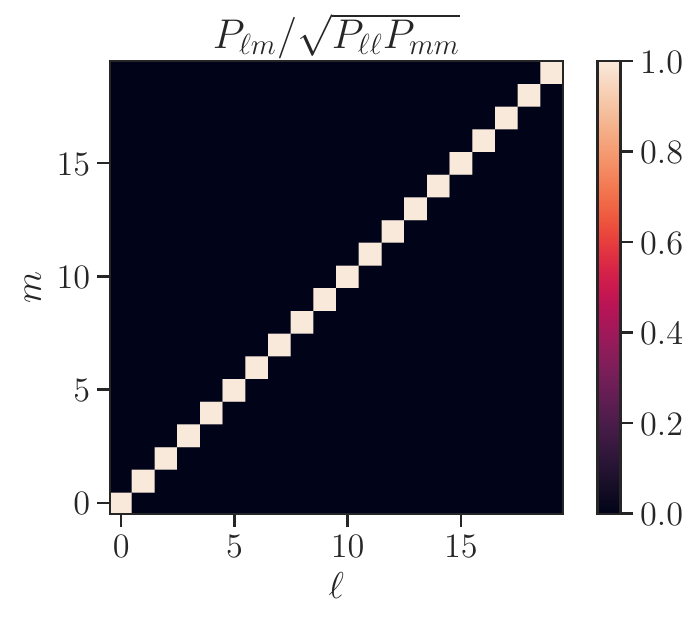}
    \includegraphics[width=0.45\linewidth]{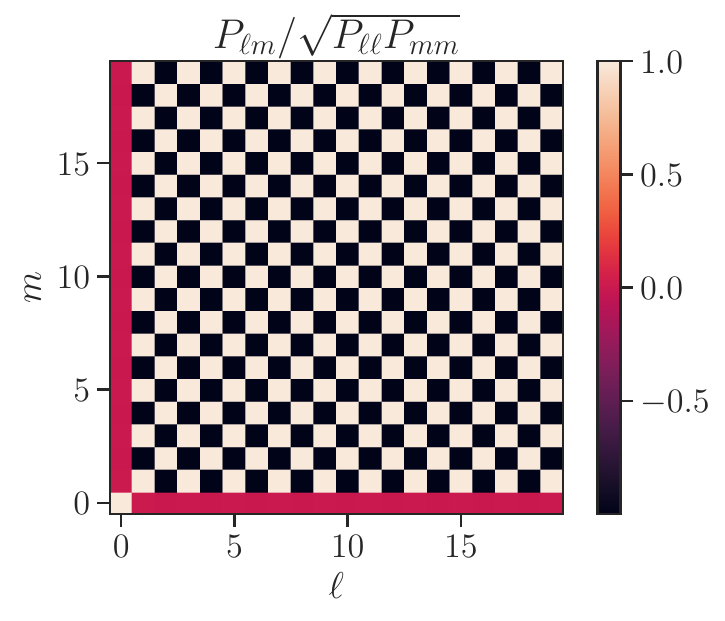}
    \caption{Same as Fig.~\ref{fig: P_lm_bandlimited}, but for the Gaussian-shaped \ac{PSD} example.}
    \label{fig: P_lm_Gaussian}
\end{figure}

For the regime $T\gg \tau$, we set $T=1$ and $\tau=10^{-2}$, while for the opposite limit, we take $T=1$ and $\tau=100$ (both in arbitrary units).
This is done for both examples. 
The results for two examples are shown in Figs.~\ref{fig: P_lm_bandlimited} and~\ref{fig: P_lm_Gaussian}, respectively. 
The first row of both figures shows the correlation matrix $P_{\ell m}/\sqrt{P_{\ell\ell}P_{mm}}$ and the relative numerical uncertainty matrix $\epsilon_{\ell m}/P_{\ell m}$ for the $T\gg \tau$ limit, while the second row shows the results for the opposite limit of $T\ll \tau$. 
Both figures indicate that when $T\gg \tau$, it is a very good approximation to ignore the off-diagonal entries of the $P_{\ell m}$ matrix, so the “$\approx$” sign that we used in \eqref{eq: correlation_discrete} is valid to very high precision. 
However, when $\tau\gg T$, the correlation between different frequency bins cannot be neglected; hence one must carefully model the entire $P_{\ell m}$ matrix.

This paper focuses only on ground-based detectors, in which case $T\gtrsim 10~\mathrm{s}\gg 10~\mathrm{ms} \simeq \tau$ is always valid. 
Thus, it suffices to restrict our attention to the diagonal entries of the $P_{\ell m}$ matrix throughout this work. We note that in practice, the estimated \ac{PSD} contains contributions from both signal and noise. Narrowband signals, such as continuous waves, have a long correlation time, which would invalidate this approximation. Since such sources are not the primary target of our isotropic \ac{SGWB} search, we leave this complication for future investigations.

\subsection{Weak-signal limit}
\label{s:weak-signal}
For all of our analyses, we will also assume that we are in the \textit{weak-signal limit}, in which the \ac{SGWB} of interest is assumed to be much weaker than the detector noise, i.e., $P_{n_a}(f)\gg P_\mathrm{gw}(f)$ for all frequencies in the analysis band. 
We stress that this assumption applies only to truly persistent stochastic signals that give rise to a \ac{SGWB}. 
Deterministic and individually resolvable signals, such as \ac{CBC} events, are not included in this context. 
When such deterministic signals are sufficiently strong, they can be detected and individually resolved, and subsequently removed from the data. 
As a result, the unresolved signal together with other \ac{SGWB} signals may still be treated collectively as a stochastic background.

\begin{figure}[!htbp]
    \centering
    \includegraphics[width=0.7\linewidth]{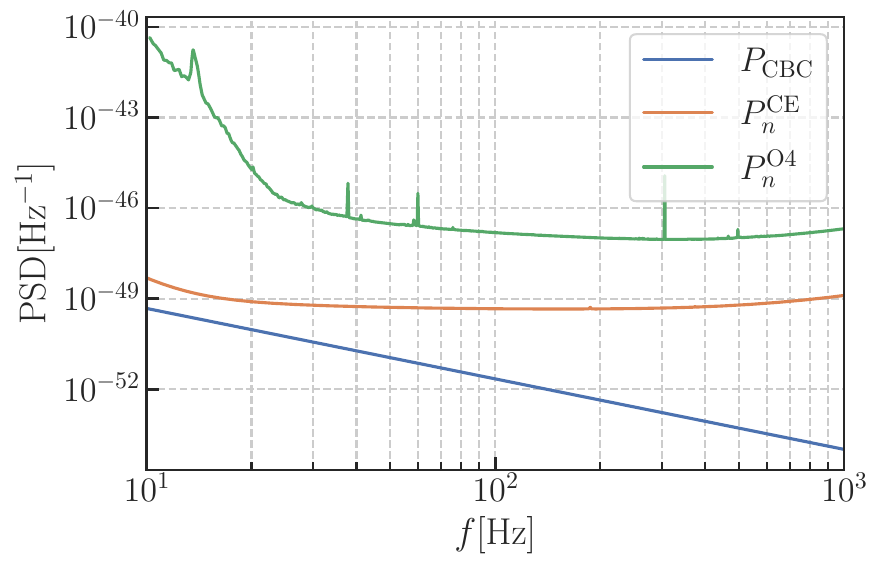}
    \caption{Noise power spectral densities for O4, Cosmic Explorer, and the unresolved background from a population of compact binary coalescence events.
    Since the power-spectral density for CBCs lies everywhere below that for the O4 and Cosmic Explorer detector noise, the weak-signal approximation is a good one.}
    \label{fig: Weak_Signal_limit}
\end{figure}

In Fig.~\ref{fig: Weak_Signal_limit}, we explicitly compare the \ac{PSD} of a \ac{CBC} background with the detector noise \ac{PSD} of a LIGO detector operating at its design sensitivity for the fourth LIGO-Virgo-KAGRA observation run (O4)~\cite{O4_PSD}, as well as the expected sensitivity of a \ac{CE} detector~\cite{CE2_PSD}.
To compute $P_\mathrm{CBC}(f)$, we assume that the energy density spectrum of the \ac{CBC} background can be modeled by a single power-law
\begin{equation}
    \Omega_\mathrm{CBC}(f)=\Omega_\mathrm{ref}\left(\frac{f}{f_\mathrm{ref}}\right)^{2/3}\,,\qquad f_\mathrm{ref}=25~\mathrm{Hz}\,.
\end{equation}
We consider $\Omega_\mathrm{ref}=6\times 10^{-10}$, which is consistent with the latest upper limit set by the \ac{LVK} Collaboration~\cite{O4aStoch}. 
One can see from Fig.~\ref{fig: Weak_Signal_limit} that for both the LIGO and CE detectors, the noise \ac{PSD} is significantly larger than that of the underlying \ac{CBC} background, and is therefore also likely to exceed the contribution from other cosmological sources~\cite{Renzini:2022alw,Caldwell:2022qsj}. 
Consequently, even for \ac{XG} detectors, it is appropriate to assume that the analysis is performed in the weak-signal limit.

We highlight that in this limit, the results we will derive for two detectors can be extended trivially to multiple detectors, because correlations between different baselines are at most of order $P_\mathrm{gw}P_n$, which is much smaller than that within the same baseline, which scales as $P_n^2$. 
See App.~\ref{s:ext-mult-det} and also footnote~36 of Ref.~\cite{Allen:1997ad} for more details.

\subsection{Narrowband estimator}
\label{s:narrowband}

The starting point for the remainder of our analyses is the narrowband estimator
\begin{equation}
    \hat{C}_\ell\equiv\frac{10}{T}\frac{\mathfrak{R}\left(\tilde{\bf{d}}_{1;\ell}\tilde{\bf{d}}^*_{2;\ell}\right)}{\gamma_{12;\ell}S_{0;\ell}}\,,
    \label{eq: C_l_def}
\end{equation}
where $\tilde{\bf{d}}_{1;\ell}$ and $\tilde{\bf{d}}_{2;\ell}$ denote the \ac{DFT} of the \textit{windowed} data $w_id_i$ evaluated at the $\ell$th frequency bin, i.e., $f_{\ell}=\ell\Delta\!f$. 
See~\eqref{e:Omega_hat_estimator} in App.~\ref{app: zero_padding} for justification for this form of the estimator~\footnote{Although App.~\ref{app: zero_padding} discusses zero-padding, we will ignore zero-padding in the main text, since it makes only minor modifications to the final formulae.}.

In a real search using ground-based detectors, it is common to divide the total observation (of duration $T_{\rm obs}$) into many segments of duration $T$.
There are several practical reasons for this including data storage efficiency, non-stationarity of the time series over long-periods, frequency resolution of the analysis, etc.
Therefore, we should replace $\hat{C}_\ell$ with $\hat{C}_{I;\ell}$ in order to explicitly label which time segment of data (denoted here with index $I$) is being used to compute the estimator.

Finally, in the above discussion, the frequency resolution of the analysis is simply determined by the duration $T$ of the segment used for the \ac{DFT}. 
However, if the power spectra are expected to be smooth over a coarser frequency resolution $\delta f=M\Delta f$ (where $M\sim\mathcal{O}(10)\ll N\sim\mathcal{O}(10^4\sim 10^5)$, with $N$ denoting the number of data points contained in each segment), then we can define a \textit{coarse-grained} spectral estimator~\cite{Matas:2020roi}%
\footnote{If $M$ is odd, then we should replace $M/2$ with $\lfloor{M}/{2}\rfloor$. $\lfloor \cdot\rfloor$ denotes the floor function. By definition, the floor function satisfies $\lfloor x\rfloor\leqslant x< \lfloor x\rfloor+1$.}
\begin{equation}
    \bar{C}_{I;\bar{\ell}}\equiv\frac{1}{M}\sum_{m=M\bar{\ell}-M/2}^{M\bar{\ell}+M/2-1}\hat{C}_{I;m}
    =\frac{1}{M}\sum_{m=M\bar{\ell}-M/2}^{M\bar{\ell}+M/2-1}\frac{10}{T}\frac{\mathfrak{R}\left(\tilde{\bf{d}}_{1;I;m}\tilde{\bf{d}}^*_{2;I;m}\right)}{\gamma_{12;m}S_{0;m}}\,,\,.
    \label{eq: C_lbar:_def}
\end{equation}
Note that to distinguish the indices for the original (fine-grained) spectrum and the new coarse-grained spectrum, we have added a bar to the estimator and the indices, where $\bar{\ell}$ can run over all integers from 0 to $\lfloor N/M\rfloor-1$. In this case, we have $f_{\barell}=\barell\delta\!f=M\barell\Delta\!f$.

\section{Roadmap of calculations}
\label{s:roadmap}
\begin{figure}[!htbp]
    \centering
    \includegraphics[width=0.95\linewidth]{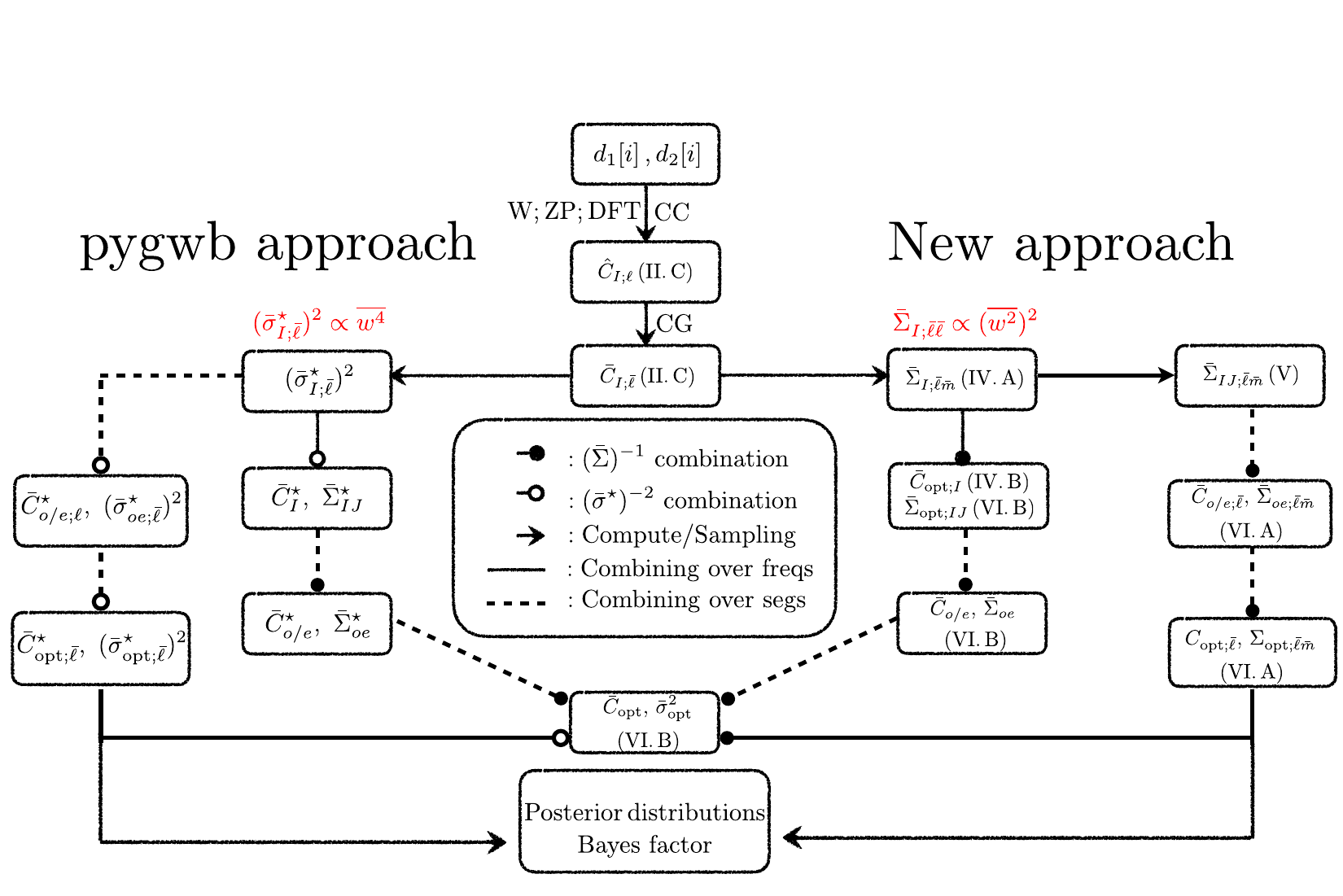}
    \caption{Flowchart comparing the approach adopted in the \texttt{pygwb} package~\cite{pygwb} with the new approach presented in this work. Quantities adopted by \texttt{pygwb} are marked with a superscript $\star$ to distinguish them from the corresponding quantities introduced here. For the quantities adopted in the new approach, we also indicate the section numbers in which corresponding quantities are defined or derived. Here, W and ZP denote windowing and zero-padding, respectively, while CC and CG correspond to cross-correlation and coarse-graining. Two relations highlighted in red indicate that the variance of the coarse-grained narrowband estimator derived in our new approach is proportional to the factor of $(\overline{w^2})^2$ (see~\eqref{eq: w2}), whereas it is proportional to $\overline{w^4}$ in \texttt{pygwb} (see (A1) of \cite{pygwb}). The definition of these two window factors, and the reason for this difference will be discussed later in the text.}
    \label{fig: flowchart}
\end{figure}


Before diving into the detailed derivations, we first present in Fig.~\ref{fig: flowchart} a “roadmap” outlining the structure of the calculations we perform in the following three sections. 
Our analysis starts from the original strain time-series $d_1$ and $d_2$ from two detectors, which are then windowed, zero-padded, discrete Fourier transformed, and cross-correlated to obtain $\hat C_{I;\ell}$. 
These quantities are subsequently coarse-grained to yield $\bar C_{I;\barell}$. 

At this stage, the flowchart branches into two paths. The left branch illustrates the procedure adopted by \texttt{pygwb}~\cite{pygwb}, which is the flagship Python package currently adopted by the \ac{LVK} collaboration for the stochastic search and corresponding data analysis, and is also extensively used by the broader community.
The quantities adopted in \texttt{pygwb} are marked by a superscript $\star$ to distinguish them from those used in our approach. 
The right branch presents our new approach, including the quantities that we will define or derive later in the text. 
For clarity, we also indicate the section numbers in which these quantities are discussed in detail.

The following three sections explain how to construct optimal cross-correlation estimators, combined over time segments, discrete frequency bins, or both.
In these sections, we will often refer to App.~\ref{app: stats} for useful definitions and results in statistics, since we will make heavy use of Isserlis's theorem (App.~\ref{app: Isserlis}) and a general method of constructing optimal estimators (App.~\ref{app: optimal_estimator}).
Although we should zero-pad the data before performing the \ac{DFT} (as described in App.~\ref{app: zero_padding}), we will ignore zero-padding in the main text, and explain in App.~\ref{s:ext-zp-data} how to adjust our key results to include the effects of zero-padding. 
Finally, some of the more lengthy technical calculations in these sections have been put into Apps.~\ref{s:single_seg_details}, \ref{s:covariance_details}, and \ref{s:multi_seg_details}.

Table~\ref{tab: listofsymbols} gives a list of symbols, together with a short definition and equation numbers for the most important quantities discussed in this paper.

\begingroup
\renewcommand{\arraystretch}{1.3}

\setlength{\LTcapwidth}{\linewidth}
\begin{longtable}[!htbp]{|p{1.35in}|p{4.83in}|p{0.78in}|}
\caption{List of symbols, including equation numbers for the most important quantities
discussed in the paper.}\\
\hline
symbol & \centering{brief description} & (equation) \\
\hline
\endfirsthead

\multicolumn{3}{l}{\tablename~\thetable\ (continued)}\\
\hline
symbol & \centering{brief description} & (equation) \\
\hline
\endhead

\hline
\multicolumn{3}{r}{\textit{continued on next page}}\\
\endfoot

\hline
\endlastfoot

$P_{n_1}(f)$, 
$P_{n_2}(f)$, 
$P_{\rm gw}(f)$
& power spectral densities (PSDs) for detector noise and detector response to the GWB &\eqref{eq: correlation_discrete} \\

$\Omega_{\rm gw}(f)$, $\Omega_{\rm ref}$, $\alpha$ 
& energy-density spectrum of the GWB, its amplitude at $f_{\rm ref}$, and spectral index &\eqref{eq: omega_gw}, \eqref{eq: omega_powerlaw} \\

$\gamma_{12}(f)$, 
$S_0(f)$ 
& overlap reduction function for the two detectors, and shape of strain PSD for $\alpha=0$ &\eqref{eq: correlation_continuous}, \eqref{eq: S_0_def}\\

$T$, $T_{\rm obs}$ 
& segment duration, duration of total observation & \\

$N$, $N_{\rm seg}$, $M$
& number of time samples per segment, number of segments,  coarse-graining factor & \\

$\Delta f$, $\delta f$
& frequency resolution for full and coarse-grained data, respectively & \\

$d_{1;I;j}$, $d_{2;I;j}$ 
& time-domain data from detectors 1 and 2, time segment $I$, time sample $j$ & \\

$\tilde d_{1;I;\ell}$, $\tilde d_{2;I;\ell}$ 
& Fourier-domain data from detectors 1 and 2, time segment $I$, freq bin $\ell$ & \\

$\tilde d_{1;I;\barell}$, $\tilde d_{2;I;\barell}$ 
& Fourier-domain data from detectors 1 and 2, time segment $I$, coarse-grained freq bin $\barell$ & \\

$w_j$, $\tilde w_\ell$, $\tilde w_\barell$ 
& same as above, but for the window function (independent of detector and time segment) & \\

$\tilde{\bf{d}}_{1;I;\ell}$, $\tilde{\bf{d}}_{2;I;\ell}$ 
& DFT of windowed time-series data from detectors 1 and 2, segment $I$, freq $\ell$ & \\

$\mathcal{K}_{j,k;M,N}$
& square of Dirichlet kernel 
& \eqref{e:K_def_app}
\\

$\overline{w^2}$, 
$\overline{w^4}$, 
$\overline{w_{\rm ovl}^4}$
& various quadratic and quartic window factors 
& \eqref{eq: w2}, \eqref{e:w4_bar}, \eqref{e:w4_ovl_def}
\\

$\Big(\overline{w_{\mathcal{K}}^2}\Big)^2$, 
$\Big(\overline{w_{\mathcal{K}, {\rm ovl}}^2}\Big)^2$
& various window factors involving $\mathcal{K}_{j,k;M,N}$ 
& \eqref{e:wK2bar2}, \eqref{e:knb}
\\

$\mathcal{W}$, $\mathcal{W}_{\mathcal{K}}$
& ratios of window factors 
& \eqref{eq: firstway}, \eqref{eq: secondway}
\\

$r_M$, $k_{\rm{nb}}$, $k_{{\rm bb}}$ 
& more ratios of window factors 
& \eqref{eq: def_r_M}, \eqref{e:knb}, \eqref{e:kbb}
\\

$P_{1;I;\ell}$, 
$P_{2;I;\ell}$, 
$P_{1;I;\barell}$
$P_{2;I;\barell}$
& total auto-correlated power in detectors 1 or 2, time segment I, freq $\ell$ or $\barell$
& \eqref{e:P12_p}
\\

$\mathcal{P}_{I;\ell}$,
$\mathcal{P}_{I;\barell}$
& quadratic combination of detector and GWB PSDs and overlap reduction function, 
for time segment $I$, and freq $\ell$ or coarse-grained freq $\barell$ 
& \eqref{eq: P_p}
\\

$\hat C_{I;\ell}$, $\hat\sigma^2_{I;\ell}$, $\hat \Sigma_{I;\ell m}$, $\hat \Sigma_{IJ;\ell m}$
& narrowband cross-corr estimator, its variance, and covariance for time segments $I$, $J$ and freq bins $\ell$, $m$ 
& \eqref{eq: C_l_def}, 
\eqref{eq: var_final}, 
\eqref{eq: hat_Sigma_pq-final}, 
\eqref{e:sighat_IJ_lm}\\

$\bar C_{I;\barell}$, $\bar\sigma^2_{I;\barell}$, $\bar\Sigma_{I;\barell\barm}$, $\bar\Sigma_{IJ;\barell\barm}$
& same as above but for coarse-grained freqs $\barell$, $\barm$ 
& \eqref{eq: C_lbar:_def}, 
\eqref{eq: def_r_M},
\eqref{eq: bar_Sigma_l_m_final},
\eqref{eq: bar_Sigma_ovl_lm_complete}
\\

$\bar C_{\opt;I}$,
$\bar\sigma^2_{\opt;I}$,
$\bar\Sigma_{\opt; IJ}$ 
& optimal broadband estimator, its variance, and covariance combined over all coarse-grained freq bins, 
for time segments $I$, and $I$, $J$, respectively 
& \eqref{eq: C_opt_var_opt_I_non_coase_grained},
\eqref{e:sigbar_opt_IJ}
\\

$\bar C_{o;\barell}$, 
$\bar C_{e;\barell}$ 
& optimal narrowband estimator for coarse-grained freq bin $\barell$, combined over all odd or even time segments, respectively 
& \eqref{eq: bar_C_o_l_bar_var_o_l_bar_C_e_l_bar_var_e_l}
\\

$\bar\sigma^2_{o;\barell}$, 
$\bar\sigma^2_{e;\barell}$, 
$\bar\sigma^2_{oe;\barell}$,
$\tilde\sigma^{-2}_\barell$
& components of the covariance matrix $\bar\Sigma_{oe;\barell}$ for $\bar C_{o;\barell}$, $\bar C_{e;\barell}$,
and useful combination of inverse variances
& \eqref{eq: bar_C_o_l_bar_var_o_l_bar_C_e_l_bar_var_e_l},
\eqref{eq: bar_var}
\\

$\bar C_{\opt; \barell}$,
$\bar\sigma^2_{\opt; \barell}$
& optimal narrowband estimator and its variance for coarse-grained freq bin $\barell$, combined over all
time segments 
& \eqref{e:final_multiseg_nb}
\\

$\bar\Sigma_{oo;\barell\barm}$,
$\bar\Sigma_{ee;\barell\barm}$,
$\bar\Sigma_{oe;\barell\barm}$ 
& components of the covariance matrix $\bar\Sigma_{\opt;\barell\barm}$ for the optimal narrowband estimator for coarse-grained freq 
bins $\barell$, $\barm$, combined over all odd, even, and odd/even time segments 
& \eqref{e:final_multiseg_nb}
\\

$\bar C_{o}$, $\bar C_{e}$ 
& optimal broadband estimators, combined over all coarse-grained freq bins and odd or even time segments, respectively 
& \eqref{e:multiseg_C_sigma_defs}
\\

$\bar\sigma^2_{o}$, 
$\bar\sigma^2_{e}$, 
$\bar\sigma^2_{oe}$, 
$\tilde\sigma^{-2}$
& components of the covariance matrix $\bar\Sigma_{oe}$ for $\bar C_o$, $\bar C_e$,
and useful combination of inverse variances
& \eqref{e:multiseg_C_sigma_defs},
\eqref{e:sigbar_oe},
\eqref{e:kbb}
\\

$\bar C_\opt$,
$\bar\sigma^2_\opt$
& optimal broadband estimator and its variance combined over all coarse-grained freq bins and time segments 
& \eqref{eq: final_broadband}
\\

$\mathcal{C}_{w;\ell m}$,
$\bar{\mathcal{C}}_{w;\barell \barm}$, 
$\mathcal{C}_{w,{\rm ovl};\ell m}$,
$\bar{\mathcal{C}}_{w,{\rm ovl};\barell \barm}$
& parts of the covariance matrices that capture the non-zero covariance between different freq bins 
(depends only on windows, independent of detectors and segments) 
& 
\eqref{e:C_wpq},
\eqref{eq: def_of_bar_C_w},
\eqref{eq: def_C_w_ovl},
\eqref{eq: def_bar_C_w_ovl}
\\

$w_I$, $\hat w_I$ 
& optimal weights and estimated weights for segment $I$ (using estimated detector PSDs) 
& \eqref{e:Copt_ideal}, \eqref{e:Copt_real}
\\

$\hat C_{\rm real}$, $\sigma^2_{\rm real}$, $\hat \sigma^2_{\rm real}$
& realistic combined estimator, its variance, and estimator of the variance 
& \eqref{e:Copt_real}
\\

$b$, $N_{\rm eff}$
& bias factor and number of effective averages for detector PSD estimation 
& \eqref{e:mean_w_hat-main}
\label{tab: listofsymbols}
\end{longtable}
\endgroup

\section{Single-segment optimal estimators}
\label{s:single_segment}

In this section, we discuss the statistical properties of optimal (i.e., linear, unbiased, and minimum-variance) narrowband and broadband estimators using single-segment narrowband estimators. 
Since, we will only be considering quantities that correspond to a single time segment $I$, we will temporarily drop the the $I$ index for notational simplicity and reintroduce it in the final ``boxed" equations, which we will often refer to in the later sections.
For example, we will use $\bar{C}_{\barell}$ as shorthand notation for $\bar{C}_{I;\barell}$.

\subsection{Narrowband estimator}
\label{s:single_seg_narrow}

\begin{figure}[!htbp]
    \centering
    \includegraphics[width=0.5\linewidth]{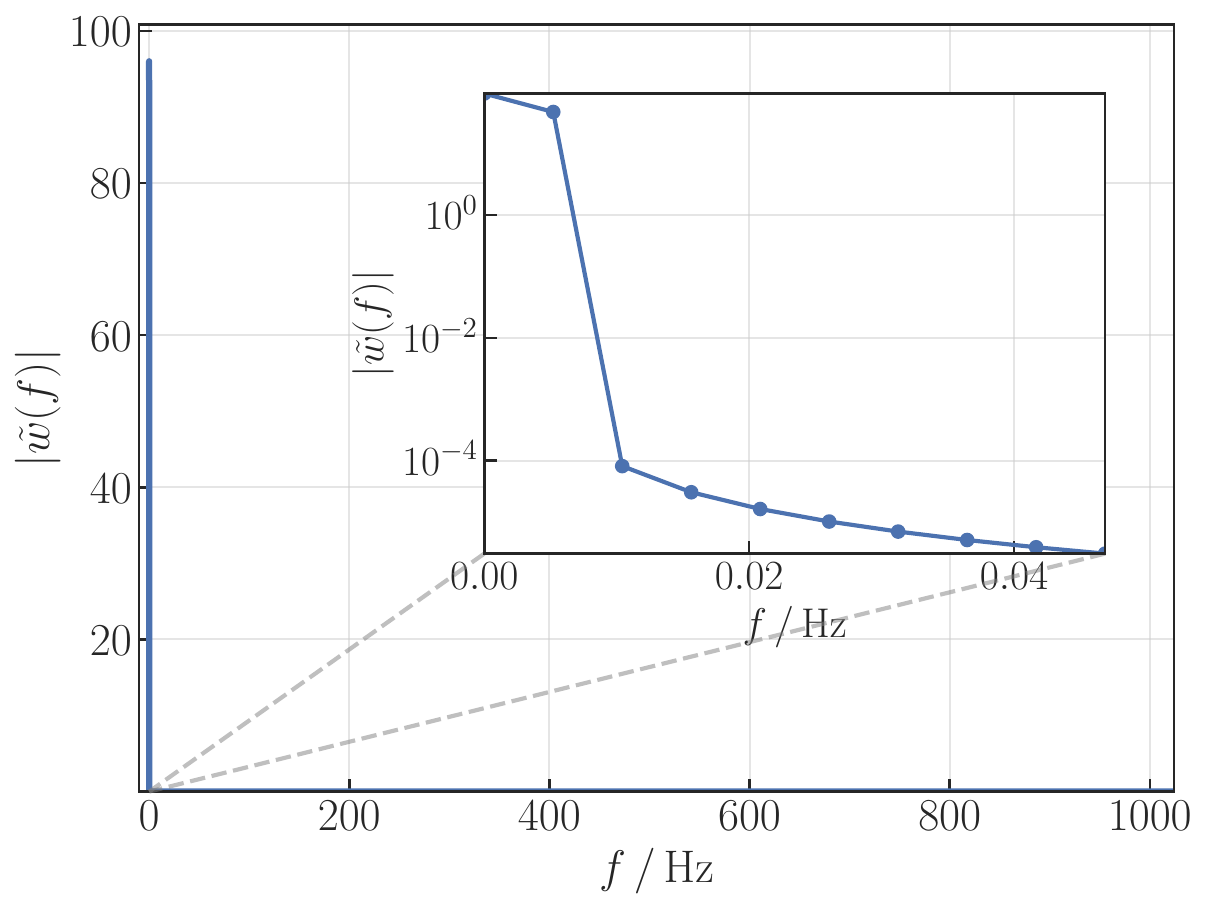}
    \caption{Frequency-domain representation of a Hann window of duration 192~sec. 
    We show the behavior of the window function for low frequencies in the inset (note the logarithmic scale), indicating that $\tilde{w}(f)$ is nonzero only when $f\approx 0$.}
    \label{fig: Hann}
\end{figure}

We start by deriving the mean of the single-segment narrowband estimator $\bar{C}_{\barell}$.
By definition
\begin{equation}
    \langle \bar{C}_{\barell}\rangle=\frac{1}{M}\sum_{m=M\bar{\ell}-\frac{M}{2}}^{M\bar{\ell}+\frac{M}{2}-1}\langle\hat{C}_{m}\rangle=\frac{1}{M}\sum_{m=M\barell-\frac{M}{2}}^{M\barell+\frac{M}{2}-1}\left\langle \frac{10}{T}\frac{\mathfrak{R}\left(\tilde{\bf{d}}_{1;m}\tilde{\bf{d}}^*_{2;m}\right)}{\gamma_{12;m}S_{0;m}}\right\rangle,
    \label{eq: Mean_bar_C_Il}
\end{equation}
which requires us to evaluate the expected value of $\mathfrak{R}\left(\tilde{\bf{d}}_{1;m}\tilde{\bf{d}}^*_{2;m}\right)$.
Recall, $\tilde{\bf{d}}_{a;\ell}$ denotes the Fourier transform of the windowed time-series data for detector $a=\{1,2\}$ in time segment $I$, with the $I$ indexed suppressed.
Using the definition of the \ac{DFT}~\eqref{eq: DFT}, it follows that 
\begin{equation}
    \tilde{\bf{d}}_{\ell}=\frac{1}{T}\sum_{j=0}^{N-1}\tilde{d}_j\tilde{w}_{\ell-j}\,,
    \label{eq: bm_d}
\end{equation}
where we use the cyclic property of the DFT to evaluate $\tilde{w}_{\ell-j}$ when the index $\ell-j$ is outside the range 0 to $N-1$.
Then, using the above relation twice, we obtain 
\begin{equation}
\begin{aligned}
    \langle \tilde{\bf{d}}_{1;\ell}\tilde
    {\bf{d}}_{2;\ell}^*\rangle &=\frac{1}{T^2} \sum_{j=0}^{N-1}\sum_{k=0}^{N-1} \langle \tilde{d}_{1;j}\tilde{d}_{2;k}^*\rangle \tilde{w}_{\ell -j}\tilde{w}^*_{\ell -k}\\
    &\approx\frac{1}{2T} \sum_{j=0}^{N-1}\gamma_{12;j}P_{\mathrm{gw};j} \tilde{w}_{\ell -j}\tilde{w}^*_{\ell -j}\,,
\end{aligned}
\label{eq: dd}
\end{equation}
where we used the Kronecker delta in the first equation of~\eqref{eq: correlation_discrete} to reduce the double sum to a single sum. 
Since the window function has significant support only when $\ell\approx j$ (see Fig.~\ref{fig: Hann} and Ref.~\cite{Note_2}) and both $\gamma_{12;j}$ and $P_{\mathrm{gw};j}$ are expected to be slowly varying functions of frequency%
\footnote{Since the characteristic time scale $\tau$ is $\sim 10~\mathrm{ms}$, we do not expect $\gamma_{12}(f)$, $P_\mathrm{gw}(f)$, and $P_{n_a}(f)$ to vary substantially over frequency intervals $\delta\!f\sim \mathcal{O}(0.01-0.1)\mathrm{Hz}$, so this should be a good approximation.}, 
we have
\begin{align}
    \langle \tilde{\bf{d}}_{1;\ell}\tilde
    {\bf{d}}_{2;\ell}^*\rangle \approx \frac{1}{2T}\gamma_{12;\ell}P_{\mathrm{gw};\ell}\sum_{j=0}^{N-1}\tilde{w}_{\ell-j}\tilde{w}^*_{\ell-j}\,.
    \label{eq:d1d2}
\end{align}
The summation of window functions in the above expression can be related to mean-squared value $\overline{w^2}$ of the window function via 
\begin{align}
    \sum_{j=0}^{N-1}\tilde{w}_{\ell-j}\tilde{w}^*_{\ell-j}= T^2\overline{w^2}\,,\qquad \overline{w^2}\equiv\frac{1}{N}\sum_{p=0}^{N-1}w^2_p\,,
    \label{eq: w2}
\end{align}
which is proved in~App.~\ref{app: proof_w2}. 
Thus,%
\footnote{If different windows were applied to the different time-series data from the two detectors, we would need to replace $\overline{w^2}$ by $\overline{w_1w_2}$. 
But since we use the same window function for both detectors for our analyses, we  have $w_1=w_2\equiv w$.}
\begin{equation}
     \langle \tilde{\bf{d}}_{1;\ell}\tilde
    {\bf{d}}_{2;\ell}^*\rangle \approx \frac{T}{2}\gamma_{12;\ell}P_{\mathrm{gw};\ell}\,\overline{w^2}
    \quad\implies \quad
    \langle \hat{C}_{\ell}\rangle \approx \Omega_{\mathrm{gw};\ell}\,\overline{w^2}\,.
    \label{eq: Exp_C}
\end{equation}
Plugging~\eqref{eq: Exp_C} back into~\eqref{eq: Mean_bar_C_Il} yields
\begin{tcolorbox}
[title=Expected value of $\bar{C}_{I;\barell}$]
    \begin{equation}
    \langle \bar{C}_{I;\barell}\rangle=\frac{1}{M}\sum_{m=M\barell-\frac{M}{2}}^{M\barell+\frac{M}{2}-1}\Omega_{\mathrm{gw};m}\overline{w^2}\approx \Omega_{\mathrm{gw};\barell}\,\overline{w^2}\,.
\end{equation}
\end{tcolorbox}

Now consider the covariance between $\bar{C}_{\barell}$ and $\bar{C}_{\barm}$, which is denoted by $\bar{\Sigma}_{\barell\barm}$. By definition:
\begin{equation}
\begin{aligned}
\bar{\Sigma}_{\barell\barm}&\equiv \langle \bar{C}_{\barell}\bar{C}_{\barm}\rangle-\langle \bar{C}_{\barell}\rangle\langle\bar{C}_{\barm}\rangle\\
&=\frac{1}{M^2}\sum_{p=M\barell-\frac{M}{2}}^{M\barell+\frac{M}{2}-1}\sum_{q=M\barm-\frac{M}{2}}^{M\barm+\frac{M}{2}-1}\left(\langle \hat{C}_p\hat{C}_q\rangle - \langle \hat{C}_p\rangle \langle \hat{C}_q\rangle\right)\\
&=\frac{1}{M^2}\sum_{p=M\barell-\frac{M}{2}}^{M\barell+\frac{M}{2}-1}\sum_{q=M\barm-\frac{M}{2}}^{M\barm+\frac{M}{2}-1}\hat{\Sigma}_{pq}\,,
\end{aligned}
\label{eq: bar_Sigma_lm}
\end{equation}
where we used the definition of the covariance $\hat{\Sigma}_{pq}$ between $\hat{C}_p$ and $\hat{C}_q$ to obtain the last line. 
Using~\eqref{eq: C_l_def} and Isserlis's theorem, one can expand $\hat{\Sigma}_{pq}$ as follows:
\begin{equation}
\begin{aligned}
    \hat{\Sigma}_{pq}=\frac{25}{T^2\gamma_{12;p}\gamma_{12;q}S_{0;p}S_{0;q}}\Bigg(&\langle\tilde{\bf{d}}_{1;p}\tilde{\bf{d}}_{2;q}^*\rangle
\langle\tilde{\bf{d}}_{2;p}^*\tilde{\bf{d}}_{1;q}\rangle
+\langle\tilde{\bf{d}}_{1;p}\tilde{\bf{d}}_{1;q}^*\rangle\langle\tilde{\bf{d}}^*_{2;p}\tilde{\bf{d}}_{2;q}\rangle\\
&+\langle\tilde{\bf{d}}^*_{1;p}\tilde{\bf{d}}_{2;q}^*\rangle\langle\tilde{\bf{d}}_{2;p}\tilde{\bf{d}}_{1;q}\rangle+\langle\tilde{\bf{d}}^*_{1;p}\tilde{\bf{d}}_{1;q}^*\rangle\langle\tilde{\bf{d}}_{2;p}\tilde{\bf{d}}_{2;q}\rangle+({\rm c.c.})\Bigg),
\end{aligned}
\label{eq: hat_Sigma_pq}
\end{equation}
where $({\rm c.c.})$ denotes complex conjugate. 
The next step is to calculate all the two-point correlations shown above, which is straightforward but lengthy. 
We calculate the first term to illustrate the method, and then simply quote the results for other terms later on. 

So, let us consider $\langle\tilde{\bf{d}}_{1;p}\tilde{\bf{d}}_{2;q}^*\rangle$.
Using the definition of the Fourier-transformed windowed data, we have
\begin{equation}
\begin{aligned}
\langle\tilde{\bf{d}}_{1;p}\tilde{\bf{d}}_{2;q}^*\rangle&=\frac{1}{T^2}\sum_{j=0}^{N-1}\sum_{k=0}^{N-1}\langle \tilde{d}_{1;j}\tilde{d}_{2;k}^*\rangle \tilde{w}_{p-j}\tilde{w}^*_{q-k}\\
&=\frac{1}{T^2}\sum_{j=0}^{N-1}\sum_{k=0}^{N-1}\frac{T}{2}\gamma_{12;j}P_{\mathrm{gw};j}\delta_{jk}\tilde{w}_{p-j}\tilde{w}^*_{q-k}\\
&\approx\frac{1}{2T}\gamma_{12;p}P_{\mathrm{gw};p}\sum_{j=0}^{N-1}\tilde{w}_{p-j}\tilde{w}^*_{q-j}\,,
\label{eq: trick}
\end{aligned}
\end{equation}
where we again used~\eqref{eq: correlation_discrete} to obtain the second line. 
We then used the property of the window function (illustrated in Fig.~\ref{fig: Hann}) to pull $\gamma_{12;j}$ and $P_{\mathrm{gw};j}$ out of the summation, replacing the index $j$ with $p$. 
The summation of two window functions above does not yield $\overline{w^2}$, as the windows now have different indices.
But, as shown in App.~\ref{app: proof_wp_wq}:
\begin{equation}
    \sum_{j=0}^{N-1}\tilde{w}_{p-j}\tilde{w}_{q-j}^*=\frac{T^2}{N}\sum_{k=0}^{N-1}w^2_ke^{-2\pi ik\frac{(p-q)}{N}}.
    \label{eq: wp_wq}
\end{equation}
Proceeding in a similar fashion, we obtain
\begin{equation}
\left\{
\begin{aligned}
\left\langle\tilde{\bf{d}}_{1;p}\tilde{\bf{d}}_{2;q}^*\right\rangle&\approx\frac{T}{2N}\gamma_{12;p}P_{\mathrm{gw};p}\sum_{k=0}^{N-1}w^2_ke^{-2\pi ik\frac{(p-q)}{N}}\,,\\
\left\langle\tilde{\bf{d}}_{1;p}\tilde{\bf{d}}_{2;q}\right\rangle&=\left\langle\tilde{\bf{d}}_{1;p}\tilde{\bf{d}}_{2;N-q}^*\right\rangle\approx \frac{T}{2N}\gamma_{12;p}P_{\mathrm{gw};p}\sum_{k=0}^{N-1}w^2_ke^{-2\pi ik\frac{(p+q)}{N}}\,,\\
\left\langle\tilde{\bf{d}}_{1;p}\tilde{\bf{d}}^*_{1;q}\right\rangle&\approx\frac{T}{2N}
P_{1;p}
\sum_{k=0}^{N-1}w^2_ke^{-2\pi ik\frac{(p-q)}{N}}\,,\\
\left\langle\tilde{\bf{d}}^*_{2;p}\tilde{\bf{d}}_{2;q}\right\rangle&\approx\frac{T}{2N}
P_{2;p}\sum_{k=0}^{N-1}w^2_ke^{2\pi ik\frac{(p-q)}{N}}\,,
\end{aligned}
\right.
\label{e:more_corrs}
\end{equation}
where
\begin{equation}
    P_{1;p}\equiv P_{n_1;p} + P_{{\rm gw};p}\,,
    \quad
    P_{2;p}\equiv P_{n_2;p} + P_{{\rm gw};p}\,,
    \label{e:P12_p}
\end{equation}
are the total auto-correlated power (noise plus GW signal) in the two detectors.
The other two-point correlations in \eqref{eq: hat_Sigma_pq} not listed in \eqref{e:more_corrs} can be obtained by simply interchanging detectors 1 and 2 and / or taking an overall complex conjugate.

Inserting all these expressions back into~\eqref{eq: hat_Sigma_pq} gives
\begin{tcolorbox}[title=Covariance between $\hat{C}_{I;p}$ and $\hat{C}_{I;q}$]
\begin{align}
&\hat{\Sigma}_{I;pq}\approx\mathcal{P}_{I;p}\left(\frac{1}{N^2}\sum_{j,k=0}^{N-1}w^2_jw^2_ke^{-2\pi i(j-k)\frac{(p-q)}{N}}+\frac{1}{N^2}\sum_{j,k=0}^{N-1}w^2_jw^2_ke^{-2\pi i(j-k)\frac{(p+q)}{N}}\right)\,,
    \label{eq: hat_Sigma_pq-final}\\
&\text{where}\hspace{0.2in}
    \mathcal{P}_{I;p}\equiv \frac{25}{2\gamma_{12;p}^2S_{0;p}^2}\left(P_{1;I;p}P_{2;I;p}+\gamma_{12;p}^2P^2_{\mathrm{gw};p}\right).
    \label{eq: P_p}
\end{align}
\end{tcolorbox}
\noindent
Comparing \eqref{eq: hat_Sigma_pq-final} and \eqref{eq: P_p} with \eqref{eq: hat_Sigma_pq}, we see that we have set $p=q$ in the factors $\gamma_{12;p}\,,\gamma_{12;q}\,,S_{0;p}\,,S_{0;q}$ to obtain \eqref{eq: hat_Sigma_pq-final}. We can do this using the same argument that we used to obtain~\eqref{eq: trick}---namely, $\hat\Sigma_{pq}$ has most of its support for $p\approx q$, and both $\gamma_{1;p}$ and $S_{0;p}$ are slowly varying functions of frequency.

Equation~\eqref{eq: hat_Sigma_pq-final} indicates that when a rectangular window is used (i.e., $w_j\equiv 1$ for all $j$), then there is no correlation between different frequency bins---i.e., $\hat{\Sigma}_{pq}=0$ for $p\ne q$. 
But if a non-rectangular window is used (e.g., a Hann window or Tukey window, etc.), then there will be non-trivial correlations between nearby frequency bins that have been overlooked in the past analyses.

For future reference, we can also rewrite $\hat\Sigma_{pq}$ by recognizing that the summations above can be rewritten in terms of \ac{DFT}s of the window function:
\begin{equation}
\begin{aligned}
     \hat{\Sigma}_{pq}&\approx \frac{\mathcal{P}_p}{T^2}\left(|\widetilde{w^2}_{p-q}|^2+|\widetilde {w^2}_{p+q}|^2\right)\\
     &=(\overline{w^2})^2\,\mathcal{P}_p\, \frac{|\widetilde{w^2}_{p-q}|^2+|\widetilde {w^2}_{p+q}|^2}{T^2(\overline{w^2})^2}\\
     &\equiv(\overline{w^2})^2\,\mathcal{P}_p\,\mathcal{C}_{w;pq}\,,
\end{aligned}
\label{eq: def_C_w}
\end{equation}
where $\widetilde{w^2}$ denotes the \ac{DFT} of $w^2$, and $\mathcal{C}_{w;pq}$ is a dimensionless matrix encoding the non-trivial correlation structure of $\hat\Sigma_{pq}$:
\begin{equation}
   \mathcal{C}_{w;pq}\equiv  \frac{|\widetilde{w^2}_{p-q}|^2+|\widetilde {w^2}_{p+q}|^2}{T^2(\overline{w^2})^2}\,.
   \label{e:C_wpq}
\end{equation}
Although we are not going to directly use this expression in subsequent derivations, this way to expressing $\hat{\Sigma}_{pq}$ isolates the correlation structure of the narrowband optimal estimator across different frequency bins to properties of the window function $w$.

Let us now consider the special case $p=q$.
Then $\hat{\Sigma}_{pp}$ reduces to the variance of $\hat{C}_p$, denoted $\hat{\sigma}^2_p$.
Using \eqref{eq: hat_Sigma_pq-final}, we can write
\begin{equation}
    \hat{\sigma}^2_p\equiv \hat{\Sigma}_{pp} \approx \frac{\mathcal{P}_p}{N^2}\left(\sum_{j,k=0}^{N-1}w^2_jw^2_k+\underbrace{\sum_{j,k=0}^{N-1}w^2_jw^2_ke^{-2\pi i(j-k)\frac{2p}{N}}}_{\approx 0\,,\quad\mathrm{unless}\,p\approx0,\frac{N}{2}}\right)\,.
\end{equation}
Note that the second term in the parentheses is negligible unless $p\approx0$ or $N/2$:
(i) $p\approx 0$ corresponds to $f\approx 0$~Hz, and
(ii) $p\approx N/2$ corresponds to $f\approx\delta\!f(N/2)=f_\mathrm{Nyquist}$. 
Since one typically filters the raw data to remove its DC component and also the components near the Nyquist frequency, we can ignore the second term above.
Thus, 
\begin{tcolorbox}[title=Variance of $\hat{C}_{I;p}$]
\begin{equation}
\hat{\sigma}^2_{I;p}\approx
\mathcal{P}_{I;p}\left(\frac{1}{N}\sum_{j=0}^{N-1}w_j^2\right)^2=(\overline{w^2})^2\,\mathcal{P}_{I;p}\,,
\qquad
\hat\Sigma_{I;pq} = \hat\sigma^2_{I;p}\,\mathcal{C}_{w;pq}\,.
\label{eq: var_final}
\end{equation}
\end{tcolorbox}

As a simple sanity check, note that if we adopt the weak-signal limit, i.e., $P_\mathrm{gw}(f)\ll P_{n_a}(f)$ where $a=1\,\mathrm{or}\,2$, then the above expression for $\hat{\sigma}^2_p$ reduces to the same quantity that the \ac{LVK} collaboration has employed in past analyses, except for the window factor $(\overline{w^2})^2$, see, e.g.,~(3) in~\cite{O3stoch}. 
(Note that the definition of $S_0$ in~\cite{O3stoch} differs from that adopted in this work by a factor of 5, leading to the factor of 25 appearing in the numerator of $\mathcal{P}_p$.)

Given $\hat{\Sigma}_{pq}$ in~\eqref{eq: hat_Sigma_pq-final}, one can return to~\eqref{eq: bar_Sigma_lm} to find (as a reminder, $\barell$ and $\barm$ denote the indices of coarse-grained frequency bins):
\begin{equation}
\begin{aligned}
\bar{\Sigma}_{\barell\barm}&=\frac{1}{M^2N^2}\sum_{p=M\barell-\frac{M}{2}}^{M\barell+\frac{M}{2}-1}\sum_{q=M\barm-\frac{M}{2}}^{M\barm+\frac{M}{2}-1}\mathcal{P}_p\left(\sum_{j,k=0}^{N-1}w^2_jw^2_ke^{-2\pi i(j-k)\frac{(p-q)}{N}}+\sum_{j,k=0}^{N-1}w^2_jw^2_ke^{-2\pi i(j-k)\frac{(p+q)}{N}}\right)\\
&\approx \frac{\mathcal{P}_{M\barell}}{M^2N^2}\sum_{j,k=0}^{N-1}w^2_jw^2_k\sum_{p=M\barell-\frac{M}{2}}^{M\barell+\frac{M}{2}-1}\sum_{q=M\barm-\frac{M}{2}}^{M\barm+\frac{M}{2}-1}\left(e^{-2\pi i(j-k)\frac{(p-q)}{N}}+e^{-2\pi i(j-k)\frac{(p+q)}{N}}\right)\\
&\equiv\frac{\mathcal{P}_{M\barell}}{N^2}\sum_{j,k=0}^{N-1}w_j^2w_k^2(\mathcal{S}^-_{jk;\barell\barm}+\mathcal{S}^+_{jk;\barell\barm})\,,
\end{aligned}
\label{eq: bar_Sigma_l_m_mid}
\end{equation}
where we interchanged the order of the summations and factored out $\mathcal{P}_p$, evaluating it at $p=M\barell$ since $\mathcal{P}$ is assumed to be a slowly-varying function over the range of $p$ appearing in the sum. 
(To avoid ambiguity, we stress that $X_{M\barell}\equiv X(f_{M\barell})=X(M\barell\Delta\!f)$, where $X$ can be $\mathcal{P}$ or $\hat\sigma^2$.)
In the final line, we defined
\begin{equation}
    \mathcal{S}^-_{jk;\barell\barm}\equiv \frac{1}{M^2}\sum_{p=M\barell-\frac{M}{2}}^{M\barell+\frac{M}{2}-1}\sum_{q=M\barm-\frac{M}{2}}^{M\barm+\frac{M}{2}-1}e^{-2\pi i(j-k)\frac{(p-q)}{N}}\,,\qquad \mathcal{S}^+_{jk;\barell\barm}\equiv \frac{1}{M^2}\sum_{p=M\barell-\frac{M}{2}}^{M\barell+\frac{M}{2}-1}\sum_{q=M\barm-\frac{M}{2}}^{M\barm+\frac{M}{2}-1}e^{-2\pi i(j-k)\frac{(p+q)}{N}}.
    \label{eq: S_-_S_+}
\end{equation}
It is then straightforward to show that (see App.~\ref{app: proof_bar_Sigma_l_m_final} for skipped details): 
\begin{tcolorbox}[title=Covariance between  $\bar{C}_{I;\barell}$ and $\bar{C}_{I;\barm}$]
\begin{equation}
    \bar{\Sigma}_{I;\barell\barm}\approx \frac{\mathcal{P}_{I;M\barell}}{N^2}\sum_{j,k=0}^{N-1}w_j^2w_k^2\mathcal{K}_{j,k;M,N}\left(e^{-2\pi i(j-k)\frac{(M\barell-M\barm)}{N}}+e^{-2\pi i(j-k)\frac{(M\barell+M\barm-1)}{N}}\right),
    \label{eq: bar_Sigma_l_m_final}
\end{equation}
\end{tcolorbox}
\noindent
where the kernel function $\mathcal{K}_{j,k;M,N}$\footnote{$\mathcal{K}_{j,k;M,N}$ is the square of the Dirichlet kernel.} is defined by
\begin{equation}
    \mathcal{K}_{j,k;M,N}\equiv \left(\frac{\sinc\left(\pi\frac{(j-k)M}{N}\right)}{\sinc\left(\pi\frac{(j-k)}{N}\right)}\right)^2.
    \label{eq: K}
\end{equation}
An alternative way to write $\bar\Sigma_{\barell\barm}$ is to substitute \eqref{eq: var_final} for $\hat\Sigma_{pq}$ into \eqref{eq: bar_Sigma_lm}, leading to:
\begin{equation}
    \bar\Sigma_{\barell\barm}=\frac{1}{M^2}\sum_{p=M\barell-\frac{M}{2}}^{M\barell+\frac{M}{2}-1}\sum_{q=M\barm-\frac{M}{2}}^{M\barm+\frac{M}{2}-1}\hat\sigma^2_p\,\mathcal{C}_{w;pq}\,.
\end{equation}
As before, we may pull $\hat\sigma^2_p$ out of the summation, evaluating it at $p=M\barell$. 
Then one has
\begin{equation}
\bar\Sigma_{\barell\barm}\approx\hat\sigma^2_{M\barell}\,\frac{1}{M^2}\sum_{p=M\barell-\frac{M}{2}}^{M\barell+\frac{M}{2}-1}\sum_{q=M\barm-\frac{M}{2}}^{M\barm+\frac{M}{2}-1}\mathcal{C}_{w;pq}\equiv\hat{\sigma}^2_{M\barell}\,\bar{\mathcal{C}}_{w;\barell\barm}\,,
\label{eq: def_of_bar_C_w}
\end{equation}
which will be useful later on.

Now let us circle back to consider the diagonal entries $\bar{\Sigma}_{\barell\barell}\equiv \bar\sigma^2_{\barell}$, which are the variances of the estimators $\bar{C}_{\barell}$. 
Similar to the reasoning that we gave right before~\eqref{eq: var_final}, we can ignore the 
second oscillating term in the summation, obtaining
\begin{equation}
\begin{aligned}
\bar{\sigma}^2_{\barell}&\equiv\bar{\Sigma}_{\barell\barell}
    \approx\frac{\mathcal{P}_{M\barell}}{N^2}\sum_{j,k=0}^{N-1}w^2_jw^2_k\mathcal{K}_{j,k;M,N}.
\end{aligned}
\end{equation}
Using \eqref{eq: var_final}, we can relate $\bar{\sigma}^2_{\barell}$ to the variance $\hat{\sigma}^2_\ell$ of the non-coarse-grained estimator:
\begin{equation}
    \bar{\sigma}^2_{\barell}=\frac{\hat{\sigma}^2_{M\barell}}{(\overline{w^2})^2}\frac{1}{N^2}\sum_{j,k=0}^{N-1}w^2_jw^2_k\mathcal{K}_{j,k;M,N}\,.
\end{equation}
The rhs of the above expression can be simplified by defining
\begin{equation}
\left(\overline{w^2_\mathcal{K}}\right)^2\equiv\frac{1}{N^2}\sum_{j,k=0}^{N-1}w^2_jw^2_k\mathcal{K}_{j,k;M,N}
\label{e:wK2bar2}
\,.
\end{equation}
Then, denoting the ratio%
\footnote{To be fully rigorous, we should denote $r_M$ as $r_{M,N}$.  
But since this ratio is determined primarily by $M$ and differs negligibly with changes to $N$, we use the simpler notation $r_M$.} 
between $\left(\overline{w^2_\mathcal{K}}\right)^2$ and $\left(\overline{w^2}\right)^2$ as $r_M$, we obtain the useful relation:
\begin{tcolorbox}[title=Relationship between $\bar\sigma^2_{I;\barell}$ and $\hat{\sigma}^2_{I;\ell}$]
\begin{equation}
\bar{\sigma}^2_{I;\barell}=r_M\hat{\sigma}^2_{I;M\barell}=r_M\hat{\sigma}^2_{I;\ell}\,,\qquad r_M\equiv\left(\overline{w^2_\mathcal{K}}\right)^2\bigg/\left(\overline{w^2}\right)^2\,.
\label{eq: def_r_M}
\end{equation}
\end{tcolorbox}

Let us now consider two instructive limiting cases:
(i) $M=1$, corresponding to no coarse-graining, for which the kernel function $\mathcal{K}_{j,k;M,N}=1$ and $\bar{\sigma}^2_{\barell}=\hat{\sigma}^2_\ell$, as expected. 
(ii) $w_j\equiv 1$, corresponding to a rectangular window, for which $r_M=1/M$, implying $\bar{\sigma}^2_{\barell}=\hat{\sigma}^2_\ell/M$. 
This result can be seen most directly by studying the first line of~\eqref{eq: bar_Sigma_l_m_mid}.

\begin{figure}[!htbp]
    \centering
    \includegraphics[width=0.8\linewidth]{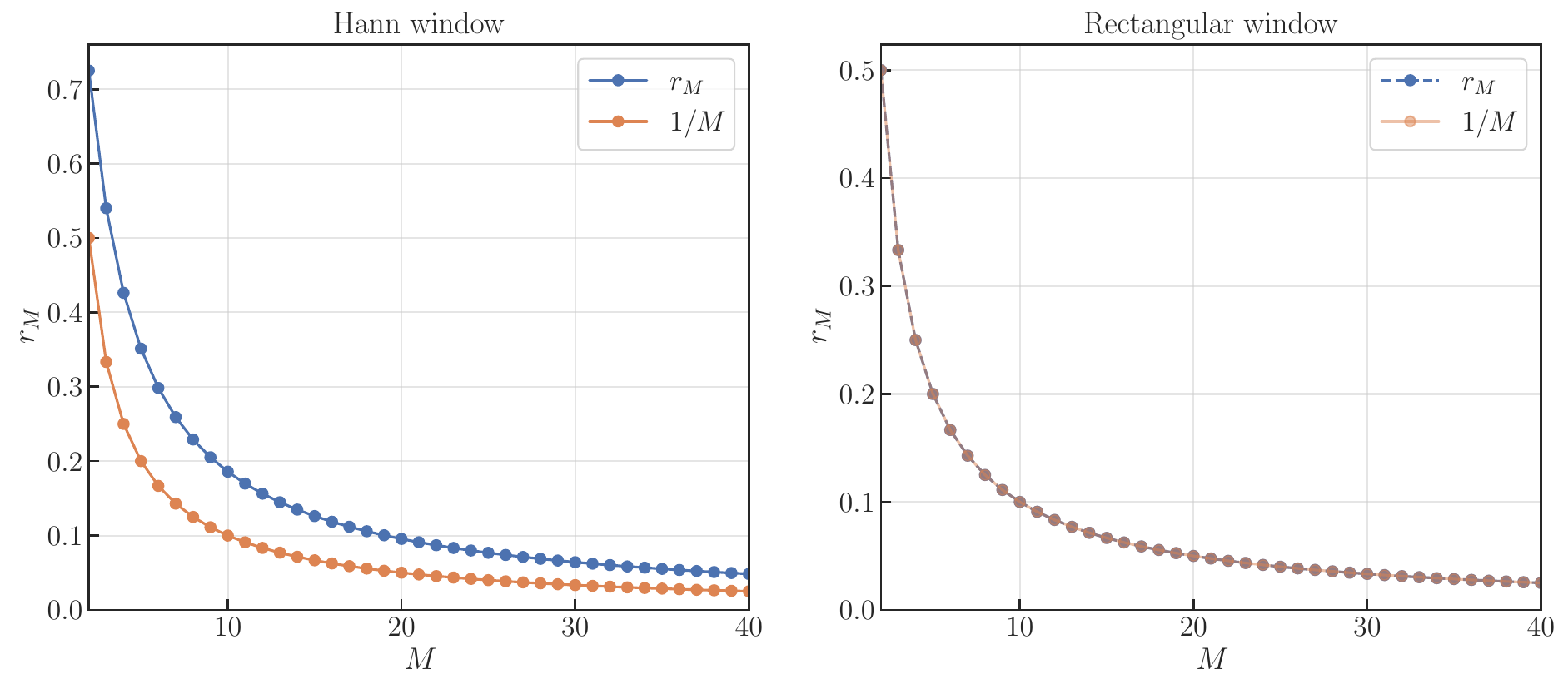}
    \caption{$r_M$ for two cases: a Hann window (left panel) and a rectangular window (right panel).
    We also plot the values of $1/M$ in orange for comparison.}
    \label{fig: r_of_M}
\end{figure}

The appearance of this $1/M$ factor explains the commonly encountered $T\delta\!f=\delta\!f/\Delta\!f=M$ factor in the expression for $\bar{\sigma}^2_{\barell}$ in the literature (see e.g.,~(6) of Ref.~\cite{pygwb} and~(6.37) of Ref.~\cite{Romano:2016dpx}). 
We emphasize, however, that the relation $r_M=1/M$ \textit{holds only for a rectangular window}, for which there are no correlations between neighboring frequency bins. 
When a different window is adopted, correlations between different bins are introduced and $r_M$ generally deviates from $1/M$. 
In Fig.~\ref{fig: r_of_M}, we show numerical evaluations of $r_M$ for two representative cases: a Hann window (left panel) and a rectangular window (right panel). 
The blue curve depicts $r_M$ and the orange curve shows the values of $1/M$, serving as reference where different frequency bins are independent.

\begin{figure}[!htbp]
    \centering
    \includegraphics[width=0.8\linewidth]{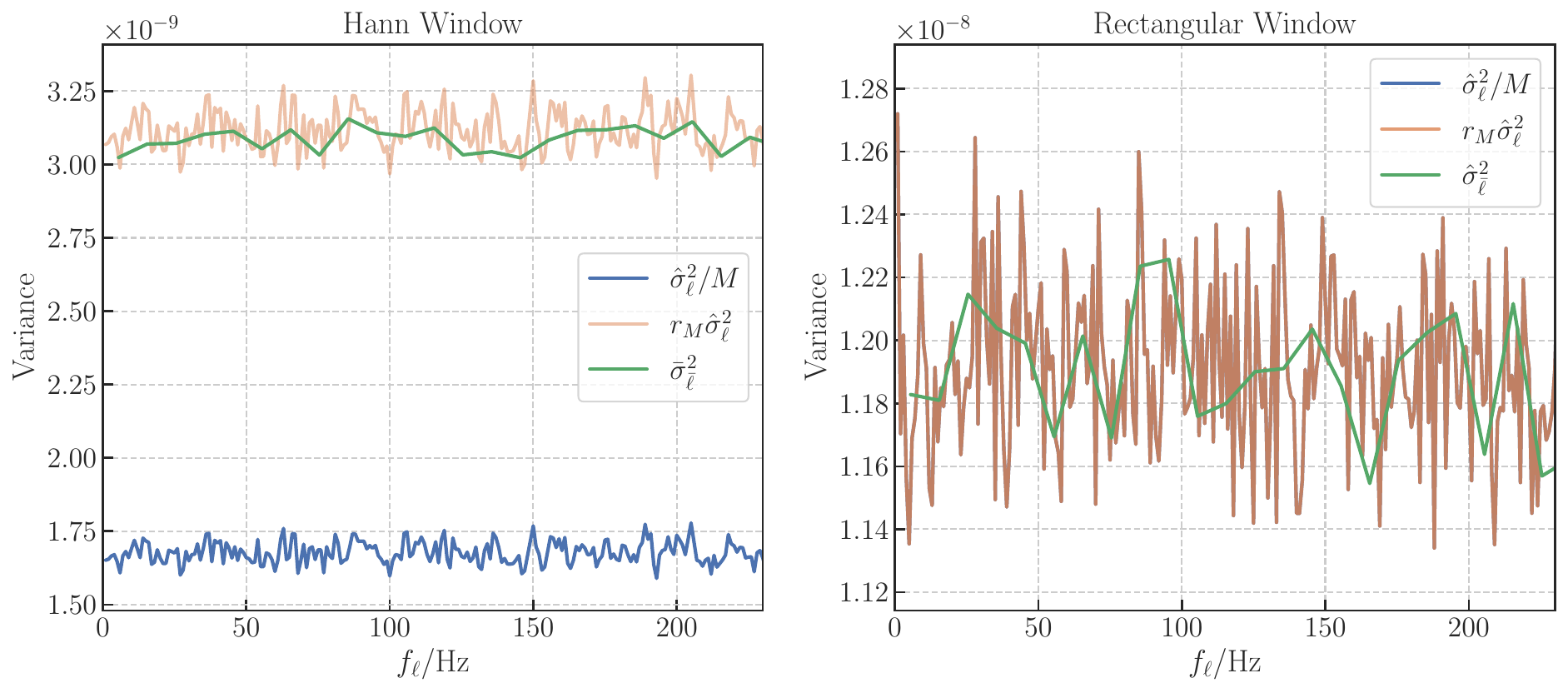}
    \caption{$\bar\sigma^2_{\barell}$ versus $\hat\sigma^2_\ell$ for two cases: a Hann window (left panel) and a rectangular window (right panel). 
    $M=10$ for both cases.}
    \label{fig: checkrofM}
\end{figure}

We also verify the behavior of $r_M$ via the following simulation. 
We simulate ten thousand realizations of 1-s long data $d_1,d_2\sim\mathcal{N}(0,1)$ sampled at 2048 Hz. 
To simplify the analysis, we define $\hat{C}_\ell=\mathfrak{R}\left(\tilde{\bf d}_{1;\ell}\tilde{\bf d}_{2;\ell}^*\right)$ and calculate its variance $\hat{\sigma}^2_\ell$ for each frequency bin. 
We then take $M=10$ and calculate the coarse-grained spectrum and corresponding variance $\bar\sigma^2_{\barell}$. 
We show the results of these simulations in the Fig.~\ref{fig: checkrofM}. 
In the left panel, we consider a Hann window.
The blue curve $\hat{\sigma}^2_\ell/M$ clearly deviates from the correct variance of the coarse-grained spectra, $\bar\sigma^2_{\barell}$, whereas the orange curve $r_M\hat{\sigma}_\ell^2$ is consistent with the estimated values of $\bar{\sigma}^2_{\barell}$. 
In the right panel, we consider a rectangular window.
For this case,  $r_M=1/M$, and all three curves are consistent with one another.

Now that we have derived the expected value of $\bar C_{\barell}$ and the covariance between $\bar C_{\barell}$ and $\bar C_{\barm}$, we quote these results again, as we will use them heavily in the subsequent analyses.
\begin{tcolorbox}[title=Two important results: Expected value of $\bar C_{I;\barell}$ and the covariance between $\bar{C}_{I;\barell}$ and $\bar C_{I;\barm}$]
\begin{equation}
\left\{
\begin{aligned}
     &\langle \bar{C}_{I;\barell}\rangle\approx \Omega_{\mathrm{gw};\barell}\,\overline{w^2}.\\
     &\bar{\Sigma}_{I;\barell\barm}\approx \frac{\mathcal{P}_{I;M\barell}}{N^2}\sum_{j,k=0}^{N-1}w_j^2w_k^2\mathcal{K}_{j,k;M,N}\left(e^{-2\pi i(j-k)\frac{(M\barell-M\barm)}{N}}+e^{-2\pi i(j-k)\frac{(M\barell+M\barm-1)}{N}}\right).
\end{aligned}
\right.
\label{eq: summary_1}
\end{equation}
\end{tcolorbox}

\subsection{Broadband estimator}
\label{s:single_seg_broad}

In the previous subsection, we derived the expected value and covariance of the  coarse-grained, single-segment narrowband estimator $\bar{C}_{\barell}$.
But, in the literature~\cite{O3stoch, O4aStoch}, one often considers a broadband estimator, $\bar C_{\rm opt}$, averaged over coarse-grained frequencies $\barell$.
In this subsection, we discuss how to construct such a broadband estimator, restricting attention to a \textit{single} time segment $I$.
In Sec.~\ref{s:multi_seg_broad}, we will extend this construction to a \textit{multi-segment} broadband estimator, which is optimally averaged over both frequency bins and time segments.

Since the narrowband estimator $\bar C_{\barell}$ captures the information of the same underlying \ac{SGWB} signal in different frequency bins, it is natural to want to combine all this information into a single broadband estimator. 
We stress, however, that how we combine the narrowband estimators is not unique; it depends on our model of the frequency-dependence of the background. 
To simplify the discussion, but without loss of generality, we will consider a single power-law model:
\begin{equation}\OGW(f;\Omega_\mathrm{ref},\alpha)=\Omega_\mathrm{ref}\left(\frac{f}{f_\mathrm{ref}}\right)^\alpha\,,
\end{equation}
which is parametrized by $\Omega_\mathrm{ref}$ (the value of $\Omega_\mathrm{gw}(f)$ at the reference frequency $f_\mathrm{ref}$, which is usually chosen to be 25 Hz), and $\alpha$ (the power-law index). 
If $\alpha$ is set to some fixed value, then one can define an estimator of $\overline{w^2}\,\Omega_\mathrm{ref}$ by multiplying $\bar{C}_{\barell}$ by a reweighting function $\mathcal{R}(f;\alpha)\equiv(f/f_\mathrm{ref})^{-\alpha}$.
For this reweighting,
\begin{equation}
    \langle \bar{C}^\mathcal{R}_{\barell}\rangle=\overline{w^2}\,\Omega_\mathrm{ref}\,,
    \quad\text{where}\quad
    \bar{C}^\mathcal{R}_{\barell}\equiv \mathcal{R}(f_{\barell};\alpha)\,\bar{C}_{\barell}\,.
    \label{eq: CR}
\end{equation}
After reweighting,  the corresponding estimator $\bar{C}^\mathcal{R}_{\barell}$ \textit{for any} $\barell$ is an estimator of $\overline{w^2}\Omega_\mathrm{ref}$ with variance $(\bar\sigma^{\mathcal{R}}_{\barell})^2\equiv\mathcal{R}^2(f_{\barell};\alpha)\,\bar\sigma^2_{\barell}$. 
We can then construct an optimal linear combination of the $\bar{C}^\mathcal{R}_{\barell}$ by assigning a weight $\lambda_{\barell}$ to each $\bar{C}^\mathcal{R}_{\barell}$, where the weights are chosen to minimize the variance of the combined estimator.
The resulting linear combination,
denoted $\bar C_{\mathrm{opt}}$, is a broadband estimator of $\overline{w^2}\,\Omega_\mathrm{ref}$. 
To further simplify our discussion, we set $\alpha=0$ in what follows.
By doing so, we no longer need to distinguish between $\bar{C}^\mathcal{R}_{\barell}$ and $\bar{C}_{\barell}$, and we have $\langle \bar{C}_{\barell}\rangle=\overline{w^2}\,\Omega_{\mathrm{gw}}(f_\ell)=\overline{w^2}\,\Omega_\mathrm{ref}$.

Adopting the Lagrange multiplier method outlined in App.~\ref{app: optimal_estimator} and Sec.~6 of~\cite{Romano:2016dpx}, we minimize the variance of $\bar{C}_{\mathrm{opt}}$ to determine the optimal weights.
(The overall normalization is fixed by the requirement that $\sum_{\barell} \lambda_{\barell}=1$, which is enforced using the Lagrange multiplier.)
More explicitly, we define\footnote{The more complete notation should be $\bar C_{\opt;I}$ and $\bar\sigma^2_{\opt;I}$. 
But since we are considering only one segment here, we temporarily drop the segment index $I$ as we did in the previous subsection.}
\begin{equation}
\bar C_{\opt}\equiv\sum_{\barell=0}^{\frac{N}{2M}-1}\lambda_{\barell}\bar C_{\barell}\,,
\end{equation}    
which has variance
\begin{equation}
\bar\sigma^2_{\rm opt}=\sum_{\barell=0}^{\frac{N}{2M}-1}\sum_{\barm=0}^{\frac{N}{2M}-1}\lambda_{\barell}\lambda_{\barm}\bar \Sigma_{\barell\barm}\,.
\label{eq: bar_sigma^2_barl}
\end{equation}
Although the minimum-variance solution for $\lambda_{\barell}$ has a closed form as indicated by~\eqref{eq: lambda_i_closedform}, we note that it is non-trivial to invert $\bar\Sigma_{\barell\barm}$ analytically. Therefore, we adopt an alternative way to determine the $\lambda_{\barell}$ and thus $\bar\sigma^2_{\rm opt}$. 

Substituting~\eqref{eq: bar_Sigma_l_m_final} for $\bar\Sigma_{\barell\barm}$ into~\eqref{eq: bar_sigma^2_barl}, we find
\begin{equation}
\begin{aligned}
\bar\sigma^2_{\rm opt}&\approx 
\sum_{\barell=0}^{\frac{N}{2M}-1}\sum_{\barm=0}^{\frac{N}{2M}-1}\lambda_{\barell}\lambda_{
\barm}\frac{\mathcal{P}_{M\barell}}{N^2}\sum_{j,k=0}^{N-1}w_j^2w_k^2\,\mathcal{K}_{j,k;M,N}\left(e^{-2\pi i(j-k)\frac{(M\barell-M\barm)}{N}}+e^{-2\pi i(j-k)\frac{(M\barell+M\barm-1)}{N}}\right)\\
    &\approx \sum_{\barell=0}^{\frac{N}{2M}-1}\sum_{\barm=0}^{\frac{N}{2M}-1}\lambda_{\barell}\lambda_{
    \barm}\frac{\mathcal{P}_{M\barell}}{N^2}\sum_{j,k=0}^{N-1}w_j^2w_k^2\,\mathcal{K}_{j,k;M,N}e^{-2\pi i(j-k)\frac{(M\barell-M\barm)}{N}}\\
    &\hspace{1in}+\sum_{\barell=0}^{\frac{N}{2M}-1}\sum_{\barm=1}^{\frac{N}{2M}}\lambda_{\barell}\lambda_{
    \barm}\frac{\mathcal{P}_{M\barell}}{N^2}\sum_{j,k=0}^{N-1}w_j^2w_k^2\,\mathcal{K}_{j,k;M,N}e^{-2\pi i(j-k)\frac{(M\barell+M\barm-1)}{N}}\\
    &\equiv \mathcal{S}_1+\mathcal{S}_2\,,
\end{aligned}
\end{equation}
where we changed the summation over $\barm$ from $\sum_{\barm=0}^{\frac{N}{2M}-1}$ to $\sum_{\barm=1}^{\frac{N}{2M}}$ to obtain the second term in the second equality.
This change in summation limits should not affect the sum since $\barm=0$ and $\barm=N/2M$ correspond to DC and the Nyquist frequency, respectively, which are typically removed from the data via high and low-pass filtering.

We now study the two summations separately. 
For $\mathcal{S}_1$, we note that the oscillating complex exponential is dominated by terms satisfying $\barell\approx\barm$ or $j\approx k$, so one can replace $\lambda_{\barm}$ by $\lambda_{\barell}$ to obtain
\begin{equation}
    \mathcal{S}_1\approx \sum_{\barell=0}^{\frac{N}{2M}-1}\lambda^2_{
    \barell}\frac{\mathcal{P}_{M\barell}}{N^2}\sum_{j,k=0}^{N-1}w_j^2w_k^2\mathcal{K}_{j,k;M,N}\sum_{\barm=0}^{\frac{N}{2M}-1}e^{-2\pi i(j-k)\frac{(M\barell-M\barm)}{N}}.
\end{equation}
A similar argument holds for $\mathcal{S}_2$, with its complex exponential dominated by terms satisfying $M\barell+M\barm-1\approx N$ or $j\approx k$. 
So, for this case, we can rewrite $\lambda_{\bar m}$ as $\lambda_{\frac{N+1}{M}-\barell}\approx\lambda_{\frac{N}{M}-\barell}$, where the last approximate equality is a consequence of $N\gg 1$. 
Since $0\leqslant \barell\leqslant \frac{N}{2M}-1$, the subscript on $\lambda_{\frac{N}{M}-\barell}$ corresponds to a negative frequency.
This means that $\lambda_{\frac{N}{M}-\barell}$ is the weight associated with $\bar C_{\frac{N}{M}-\barell}$, but $\bar C_{\frac{N}{M}-\barell}=\bar C_{\barell}$. Therefore, symmetry implies that we must have $\lambda_{\frac{N}{M}-\barell}=\lambda_{\barell}$. Using this result,  $\mathcal{S}_2$ simplifies to 
\begin{equation}
    \mathcal{S}_2\approx \sum_{\barell=0}^{\frac{N}{2M}-1}\lambda^2_{
    \barell}\frac{\mathcal{P}_{M\barell}}{N^2}\sum_{j,k=0}^{N-1}w_j^2w_k^2\mathcal{K}_{j,k;M,N}\sum_{\barm=1}^{\frac{N}{2M}}e^{-2\pi i(j-k)\frac{(M\barell+M\barm-1)}{N}}.
\end{equation}
In addition, we can change the variable $\bar m$ in the above summation to $\tilde m\equiv \frac{N}{M}-\barm$, leading to 
\begin{equation}
    \mathcal{S}_2\approx \sum_{\barell=0}^{\frac{N}{2M}-1}\lambda^2_{\barell}\frac{\mathcal{P}_{M\barell}}{N^2}\sum_{j,k=0}^{N-1}w_j^2w_k^2\mathcal{K}_{j,k;M,N}\sum_{\tilde m=\frac{N}{2M}}^{\frac{N}{M}-1}e^{-2\pi i(j-k)\frac{(M\barell-M\tilde m-1)}{N}}\,.
\end{equation}
Finally, we can drop the “$-1$” term in the numerator of the exponential because $M\tilde m\geqslant \frac{N}{2}\gg1$. 
This allows us to combine $\mathcal{S}_1$ and $\mathcal{S}_2$, yielding
\begin{equation}
\begin{aligned}
     \bar\sigma^2_{\rm opt}&\approx\sum_{\barell=0}^{\frac{N}{2M}-1}\lambda^2_{\barell}\frac{\mathcal{P}_{M\barell}}{N^2}\sum_{j,k=0}^{N-1}w^2_jw^2_k\mathcal{K}_{j,k;M,N}\sum_{\barm=0}^{\frac{N}{M}-1}e^{-2\pi i(j-k)\frac{(M\barell-M\barm)}{N}}\\
     &=\sum_{\barell=0}^{\frac{N}{2M}-1}\lambda^2_{\barell}\frac{\mathcal{P}_{M\barell}}{N^2}\sum_{j,k=0}^{N-1}w^2_jw^2_k\mathcal{K}_{j,k;M,N}\frac{N}{M}\delta_{jk}\\
     &=\frac{1}{M}\sum_{\barell=0}^{\frac{N}{2M}-1}\lambda^2_{\barell}\mathcal{P}_{M\barell}\left(\frac{1}{N}\sum_{j=0}^{N-1}w^4_j\right)\\
     &= \frac{\overline{w^4}}{M}\,\sum_{\barell=0}^{\frac{N}{2M}-1}\lambda^2_{\barell}\mathcal{P}_{M\barell}\,,\quad{\rm where}\quad \overline{w^4}\equiv\frac{1}{N}\sum_{j=0}^{N-1}w_j^4\,.
\label{e:w4_bar}
\end{aligned}
\end{equation}
Using~\eqref{eq: var_final} and~\eqref{eq: def_r_M}, we can write
\begin{equation}
    \bar\sigma^2_{\rm opt}=\frac{\mathcal{W}}{M}\,\sum_{\barell=0}^{\frac{N}{2M}-1}\lambda^2_{\barell}\hat{\sigma}^2_{M\barell}\,,\qquad \mathcal{W}\equiv \frac{\overline{w^4}}{\left(\overline{w^2}\right)^2}\,,\label{eq: firstway}
\end{equation}
or, alternatively,
\begin{equation}
    \bar\sigma^2_{\rm opt}=\frac{\mathcal{W}_\mathcal{K}}{M}\,\sum_{\barell=0}^{\frac{N}{2M}-1}\lambda^2_{\barell}\bar{\sigma}^2_{\barell}\,,\qquad \mathcal{W}_\mathcal{K}\equiv \frac{\overline{w^4}}{\left(\overline{w^2_\mathcal{K}}\right)^2}\,.\label{eq: secondway}
\end{equation}
To obtain~\eqref{eq: firstway} we rewrote $\mathcal{P}_{M\barell}$ in terms of the variance $\hat{\sigma}^2_{M\barell}$ of the non-coarse-grained spectrum, whereas to obtain~\eqref{eq: secondway} we additionally used the relationship  
\begin{equation}
\mathcal{W}\hat\sigma^2_{M\barell}=\mathcal{W_{\mathcal{K}}}\bar\sigma^2_{\barell} 
\label{e:Wsigma_relation}
\end{equation}
between $\hat{\sigma}^2_{M\barell}$ and $\bar\sigma^2_{\barell}$. Equation~\eqref{eq: firstway} explains why approximating the variance of the coarse-grained spectrum as the variance of the original spectrum divided by $M$ yields a correct result, which has been widely used in the literature. 

Importantly,~\eqref{eq: firstway} and~\eqref{eq: bar_sigma^2_barl} together indicate that it is not necessary to use the full \textit{non-diagonal} covariance matrix, $\bar\Sigma_{\barell\barm}$, to perform the optimal combination of the narrowband estimators.
Instead, we can define an effective \textit{diagonal} matrix 
\begin{equation}
\bar\Lambda_{\barell\barm}\equiv\frac{\mathcal{W}}{M}\,\mathrm{diag}[\hat{\sigma}^2_{M0},\hat{\sigma}^2_{M1},...]
\equiv\frac{\mathcal{W}}{M}\delta_{\barell\barm}\,\hat{\sigma}^2_{M\barell}
\end{equation} 
to simplify the calculation. 
Substituting $\bar\Lambda_{\barell\barm}$
into~\eqref{eq: bar_sigma^2_barl}, we obtain the main result:
\begin{tcolorbox}[title=Optimal broadband estimator and its variance for a single segment]
\begin{equation}
\bar{C}_{\mathrm{opt};I}=\frac{\displaystyle{\sum_{\barell=0}^{\frac{N}{2M}-1}\hat\sigma^{-2}_{I;M\barell}}\,\bar C_{I;\barell}}{\displaystyle{\sum_{\barell=0}^{\frac{N}{2M}-1}\hat\sigma^{-2}_{I;M\barell}}}\,,\qquad \bar\sigma^2_{\mathrm{opt};I}=\frac{\mathcal{W}/M}{\displaystyle{\sum_{\barell=0}^{\frac{N}{2M}-1}\hat\sigma^{-2}_{I;M\barell}}}\,,\qquad \mathcal{W}\equiv \frac{\overline{w^4}}{\left(\overline{w^2}\right)^2}\,,
\label{eq: C_opt_var_opt_I_non_coase_grained}
\end{equation}
\end{tcolorbox}
\noindent
or similar expressions for $\bar C_{I;\opt}$ and $\bar\sigma^2_{\opt;I}$ in terms of $\mathcal{W_{\mathcal{K}}}\bar\sigma^2_{I;\barell}$ 
using~\eqref{e:Wsigma_relation}.

Before moving on to combining data across different time segments, we briefly comment on the relation between our results and previous work.
A direct comparison shows that the expressions for the optimal single-segment broadband estimator and its variance derived here are consistent with those given in Eqs.~(17) and (26) of Ref.~\cite{Note_1} and Eqs.~(4.6) and (4.12) of Ref.~\cite{Note_2}. 
In those two unpublished notes, however, the authors aim to construct a broadband estimator directly, and therefore do not introduce a narrowband estimator for $\Omega_{\mathrm{gw};\barell}$. 

As shown in~\eqref{eq: C_opt_var_opt_I_non_coase_grained}, the variance $\bar\sigma^2_{\rm opt}$ of the broadband estimator is proportional to $\mathcal{W}\propto(\overline{w^4})$, whereas the variance $\bar\sigma^2_{\barell}$ of the narrowband estimator  is proportional to $(\overline{w_{\mathcal{K}}^2})^2$, which is not equal, in general, to $(\overline{w^4})$. 
In~\cite{pygwb}, the authors incorrectly extrapolate the results that  strictly hold only for the broadband estimator, adopting the following two assumptions:
(i) they assume $\bar\sigma_{\barell}^2\propto(\overline{w^4})$, and 
(ii) they assume that the covariance between $\bar C_{\barell}$ and $\bar C_{\barm}$ is zero when $\barell\neq\barm$. 
Based on the derivations presented above, neither of these two assumptions holds in general. 
Nevertheless, under these two assumptions, one still recovers the correct single-segment broadband estimator and its variance. 
This is because combining narrowband estimators exploiting the full covariance matrix, whose diagonal entries are proportional to $(\overline{w^2_{\mathcal{K}}})^2$, is mathematically equivalent to performing the same combination with a rescaled diagonal matrix proportional to $(\overline{w^4})$. 
Although these two procedures are formally different, they yield the same broadband result.

We emphasize, however, that the correct expressions of the mean and covariance of the narrowband estimator are those given in~\eqref{eq: summary_1}, which should be used whenever narrowband quantities and/or their correlations are of interest.

\section{Covariance across segments and frequencies}
\label{s:covariance}

In Sec.~\ref{s:single_segment}, we considered narrowband and broadband estimators and their corresponding variances, restricting attention to a single segment of data. 
In particular, we derived the expected value and variance of the narrowband estimator $\bar{C}_{I;\barell}$, and showed how to combine $\bar C_{I;\barell}$ over frequency indices $\barell$ to construct a broadband estimator $\bar{C}_{\mathrm{opt};I}$. 
But, in practice, an analysis typically involves many data segments, which can be further combined to produce a \textit{final} narrowband estimator  $\bar{C}_{\mathrm{opt};\barell}$ (a spectrum)
and a \textit{final} broadband estimator $\bar{C}_\mathrm{opt}$ (an overall amplitude) involving all available data.
(These are \textit{optimal} estimators in the sense that they are minimum-variance, unbiased estimators of the GW spectrum or overall amplitude.)
Achieving this goal requires a careful treatment of the covariance between estimators from different segments, namely between $\bar{C}_{I;\barell}$ and $\bar{C}_{J;\barm}$.

\begin{figure}[!htbp]
    \centering
    \includegraphics[width=0.95\linewidth]{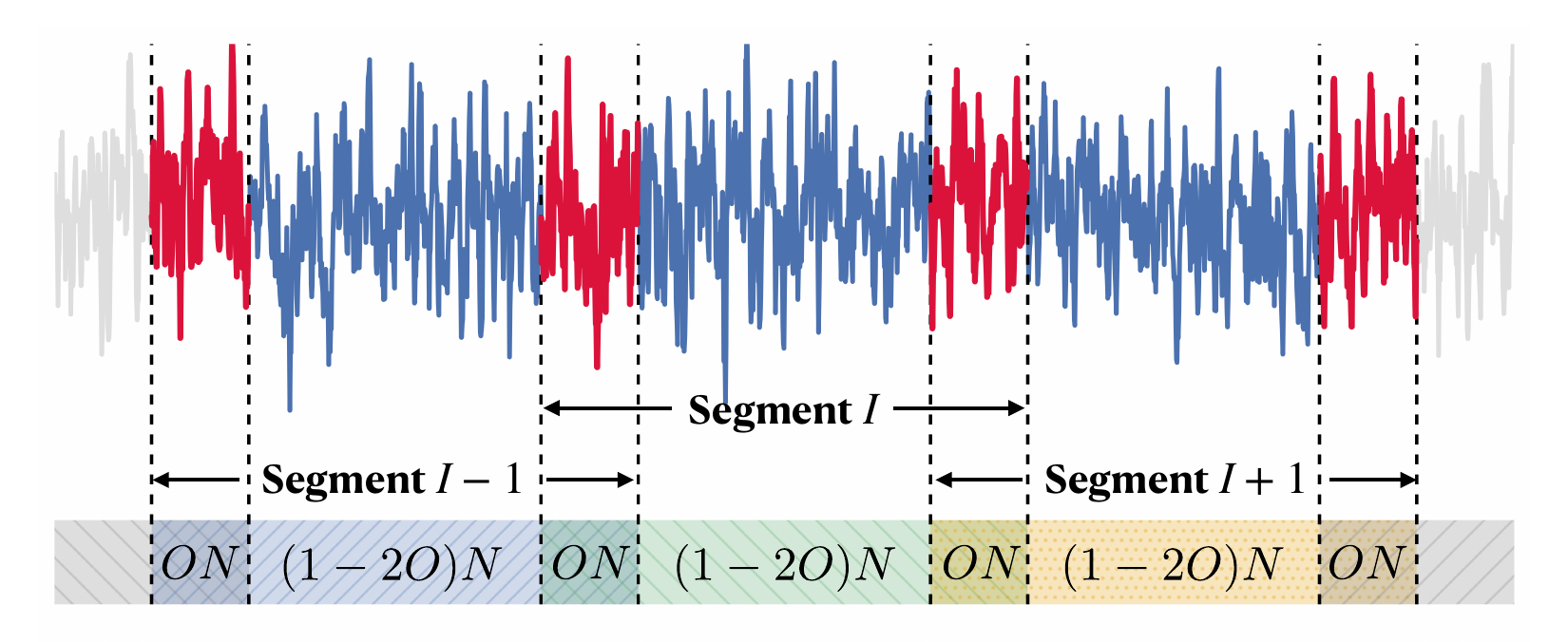}
    \caption{A schematic illustration of how the time-domain data are segmented with overlap between neighboring segments. 
    The blue curve represents the original time series. The data are divided into segments, each containing $N$ data points, while adjacent segments share an overlap of $ON$ data points. The samples within the overlap regions are highlighted in red.}
    \label{fig: overlap_sketch}
\end{figure}

In the absence of windowing and overlapping of data, $\bar{C}_{I;\barell}$ and $\bar{C}_{J;\barm}$ are strictly independent, as implied by~\eqref{eq: correlation_discrete}.
However, the application of a window function and overlapping data segments introduces nontrivial correlations between the the estimators.
In what follows, we will assume that adjacent segments overlap by $O\times 100\%$, where $O$, the so-called \textit{overlap factor} can vary from 0 to 1. 
But for our analyses, we will restrict $O$ to lie between 0 and 0.5.
Otherwise, one segment will overlap with multiple segments, which will make the subsequent calculation much more complex. Given that assumption, we need only study the correlation of estimators between adjacent segments (i.e. $J=I\pm1$). 
We also note that for the analyses~\cite{pygwb,O3stoch, O4aStoch} conducted by the \ac{LVK} collaboration, one usually overlaps the data by 50\%, so the treatment presented here is sufficient for current applications.
For clarity, Fig.~\ref{fig: overlap_sketch} illustrates schematically how the data are divided into overlapping segments.

Let us denote the covariance between $\bar{C}_{I;\barell}$ and $\bar{C}_{J;\barm}$ by $\bar\Sigma_{IJ;\barell\barm}$.
Then, by definition, 
\begin{equation}
\bar\Sigma_{IJ;\barell\barm}=\langle \bar{C}_{I;\barell}\bar{C}_{J;\barm}\rangle-\langle \bar{C}_{I;\barell}\rangle\langle \bar{C}_{J;\barm}\rangle=\frac{1}{M^2}\sum_{p=M\barell-\frac{M}{2}}^{M\barell+\frac{M}{2}-1}\sum_{q=M\barm-\frac{M}{2}}^{M\barm+\frac{M}{2}-1}\hat{\Sigma}_{IJ;pq}\,,
\label{eq: bar_Sigma_ovl_lm}
\end{equation}
where
\begin{equation}
\hat{\Sigma}_{IJ;pq}\equiv\langle\hat{C}_{I;p}\hat{C}_{J;q}\rangle-\langle\hat{C}_{I;p}\rangle\langle\hat{C}_{J;q}\rangle\,.
\label{eq: hat_Sigma_ovl_pq}
\end{equation}
As we did previously to obtain~\eqref{eq: hat_Sigma_pq}, we use Isserlis's theorem to  expand the rhs of the above expression for $\hat{\Sigma}_{IJ;pq}$ as
\begin{equation}
\begin{aligned}
\hat\Sigma_{IJ;pq}=\frac{25}{T^2\gamma_{12;p}^2S_{0;p}^2}  \Bigg[&\left\langle\tilde{\bf{d}}_{1;I;p}\tilde{\bf{d}}_{2;J;q}^*\right\rangle\left\langle\tilde{\bf{d}}_{2;I;p}^*\tilde{\bf{d}}_{1;J;q}\right\rangle+\left\langle\tilde{\bf{d}}_{1;I;p}\tilde{\bf{d}}_{1;J;q}^*\right\rangle\left\langle\tilde{\bf{d}}_{2;I;p}^*\tilde{\bf{d}}_{2;J;q}\right\rangle\\
&\hspace{0.5in}+\left\langle\tilde{\bf{d}}_{1;I;p}^*\tilde{\bf{d}}_{2;J;q}^*\right\rangle\left\langle\tilde{\bf{d}}_{2;I;p}\tilde{\bf{d}}_{1;J;q}\right\rangle+\left\langle\tilde{\bf{d}}_{1;I;p}^*\tilde{\bf{d}}_{1;J;q}^*\right\rangle\left\langle\tilde{\bf{d}}_{2;I;p}\tilde{\bf{d}}_{2;J;q}\right\rangle+({\rm c.c.})\Bigg].
\end{aligned}
\label{eq: cov_ovl}
\end{equation}
To proceed, we need to compute terms like $\left\langle\tilde{\bf{d}}_{1;I;p}\tilde{\bf{d}}_{1;J;q}^*\right\rangle$, which require careful consideration. 
Explicitly, this two-point correlation can be expanded as
\begin{equation}
\left\langle\tilde{\bf{d}}_{1;I;p}\tilde{\bf{d}}_{1;J;q}^*\right\rangle =\frac{1}{T^2} \sum_{j=0}^{N-1}\sum_{k=0}^{N-1} \left\langle \tilde{d}_{1;I;j}\tilde{d}_{1;J;k}^*\right\rangle \tilde{w}_{I;p -j}\tilde{w}_{J;q-k}^*\,,
\label{eq: dd_ww}
\end{equation}
where we separated the DFTs of the data and the windows.
Note that similar results apply as well when the data are associated with two different detectors, i.e., 1 and 2.

We emphasize that $\langle \tilde{d}_{1;I;j}\tilde d^*_{1;J;k}\rangle$ is strictly zero when $I=J$ and $j\neq k$. Nevertheless, for $I\neq J$, this term can be nonzero even when $j\neq k$, due to the \textit{phase shift} introduced by the time offset between two segments. 
The reason for explicitly attaching subscripts $I$ and $J$ to the window functions is that the Fourier modes of the data are intrinsically tied to the specific time segments from which they are computed. In most applications, one multiplies  Fourier modes derived from data segments having identical start and end times, in which case this subtlety does not arise. 
In the present scenario, however,  the windows function $w_{I}$ and $w_{J}$ correspond to segments with different time origins. 
As a result, a constant time shift between the segments must be carefully incorporated into the analysis.

To address the timeshift that comes from the different time origins corresponding to different segments, we will restrict attention in what follows to the case $J=I+1$, recalling that we need $J=I\pm 1$ for there to be a non-zero correlation.
The relevant quadratic expectation value in \eqref{eq: dd_ww} then becomes:
\begin{equation}
    \langle \tilde{d}_{1;I;j}\tilde{d}_{1;I+1;k}^*\rangle=(\Delta t)^2\sum_{r=0}^{N-1}\sum_{s=0}^{N-1}\langle d_{1;I;r}d_{1;I+1;s}\rangle\, e^{-2\pi i \frac{rj}{N}}e^{2\pi i\frac{sk}{N}}\, \underbrace{e^{2\pi i\frac{(1-O)Nk}{N}}}_{\mathrm{phase~shift}}\,,
\label{eq: d_1Id_1J}
\end{equation}
where the last exponential term encodes the phase shift mentioned above. 
To remove the offset associated with different time origins, we will define an ``extended" time-series $d_{1;t}^{I^+}$ with index $t$ running from the start of segment $I$ to the end of segment $I+1$:%
\footnote{We can define a similar expression $d^{I^-}_{1;t}$ for the case $J=I-1$, where $t$ runs from the start of segment $I-1$ to the end of segment $I$. 
We can also change the detector index from 1 to 2 if we need to evaluate the quadratic expectation value of the product of data from two different detectors.}
\begin{equation}
    d^{I^+}_{1;t}:=
    \begin{cases}
    d_{1;I;t}, &0\leqslant t\leqslant (1-O)N-1\\
    d_{1;I+1;t-(1-O)N}, &(1-O)N\leqslant t\leqslant (2-O)N-1
    \end{cases}
    \label{e:dI+_def}
\end{equation}
Using this definition to rewrite~\eqref{eq: d_1Id_1J}, we find (see App.~\ref{app: proof of d1Ip_d1Jq} for details):
\begin{equation}
\begin{aligned}
 \langle \tilde{d}_{1;I;j}\tilde{d}_{1;I+1;k}^*\rangle=(\Delta t)^2\sum_{r=0}^{N-1}\sum_{s=(1-O)N}^{(2-O)N-1}\langle d^{I^+}_{1;r}d^{I^+}_{1;s}\rangle e^{-2\pi i \frac{(rj-sk)}{N}}.
\end{aligned}
\label{eq: d1Ip_d1Jq_begin}
\end{equation}
Note that $\langle d^{I^+}_{1;r}d^{I^+}_{1;s}\rangle$ is the auto-correlation function $\sigma^2_1 R_{1;|r-s|}$ of the time series $d^{I^+}_{1;t}$, whose domain $|r-s|$ can be any integers between 0 and $+\infty$. 
But, in practice, we use only a segment's worth of data (of duration $T$) to compute the auto-correlation function and its Fourier transform, the \ac{PSD}. 
This means that we require $R_{1;|r-s|}=0$ when $|r-s|\geqslant N$, leading to a change in the upper limit of the summation over $s$ from 
$s=(2-O)N-1$ to
$s={\rm min} \big\{r+N-1, (2-O)N-1\big\}$.
As shown in detail in App.~\ref{app: proof of d1Ip_d1Jq_mid}, this implies that~\eqref{eq: d1Ip_d1Jq_begin} reduces to
\begin{equation}
 \langle \tilde{d}_{1;I;j}\tilde{d}_{1;I+1;k}^*\rangle\approx\frac{T}{2N}P_{1;j}\sum_{t=(1-O)N}^{N-1}e^{-2\pi i\frac{t(j-k)}{N}}.
\label{eq: d1Ip_d1Jq_mid}
\end{equation}
Now, the window functions that appear in~\eqref{eq: dd_ww} can be written in a fashion similar to~\eqref{eq: d_1Id_1J}:
\begin{equation}
    \tilde w_{I;p-j}\tilde w^*_{I+1;q-k}=(\Delta t)^2\sum_{r=0}^{N-1}\sum_{s=0}^{N-1}w_rw_s e^{-2\pi i\frac{r(p-j)}{N}}e^{2\pi i\frac{s(q-k)}{N}} e^{2\pi i\frac{(1-O)N(q-k)}{N}}.
    \label{eq: ww}
\end{equation}
So, if we substitute~\eqref{eq: d1Ip_d1Jq_mid} and~\eqref{eq: ww} back into~\eqref{eq: dd_ww} we find (again, see App.~\ref{app: proof of d1Ip_d1Jq_end} for details):
\begin{equation}
\left\langle\tilde{\bf{d}}_{1;I;p}\tilde{\bf{d}}_{1;I+1;q}^*\right\rangle\approx \frac{T}{2N}P_{1;p}\sum_{t=(1-O)N}^{N-1}w_t w_{t-(1-O)N}e^{-2\pi i\frac{t(p-q)}{N}}.
\label{eq: d1Ip_d1Jq_end}
\end{equation}
Although the above result was derived for the case $J=I+1$, it is also valid for for $J=I-1$.

A similar calculation to~\eqref{eq: d1Ip_d1Jq_end} above needs to be performed for all of the quadratic expectation values in~\eqref{eq: cov_ovl}.
Doing so, and combining the results, allow us to compute the covariance $\hat\Sigma_{IJ;pq}$:
\begin{equation}
\begin{aligned}
 \hat\Sigma_{IJ;pq}\approx \delta_{I,J=I\pm 1}\frac{\mathcal{P}_{p}}{N^2}\sum_{j=(1-O)N}^{N-1}\sum_{k=(1-O)N}^{N-1}w_jw_{j-(1-O)N}w_kw_{k-(1-O)N} \left(e^{2\pi i(j-k)\frac{(p-q)}{N}}+e^{2\pi i(j-k)\frac{(p+q)}{N}}\right)\,,
\label{e:sighat_IJ_lm}
\end{aligned}
\end{equation}
which is non-zero only for $J=I\pm 1$.
We note that, in practice, $\mathcal{P}_p$ is not known \textit{a priori}, but needs to be estimated from the data. 
For this case, where we have segments $I$ and $J=I\pm 1$, it is natural to estimate $\mathcal{P}_{p}$ by $(\mathcal{P}_{I;p}+\mathcal{P}_{J;p})/2$, which is also the choice adopted in~\cite{Note_1}. 
So making this change to $\mathcal{P}_{p}$ 
and substituting the resulting expression back into~\eqref{eq: bar_Sigma_ovl_lm}, we obtain:
\begin{tcolorbox}[ title=Covariance between $\bar C_{I;\barell}$ and $\bar C_{J;\barm}$]
\begin{equation}
\begin{aligned}
\bar\Sigma_{IJ;\barell\barm}
&=\delta_{I,J=I\pm 1}\sum_{p=M\barell-\frac{M}{2}}^{M\barell+\frac{M}{2}-1}\sum_{q=M\barm-\frac{M}{2}}^{M\barm+\frac{M}{2}-1}\frac{\mathcal{P}_{I;p}+\mathcal{P}_{J;p}}{2M^2N^2}\\
&\hspace{0.1in}\times\sum_{j,k=(1-O)N}^{N-1}w_jw_{j-(1-O)N}w_kw_{k-(1-O)N} \left(e^{2\pi i(j-k)\frac{(p-q)}{N}}+e^{2\pi i(j-k)\frac{(p+q)}{N}}\right). 
\label{eq: bar_Sigma_ovl_lm_complete}
\end{aligned}
\end{equation}
\end{tcolorbox}

For later convenience, we calculate the diagonal entries of the covariance matrix:
\begin{equation}
\bar\Sigma_{IJ;\barell\barell}=
\delta_{I,J=I\pm 1}\sum_{p,q=M\barell-\frac{M}{2}}^{M\barell+\frac{M}{2}-1}\frac{\mathcal{P}_{I;p}+\mathcal{P}_{J;p}}{2M^2N^2}\sum_{j,k=(1-O)N}^{N-1}w_jw_{j-(1-O)N}w_kw_{k-(1-O)N} \left(e^{2\pi i(j-k)\frac{(p-q)}{N}}+e^{2\pi i(j-k)\frac{(p+q)}{N}}\right)\,.
\end{equation}
Similar to previous calculations presented in Sec.~\ref{s:single_segment}, we can pull $\mathcal{P}_{I;p}$ and $\mathcal{P}_{J;p}$ out of the summation over $p$ (replacing the index $p$ by $M\barell$), which also allows us to drop the contribution coming from the second exponential term in parentheses.
This leads to 
\begin{equation}
\bar\Sigma_{IJ;\barell\barell}\approx
\delta_{I,J=I\pm 1}\frac{\mathcal{P}_{I;M\barell}+\mathcal{P}_{J;M\barell}}{2N^2}\sum_{j,k=(1-O)N}^{N-1}w_jw_{j-(1-O)N}w_kw_{k-(1-O)N}\,\frac{1}{M^2}\sum_{p,q=M\barell-\frac{M}{2}}^{M\barell+\frac{M}{2}-1}e^{2\pi i(j-k)\frac{(p-q)}{N}}\,.
\end{equation}
The summations over $p$ and $q$ of the exponential term then reduces to $\mathcal{K}_{j,k;M,N}$ (by the same calculation given in in App.~\ref{app: proof_bar_Sigma_l_m_final}), leading to:
\begin{equation}    \bar\Sigma_{IJ;\barell\barell}\approx\delta_{I,J=I\pm 1}\frac{O^2}{2}\frac{(\overline{w^2_{\mathcal{K},\ovl}})^2}{(\overline{w^2})^2}(\hat{\sigma}^2_{I;M\barell}+\hat{\sigma}^2_{J;M\barell})\equiv\delta_{I,J=I\pm 1}\frac{O^2}{2}k_{\rm nb}(\bar{\sigma}^2_{I;\barell}+\bar{\sigma}^2_{J;\barell})\,,
    \label{eq: bar_Sigma_ovl_barlbarl}
\end{equation}
where we defined 
\begin{equation}
(\overline{w^2_{\mathcal{K},\ovl}})^2\equiv \frac{1}{O^2N^2}\sum_{j,k=(1-O)N}^{N-1}w_jw_{j-(1-O)N}w_kw_{k-(1-O)N}\mathcal{K}_{j,k;M,N}\,,\qquad  k_{\rm nb}\equiv \frac{(\overline{w^2_{\mathcal{K},\ovl}})^2}{(\overline{w_\mathcal{K}^2})^2}\,,
\label{e:knb}
\end{equation}
with the “nb” subscript on $k_{\rm nb}$ indicating that this factor arises in the context of narrowband estimators.

To finish this section, we note that we can rewrite $\hat{\Sigma}_{IJ;pq}$ and $\bar{\Sigma}_{IJ;\barell\barm}$ in the same manner as~\eqref{eq: def_C_w} and~\eqref{eq: def_of_bar_C_w}. 
To do so, we first define an auxiliary function $W_j$, such that
\begin{equation}
    W_j\equiv
    \begin{cases}
        0\,,&\qquad 0\leqslant j\leqslant (1-O)N-1\\
        w_jw_{j-(1-O)N}\,,&\qquad (1-O)N\leqslant j\leqslant N-1
    \end{cases}
\end{equation}
and denote its \ac{DFT} as $\widetilde{w^2}_{\ovl;p}$. We then have
\begin{equation}
\hat\Sigma_{IJ;pq}\approx\delta_{I,J=I\pm 1}\left(\frac{\hat\sigma_{I;p}^2+\hat\sigma_{J;p}^2}{2}\right)\frac{|\widetilde{w^2}_{\ovl;p-q}|^2+|\widetilde{w^2}_{\ovl;p+q}|^2}{T^2(\overline{w^2})^2}\equiv 
\delta_{I,J=I\pm 1}\left(\frac{\hat\sigma_{I;p}^2+\hat\sigma_{J;p}^2}{2}\right)\mathcal{C}_{w,\ovl;pq}\,,
\label{eq: def_C_w_ovl}
\end{equation}
and
\begin{equation}
    \bar\Sigma_{IJ;\barell\barm}\approx\delta_{I,J=I\pm 1}\left(\frac{\hat\sigma^2_{I;M\barell}+\hat\sigma^2_{J;M\barell}}{2}\right)\frac{1}{M^2}\sum_{p=M\barell-\frac{M}{2}}^{M\barell+\frac{M}{2}-1}\sum_{q=M\barm-\frac{M}{2}}^{M\barm+\frac{M}{2}-1}\mathcal{C}_{w,\ovl;pq}\equiv \delta_{I,J=I\pm 1}\left(\frac{\hat\sigma^2_{I;M\barell}+\hat\sigma^2_{J;M\barell}}{2}\right)\bar{\mathcal{C}}_{w,\ovl;\barell\barm}\,.
    \label{eq: def_bar_C_w_ovl}
\end{equation}

\section{Multi-segment optimal estimators}
\label{s:multi_segment}

Given the results from the previous sections, we are now ready to combine all the available data (both over coarse-grained frequency bins and time segments) to obtain final optimal broadband and narrowband estimators. 
Since the combination procedure is linear, the order of combining the data (e.g., over frequencies first then segments, or vice versa) does not matter as long as we take into account the nonzero correlations between the resultant estimators. 
That said, we can choose the simplest “path” to realize the combination.
Since the “path” adopted in Refs.~\cite{Note_1, pygwb} is fairly simple, we also adopt this path in our work.
So, we will first separately combine estimators corresponding to odd and even data segments.
Then given the covariance between these two intermediate estimators, we optimally combine them to form our final estimator. 
We will discuss the optimal narrowband estimator first, and then discuss the optimal broadband estimator.

\subsection{Narrowband estimator}
\label{s:multi_seg_narrow}

We will combine the estimators $\bar C_{I;\barell}$ for all odd segments ($I=1, 3, 5,...$) to obtain $\bar C_{o;\barell}$, and for all even segments ($I=0, 2, 4,...$) to obtain $\bar C_{e;\barell}$. 
In what follows, we will use $I\in o$
($I\in e$) to denote that $I$ labels an odd (even) segment.
To facilitate comparison with the results reported in the literature~\cite{pygwb,O3stoch, O4aStoch}, hereafter we will further restrict $O$ to $0.5$. For $O=0.5$, odd (even) segments do not overlap with other odd (even) segments.
Thus, we can use the standard inverse-variance weighting to obtain the optimal combinations $\bar C_{o;\barell}$ and $\bar C_{e;\barell}$ and their corresponding variances $\bar\sigma^2_{o;\barell}$ and $\bar\sigma^2_{e;\barell}$:
\begin{equation}\begin{aligned}
&\bar C_{o;\barell}=\frac{\sum_{I\in o}\bar\sigma^{-2}_{I;\barell}\bar C_{I;\barell}}{\sum_{I\in o}\bar\sigma^{-2}_{I;\barell}}\,,\qquad \bar\sigma^2_{o;\barell}=\frac{1}{\sum_{I\in o}\bar\sigma^{-2}_{I;\barell}}\,,\\
& \bar C_{e;\barell}=\frac{\sum_{I\in e}\bar\sigma^{-2}_{I;\barell}\bar C_{I;\barell}}{\sum_{I\in e}\bar\sigma^{-2}_{I;\barell}}\,,\qquad \bar\sigma^2_{e;\barell}=\frac{1}{\sum_{I\in e}\bar\sigma^{-2}_{I;\barell}}\,.
\label{eq: bar_C_o_l_bar_var_o_l_bar_C_e_l_bar_var_e_l}
\end{aligned}
\end{equation}

In order to further combine these two spectra to obtain the final narrowband estimator, we need to study their covariance:
\begin{equation}
\bar\Sigma_{oe;\barell}\equiv
\begin{pmatrix}        \bar\sigma_{o;\barell}^2&\bar\sigma_{oe;\barell}^2\\    \bar\sigma_{oe;\barell}^2&\sigma_{e;\barell}^2
\end{pmatrix},
\end{equation}
where $\bar\sigma^2_{oe;\barell}$ is given by%
\footnote{Although we write the off-diagonal element $\bar\sigma^2_{oe;\barell}$ as a ``squared" quantity, it can take on  positive or negative values.}
\begin{equation}
\begin{aligned}
 \bar\sigma_{oe;\barell}^2&=\langle \bar{C}_{o;\barell}\bar{C}_{e;\barell}\rangle -\langle \bar{C}_{o;\barell}\rangle\langle \bar{C}_{e;\barell}\rangle\\
 &=\frac{\sum_{I\in o}\sum_{J\in e}\bar\sigma^{-2}_{I;\barell}\bar\sigma^{-2}_{J;\barell}\Big(\langle \bar{C}_{I;\barell}\bar{C}_{J;\barell}\rangle-\langle \bar{C}_{I;\barell}\rangle \langle \bar{C}_{J;\barell}\rangle\Big)}{\sum_{I\in o}\sum_{J\in e}\bar\sigma^{-2}_{I;\barell}\bar\sigma^{-2}_{J;\barell}}\\
&=\bar\sigma_{o;\barell}^2\bar\sigma_{e;\barell}^2\sum_{I\in o}\sum_{J\in e}\bar\sigma_{I;\barell}^{-2}\bar\sigma_{J;\barell}^{-2}\,\bar\Sigma_{IJ;\barell\barell}\,.
\end{aligned}
\label{eq: var_oe_l}
\end{equation}
Substituting~\eqref{eq: bar_Sigma_ovl_barlbarl} into the above expression, we obtain
\begin{equation}
    \bar\sigma^2_{oe;\barell}=\frac{}{}\bar\sigma^2_{o;\barell}\bar\sigma^2_{e;\barell}\sum_{I\in o}\sum_{J=I\pm1}\frac{O^2}{2}k_{\rm nb}(\bar{\sigma}^{-2}_{I;\barell}+\bar{\sigma}^{-2}_{J;\barell})\,.
\end{equation}
Expanding the summations over $I$ and $J$ yields (see App.~\ref{app: proof_bar_var} for intermediate steps):
\begin{equation}
    \bar\sigma^2_{oe;\barell}=O^2k_{\rm nb}\bar\sigma^2_{o;\barell}\bar\sigma^2_{e;\barell}\tilde\sigma^{-2}_{\barell}\,,\qquad  \tilde{\sigma}^{-2}_{\barell}\equiv\left(\bar\sigma^{-2}_{o;\barell}+\bar\sigma^{-2}_{e;\barell}-\frac{1}{2}\left(\bar\sigma^{-2}_{0;\barell}+\sigma^{-2}_{-1;\barell}\right)\right)\,.
    \label{eq: bar_var}
\end{equation}
The $-1$ subscript in the last equation corresponds to the index of the last segment---i.e., $I=N_\mathrm{seg}-1$, consistent with the syntax of Python.

We now perform the Lagrange multiplier procedure to construct the optimal (i.e., minimum variance, unbiased) narrowband estimator combining all available data:
\begin{equation}
\bar{C}_{\opt;\barell} = \frac{\lambda_{o;\barell}\bar{C}_{o;\barell}+\lambda_{e;\barell} \bar{C}_{e;\barell}}{\lambda_{o;\barell}+\lambda_{e;\barell}}\,,\qquad
\bar\sigma_{\opt;\barell}^2=\frac{\lambda_{o;\barell}^2\bar{\sigma}_{o;\barell}^2+2\lambda_{o;\barell}\lambda_{e;\barell}\bar\sigma_{oe;\barell}^2+\lambda_{e;\barell}^2\bar\sigma_{e;\barell}^2}{(\lambda_{o;\barell}+\lambda_{e;\barell})^2}\,,
\label{eq: combine_narrowband}
\end{equation}
with weights $\lambda_{o;\barell}$ and $\lambda_{e;\barell}$ given by
\begin{equation}
\lambda_{o;\barell}=\frac{(\bar\sigma_{e;\barell}^2-\bar\sigma_{oe;\barell}^2)^2}{\bar\sigma_{o;\barell}^2\bar\sigma_{e;\barell}^2-\bar\sigma_{oe;\barell}^4}\,,\qquad \lambda_{e;\barell}=\frac{(\bar\sigma_{o;\barell}^2-\bar\sigma_{oe;\barell}^2)^2}{\bar\sigma_{o;\barell}^2\bar\sigma_{e;\barell}^2-\bar\sigma_{oe;\barell}^4}\,.
\end{equation}
Substituting these expressions for $\lambda_{o;\barell}$ and $\lambda_{e;\barell}$ back into $\bar C_{\opt;\barell}$ and $\bar\sigma^2_{\opt;\barell}$ yields
\begin{equation}
\begin{aligned}
 \bar C_{\opt;\barell}&\equiv A_{\barell}\bar C_{o;\barell} + B_{\barell}\bar C_{e;\barell}
\\
&=\frac{\bar\sigma_{o;\barell}^{-2}\left(1- O^2k_{\rm nb}\sigma_{o;\barell}^2\tilde{\sigma}^{-2}_{\barell}\right)}{\bar\sigma_{o;\barell}^{-2}+\sigma_{e;\barell}^{-2}-2O^2k_{\rm nb}\tilde{\sigma}^{-2}_{\barell}}\bar C_{o;\barell} + \frac{\bar\sigma_{e;\barell}^{-2}\Big(1- O^2k_{\rm nb}\bar\sigma_{e;\barell}^2\tilde{\sigma}^{-2}_{\barell}\Big)}{\bar\sigma_{o;\barell}^{-2}+\bar\sigma_{e;\barell}^{-2}-2O^2k_{\rm nb} \bar{\sigma}^{-2}_{\barell}}\bar C_{e;\barell}\\
\end{aligned}
\label{eq: bar_Copt}
\end{equation}
and
\begin{equation}
\bar\sigma^2_{\opt;\barell}=\frac{1- O^4k_{\rm nb}^2\bar\sigma_{o;\barell}^2\bar\sigma_{e;\barell}^2\tilde{\sigma}^{-4}_{\barell}}{\bar\sigma_{o;\barell}^{-2}+\bar\sigma_{e;\barell}^{-2}-2 O^2k_{\rm nb}\tilde{\sigma}^{-2}_{\barell}}\,.
    \label{eq: bar_varopt}
\end{equation}
We note that these two expressions are different from $\bar C_{\opt;\barell}^\star$ and $(\bar\sigma^\star_{\opt;\barell})^2$ adopted in the \texttt{pygwb} package (see  Eqs.~(25) and (26) in~\cite{pygwb}).
We will discuss these differences in more detail later on in Sec.~\ref{s:analytical_comparison}. 

Finally, we consider the covariance between $\bar C_{\opt;\barell}$ and $\bar C_{\opt;\barm}$. 
By definition, we have
\begin{equation}
\begin{aligned}
\bar\Sigma_{\opt;\barell\barm}&\equiv\langle\bar C_{\opt;\barell}\bar C_{\opt;\barm} \rangle-\langle \bar C_{\opt;\barell}\rangle\langle \bar C_{\opt;\barm}\rangle\\
 &=\langle(A_{\barell}\bar C_{o;\barell} + B_{\barell}\bar C_{e;\barell})(A_{\barm}\bar C_{o;\barm} + B_{\barm}\bar C_{e;\barm}) \rangle-\langle A_{\barell}\bar C_{o;\barell} + B_{\barell}\bar C_{e;\barell}\rangle\langle A_{\barm}\bar C_{o;\barm} + B_{\barm}\bar C_{e;\barm}\rangle\\
 &=A_{\barell}A_{\barm}\left(\langle\bar C_{o;\barell}\bar C_{o;\barm} \rangle-\langle\bar C_{o;\barell}\rangle\langle\bar C_{o;\barm}\rangle\right)+B_{\barell}B_{\barm}\left(\langle\bar C_{e;\barell}\bar C_{e;\barm} \rangle-\langle\bar C_{e;\barell}\rangle\langle\bar C_{e;\barm}\rangle\right)\\
&\hspace{0.5in}+A_{\barell}B_{\barm}\left(\langle\bar C_{o;\barell}\bar C_{e;\barm} \rangle-\langle\bar C_{o;\barell}\rangle\langle\bar C_{e;\barm}\rangle\right)+A_{\barm}B_{\barell}\left(\langle\bar C_{o;\barm}\bar C_{e;\barell} \rangle-\langle\bar C_{o;\barm}\rangle\langle\bar C_{e;\barell}\rangle\right)\\
 &\equiv A_{\barell}A_{\barm}\bar\Sigma_{oo;\barell\barm}+B_{\barell}B_{\barm}\bar\Sigma_{ee;\barell\barm}+A_{\barell}B_{\barm}\bar\Sigma_{oe;\barell\barm}+A_{\barm}B_{\barell}\bar\Sigma_{oe;\barm\barell}\,.
\end{aligned}
\label{eq: bar_Sigma_opt_barlbarm}
\end{equation}
The calculations of $\bar\Sigma_{oo;\barell\barm}$, $\bar\Sigma_{oe;\barell\barm}$, and $\bar\Sigma_{ee;\barell\barm}$ using~\eqref{eq: bar_C_o_l_bar_var_o_l_bar_C_e_l_bar_var_e_l} are straightforward but tedious.
Details are given in App.~\ref{app: bar_Sigma_oo_oe_ee}, with the final results:
\begin{equation}
\begin{aligned}
\bar\Sigma_{oo;\barell\barm}\approx \frac{1}{r_M}\bar\sigma^{2}_{o;\barell}\,{\bar{\mathcal{C}}_{w;\barell\barm}}\,, \qquad
\bar\Sigma_{ee;\barell\barm}\approx \frac{1}{r_M}\bar\sigma^{2}_{e;\barell}\,{\bar{\mathcal{C}}_{w;\barell\barm}}\,, \qquad
\bar\Sigma_{oe;\barell\barm} \approx\frac{1}{r_M}\bar\sigma^2_{o;\barell}\bar\sigma^2_{e;\barell}\tilde{\sigma}^{-2}_{\barell}\,{\bar{\mathcal{C}}_{w,\ovl;\barell\barm}}\,.
\label{e:Sigma_oo_ee_oe}
\end{aligned}
\end{equation}
To summarize, we gather together the main results:
\begin{tcolorbox}[ title=Multi-segment narrowband estimator $\bar C_{{\rm opt};\barell}$ and covariance matrix components $\bar \Sigma_{{\rm opt};\barell\barm}$]
\begin{equation}
\begin{aligned}
&\bar C_{\opt;\barell}\equiv A_{\barell}\bar C_{o;\barell} + B_{\barell}\bar C_{e;\barell}\,,
\qquad
\bar\sigma^2_{\opt;\barell}=\frac{1- O^4k_{\rm nb}^2\bar\sigma_{o;\barell}^2\bar\sigma_{e;\barell}^2\tilde{\sigma}^{-4}_{\barell}}{\bar\sigma_{o;\barell}^{-2}+\bar\sigma_{e;\barell}^{-2}-2 O^2k_{\rm nb}\tilde{\sigma}^{-2}_{\barell}}\,,
\\
&\bar\Sigma_{{\rm opt};\barell\barm}\equiv A_{\barell}A_{\barm}\bar\Sigma_{oo;\barell\barm}+B_{\barell}B_{\barm}\bar\Sigma_{ee;\barell\barm}+A_{\barell}B_{\barm}\bar\Sigma_{oe;\barell\barm}+A_{\barm}B_{\barell}\bar\Sigma_{oe;\barm\barell}\,,
\\
&\text{where}\quad
k_{\rm nb}\equiv \frac{(\overline{w^2_{\mathcal{K},\ovl}})^2}{(\overline{w_\mathcal{K}^2})^2}\,,
\qquad
\tilde{\sigma}^{-2}_{\barell}\equiv\left(\bar\sigma^{-2}_{o;\barell}+\bar\sigma^{-2}_{e;\barell}-\frac{1}{2}\left(\bar\sigma^{-2}_{0;\barell}+\sigma^{-2}_{-1;\barell}\right)\right)\,,
\\
&\text{and}\hspace{.2in}
A_{\barell}\equiv\frac{\bar\sigma_{o;\barell}^{-2}\left(1- O^2k_{\rm nb}\sigma_{o;\barell}^2\tilde{\sigma}^{-2}_{\barell}\right)}{\bar\sigma_{o;\barell}^{-2}+\sigma_{e;\barell}^{-2}-2O^2k_{\rm nb}\tilde{\sigma}^{-2}_{\barell}}\,,\qquad
B_{\barell}\equiv\frac{\bar\sigma_{e;\barell}^{-2}\Big(1- O^2k_{\rm nb}\bar\sigma_{e;\barell}^2\tilde{\sigma}^{-2}_{\barell}\Big)}{\bar\sigma_{o;\barell}^{-2}+\bar\sigma_{e;\barell}^{-2}-2O^2k_{\rm nb} \bar{\sigma}^{-2}_{\barell}}\,,
\\
&\text{and}\hspace{.01in}\quad\bar\Sigma_{oo;\barell\barm}\approx \frac{1}{r_M}\bar\sigma^{2}_{o;\barell}\,{\bar{\mathcal{C}}_{w;\barell\barm}}\,, \quad
\bar\Sigma_{ee;\barell\barm}\approx \frac{1}{r_M}\bar\sigma^{2}_{e;\barell}\,{\bar{\mathcal{C}}_{w;\barell\barm}}\,, \quad
\bar\Sigma_{oe;\barell\barm} \approx\frac{1}{r_M}\bar\sigma^2_{o;\barell}\bar\sigma^2_{e;\barell}\tilde{\sigma}^{-2}_{\barell}\,{\bar{\mathcal{C}}_{w,\ovl;\barell\barm}}\,.    
\label{e:final_multiseg_nb}
\end{aligned}
\end{equation}
\end{tcolorbox}

\subsection{Broadband estimator}
\label{s:multi_seg_broad}
In this subsection, we will construct the optimal broadband estimator $\bar C_{\rm opt}$. 
We could start with the results that we just derived for the narrowband estimators, i.e.,~\eqref{eq: bar_Copt} and~\eqref{eq: bar_Sigma_opt_barlbarm}, and then optimally combine those over frequencies.
But that turns out to be rather difficult given the complicated expression for the covariance matrix $\bar\Sigma_{\opt;\barell\barm}$. 
So, we will adopt an alternative strategy that was also used in~\cite{pygwb}.

Suppose one has computed the optimal broadband estimators $\bar C_{\opt;I}$ and their variances $\bar\sigma^2_{\opt;I}$ for all individual segments $I$.
Then as we did in the previous subsection, we can combine all the odd (or even) segments by the normal inverse-variance weighed average to get
\begin{equation}
\begin{aligned}
&\bar C_{o}=\frac{\sum_{I\in o}\bar\sigma^{-2}_{\opt;I}\bar C_{\opt;I}}{\sum_{I\in o}\bar\sigma^{-2}_{\opt}}\,,\qquad\bar\sigma^{2}_{o}=\frac{1}{\sum_{I\in o}\bar\sigma^{-2}_{\opt}}\,,\\
&\bar C_{e}=\frac{\sum_{I\in e}\bar\sigma^{-2}_{\opt;I}\bar C_{\opt;I}}{\sum_{I\in e}\bar\sigma^{-2}_{\opt}}\,,\qquad\bar\sigma^{2}_{e}=\frac{1}{\sum_{I\in e}\bar\sigma^{-2}_{\opt}}\,.
\label{e:multiseg_C_sigma_defs}
\end{aligned}
\end{equation}
As before, we also need to evaluate the covariance matrix:
\begin{equation}
    \bar\Sigma_{oe}=\begin{pmatrix}
        \bar\sigma^2_o&\bar\sigma^2_{oe}\\
        \bar\sigma^2_{oe}&\bar\sigma^2_e
    \end{pmatrix}\,.
    \label{e:oe_cov_broad}
\end{equation}
The diagonal terms have already been calculated, and the off-diagonal term can be written as
\begin{equation}
 \bar\sigma^2_{oe}\equiv\langle \bar{C}_o\bar{C}_e\rangle-\langle\bar{C}_o\rangle\langle \bar{C}_e\rangle=\bar\sigma^2_o\bar\sigma^2_e\sum_{I\in o}\sum_{J\in e}\bar\sigma^{-2}_{\opt;I}\bar\sigma^{-2}_{\opt;J}\bar\Sigma_{\opt;IJ}\,,
\end{equation}
where
\begin{equation}
 \bar\Sigma_{\opt;IJ}\equiv\langle\bar C_{\opt;I}\bar C_{\opt;J}\rangle-\langle\bar C_{\opt;I}\rangle\langle\bar C_{\opt;J}\rangle\,.
 \end{equation}
Using~\eqref{eq: C_opt_var_opt_I_non_coase_grained} expressed in terms of $\mathcal{W}_{\mathcal{K}}\bar\sigma^2_{\barell}$, it is straightforward to show that
\begin{equation}
    \bar\Sigma_{\opt;IJ}=\frac{M^2}{\mathcal{W}_\mathcal{K}^2}\bar\sigma^2_{\opt;I}\bar\sigma^2_{\opt;J}\sum_{\barell=0}^{\frac{N}{2M}-1}\sum_{\barm=0}^{\frac{N}{2M}-1}\bar\sigma^{-2}_{I;\barell}\bar\sigma^{-2}_{J;\barm}\bar\Sigma_{IJ;\barell\barm}\,.
\label{e:sigbar_opt_IJ}
\end{equation}
Then substituting~\eqref{eq: bar_Sigma_ovl_lm_complete}
for $\bar\Sigma_{IJ;\barell\barm}$, we obtain (see App.~\ref{app: proof_bar_Sigma_IJ} for details):
\begin{equation}
\sum_{\barell=0}^{\frac{N}{2M}-1}\sum_{\barm=0}^{\frac{N}{2M}-1}\bar\sigma^{-2}_{I;\barell}\bar\sigma^{-2}_{J;\barm}\bar\Sigma_{IJ;\barell\barm}
=\delta_{I,J=I\pm1}\frac{O}{2M^2}\frac{\overline{w^4_{\ovl}}}{(\overline{w^2_\mathcal{K}})^2}\mathcal{W}_\mathcal{K}(\bar\sigma^{-2}_{\opt;I}+\bar\sigma^{-2}_{\opt;J})\,.
    \label{eq: bar_sigma^2_IJ}
\end{equation}
where
\begin{equation}
\overline{w^4_{\ovl}}\equiv \frac{1}{ON}\sum_{j=(1-O)N}^{N-1}w^2_jw^2_{j-(1-O)N}\,.
\label{e:w4_ovl_def}
\end{equation}
Given these results, we can now rewrite $\bar\sigma_{oe}^2$ as
\begin{equation}
\begin{aligned}
    \bar\sigma^2_{oe}&=\frac{M^2}{\mathcal{W}^2_\mathcal{K}}\bar\sigma^2_o\bar\sigma^2_e\sum_{I\in o}\sum_{J=I\pm1}\frac{O}{2M^2}\frac{\overline{w^4_{\ovl}}}{(\overline{w^2_\mathcal{K}})^2}\mathcal{W}_\mathcal{K}(\bar\sigma^{-2}_{\opt;I}+\bar\sigma^{-2}_{\opt;J})\\
    &=\frac{O}{2}\bar\sigma^2_o\bar\sigma^2_e\frac{\overline{w^4_{\ovl}}}{\overline{w^4}}\sum_{I\in o}\sum_{J=I\pm1}(\bar\sigma^{-2}_{\opt;I}+\bar\sigma^{-2}_{\opt;J})\\
    &=O\bar\sigma^2_o\bar\sigma^2_e\frac{\overline{w^4_{\ovl}}}{\overline{w^4}}\left(\bar\sigma^{-2}_o+\bar\sigma^{-2}_e-\frac{1}{2}(\bar\sigma_1^{-2}+\bar\sigma^{-2}_{-1})\right)\\
    &= Ok_{\rm bb}\bar\sigma^2_o\bar\sigma_e^2\tilde\sigma^{-2}\,,
    \label{e:sigbar_oe}
\end{aligned}
\end{equation}
where
\begin{equation}
k_{\rm bb}\equiv\frac{\overline{w^4_{\ovl}}}{\overline{w^4}}\,,
\qquad
\tilde\sigma^{-2}\equiv \left(\bar\sigma^{-2}_o+\bar\sigma^{-2}_e-\frac{1}{2}(\bar\sigma_1^{-2}+\bar\sigma^{-2}_{-1})\right)\,,
\label{e:kbb}
\end{equation}
with the “bb” subscript on $k_{\rm bb}$ indicating that this factor arises in the context of the broadband estimator.

Finally, now that we know all the elements of $\bar\Sigma_{oe}$, we can combine $\bar C_o$ and $\bar C_e$ to obtain $\bar C_{\rm opt}$ and $\bar\sigma^2_{\rm opt}$, similar to what we did in~\eqref{eq: combine_narrowband}:
\begin{equation}
    \bar C_{\opt}\equiv\frac{\lambda_o\bar C_o+\lambda_e\bar C_e}{\lambda_o+\lambda_e}\,,\qquad \bar\sigma^2_{\opt}=\frac{\lambda_o^2\bar\sigma^2_o+2\lambda_o\lambda_e\bar\sigma^2_{oe}+\lambda_e^2\bar\sigma^2_e}{(\lambda_o+\lambda_e)^2},
\end{equation}
where the weights $\lambda_o$ and $\lambda_e$ are given by
\begin{equation}
    \lambda_o=\frac{(\bar\sigma_e^2-\bar\sigma_{oe}^2)^2}{\bar\sigma_o^2\bar\sigma_e^2-\bar\sigma_{oe}^4}\,,\qquad\lambda_e=\frac{(\bar\sigma_o^2-\bar\sigma_{oe}^2)^2}{\bar\sigma_o^2\bar\sigma_e^2-\bar\sigma_{oe}^4}\,.
\end{equation}
After some algebra, one obtains the following final expressions:
\begin{tcolorbox}[ title=Multi-segment broadband estimator $\bar C_{\rm opt}$ and variance $\bar \sigma^2_{\rm opt}$]
\begin{equation}
\begin{aligned}
    &\bar C_{\opt}=\frac{\bar C_o\bar\sigma^{-2}_o+\bar C_e\bar\sigma^{-2}_e-(\bar C_o+\bar C_e)k_{\rm bb}O\tilde\sigma^{-2}}{\bar\sigma^{-2}_o+\bar\sigma^{-2}_e-2k_{\rm bb}O\tilde{\sigma}^{-2}}\,,\qquad \bar\sigma^2_{\opt}=\frac{1-k_{\rm bb}^2O^2\bar\sigma^2_o\bar\sigma^2_e\tilde{\sigma}^{-4}}{\bar\sigma^{-2}_o+\bar\sigma^{-2}_e-2k_{\rm bb}O\tilde{\sigma}^{-2}}\,,\\
    &\text{where}\quad k_{\rm bb}\equiv\frac{\overline{w^4_{\ovl}}}{\overline{w^4}}\,,
\qquad
\tilde\sigma^{-2}\equiv \left(\bar\sigma^{-2}_o+\bar\sigma^{-2}_e-\frac{1}{2}(\bar\sigma_1^{-2}+\bar\sigma^{-2}_{-1})\right)\,.
    \label{eq: final_broadband}
    \end{aligned}
\end{equation}
\end{tcolorbox}
\noindent
We highlight that these expressions for the broadband estimator $\bar C_{\opt}$ and its variance $\bar\sigma^2_{\opt}$ agree with those found in Eqs.~(25)--(28) in~\cite{pygwb}.
\section{Comparisons and simulations}
\label{s:comparisons}
In this section, we compare the results derived in previous sections with those in \texttt{pygwb}.
We do so both in terms of the analytic expressions (Sec.~\ref{s:analytical_comparison}) and via numerical simulations (Sec.~\ref{s:simulations}).

\subsection{Comparison with \texttt{pygwb}}
\label{s:analytical_comparison}

As we have already mentioned, both the single-segment broadband estimator $\bar C_{\opt;I}$ (given by~\eqref{eq: C_opt_var_opt_I_non_coase_grained}) and the multi-segment broadband estimator $\bar C_{\opt}$ (given by~\eqref{eq: final_broadband}) agree with those in \texttt{pygwb}~\cite{pygwb}, Eqs.~(14), (25), and (28). 
Thus, the published LVK SGWB analyses~\cite{O3stoch,O4aStoch} related to broadband estimators are reliable. 

However, our results for the variance $\bar\sigma^2_{I;\barell}$ of the single-segment narrowband estimator (given by \eqref{eq: def_r_M}) and $\bar\sigma^2_{\opt;\barell}$ for the multi-segment narrowband estimator (given by \eqref{eq: bar_varopt}), differ from those in~\cite{pygwb} (cf.~Eqs.~(25) and (26)). 
In addition, even our multi-segment narrowband estimator $\bar C_{\opt;\barell}$ differs from that in~\cite{pygwb}. These differences can be attributed to the following items:

\begin{enumerate}
    \item In Ref.~\cite{pygwb}, the authors adopt results for the narrowband estimators that are strictly valid only for the broadband estimator derived in~\cite{Note_1,Note_2}. 
    So although the broadband results are valid, the corresponding narrowband quantities are not.
    
    \item In Ref.~\cite{pygwb}, the authors ignore the non-zero correlations between different frequency bins of the single-segment narrowband estimator, which are introduced by the window function.
    In other words, they \textit{incorrectly} assume that $\bar\Sigma_{I;\barell\barm}=0$ for $\barell\neq\barm$.
    
    \item As a consequence of the previous item, the authors of Ref.~\cite{pygwb} assume that the standard inverse-variance-weighted average remains valid when combining narrowband estimators across different frequency bins. 
    In effect, the authors use the correct expression for the broadband estimators to “solve” for the narrowband estimators, assuming that the narrowband estimators obtained in this manner are optimal. 
    However, this optimality assumption is also not correct.
\end{enumerate}

In summary, one cannot regard $\bar C_{\opt;\barell}^\star$ (Eq.~(26) in Ref.~\cite{pygwb}) and $(\bar\sigma^\star_{\opt;\barell})^2$ (Eq.~(25) in Ref.~\cite{pygwb}) as the optimal narrowband estimator and the corresponding variance.
The correct expressions are given above in equations~\eqref{eq: bar_Copt}--\eqref{eq: bar_Sigma_opt_barlbarm}. 
One may be concerned that results of the analyses based on the  \texttt{pygwb} narrowband estimators are also not correct.
Nonetheless, we find that although $\bar C_{\opt;\barell}^\star$ and $(\bar\sigma^\star_{\opt;\barell})^2$ are \textit{not} the correct optimal narrowband estimator and variance, using them in the usual Gaussian likelihood to perform \ac{PE} and to compute Bayes factors for model selection \textit{does} give correct results. 
We demonstrate this claim in detail in Sec.~\ref{sec: PE}.

\subsection{Numerical simulations}
\label{s:simulations}

We conclude this section by verifying several results from the previous sections using numerical simulations.
These simulations consist of 2,000 noise-only \ac{GW} frames, each containing 2048~s worth of data.
The \ac{GW} frames can be thought of 2,000 independent realizations of time-series data, satisfying the weak-signal limit. 
(More specific details regarding the simulations can be found in App.~\ref{app: simulation}.)

\begin{figure}[!htbp]
\centering
\includegraphics[width=0.7\linewidth]{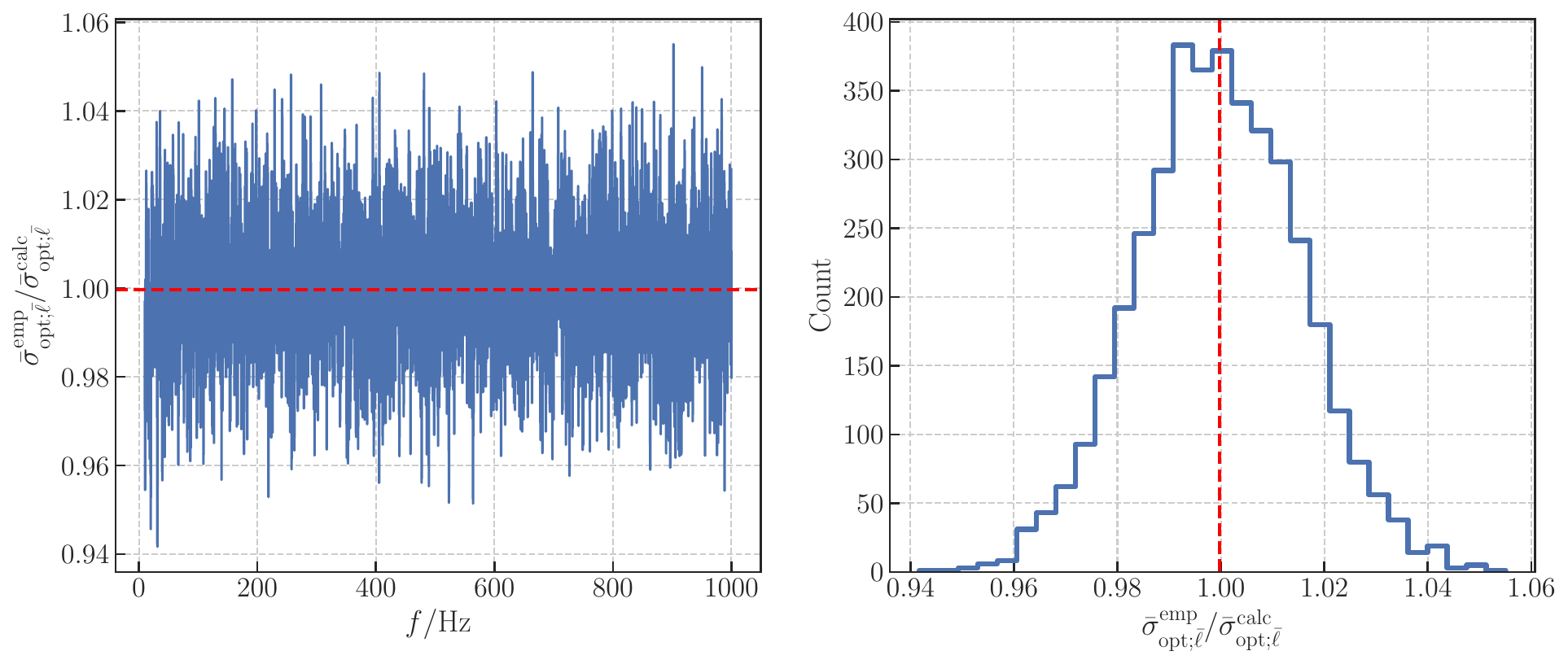}
    \caption{Left panel: Ratio between the empirically estimated standard deviation of the optimal narrowband estimator and its theoretical prediction, plotted as a function of frequency. 
    Right panel: Histogram obtained by stacking the data from the left panel. 
    The red dashed line represents the mean for the distribution for both panels. 
    (For these simulations, we took $T=4~{\rm s}$ and $\delta\!f=1/4$ Hz.)}
    \label{fig: bias_4}
\end{figure}

\begin{figure}[!htbp]
    \centering   \includegraphics[width=0.7\linewidth]{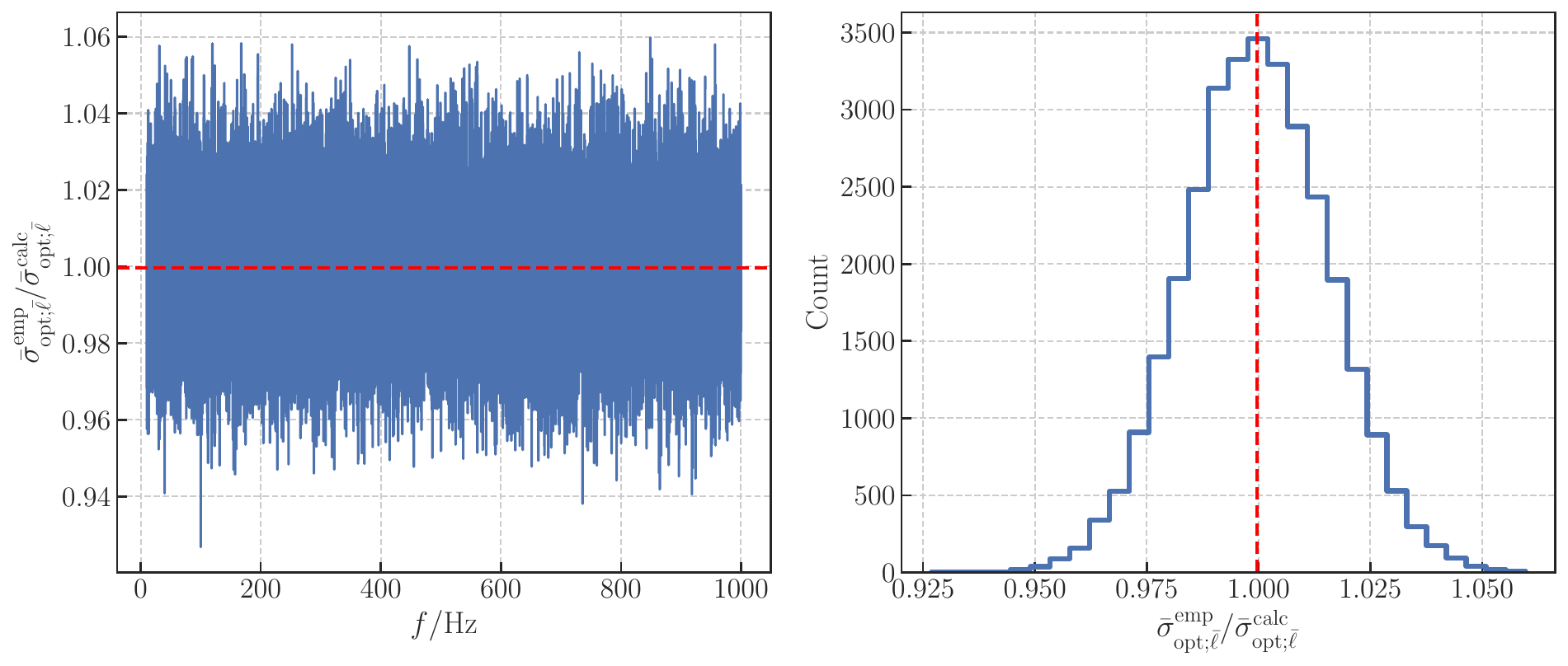}
    \caption{Same as Fig.~\ref{fig: bias_192} but for $T=192~{\rm s}$ and $\delta\!f=1/192$ Hz.}
    \label{fig: bias_192}
\end{figure}

In Fig.~\ref{fig: bias_192}, we compare the empirical standard deviation  $\bar\sigma^\mathrm{emp}_{\opt;\barell}$ of the narrowband estimator $\bar C_{\opt;\barell}$ to its theoretical value $\bar\sigma^\mathrm{calc}_{\opt;\barell}$ for $T=192$~s segments and $\delta\!f=1/32$~Hz frequency resolution. 
We calculate $\bar\sigma^\mathrm{emp}_{\opt;\barell}$ by taking the standard deviation of the 2,000 simulated narrowband estimators $\bar C_{\opt;\barell}$. 
We calculate $\bar\sigma^\mathrm{calc}_{\opt;\barell}$ by taking the average of the 2,000 theoretical narrowband uncertainties $\bar\sigma_{\opt;\barell}$. 
We plot the ratio of these two quantities as a function of frequency in the left panel of Fig.~\ref{fig: bias_192}.
In the right panel, we show a histogram constructed by stacking together all the data from the left panel. 
Both panels indicate that the empirically estimated standard deviation of the narrowband estimator is consistent with our theoretical expectation. 
Figure~\ref{fig: bias_4} is similar to Fig.~\ref{fig: bias_192}, but for $T=4$~s segments and a $\delta\!f=1/4$~Hz frequency resolution.

Similarly, we compare the empirical standard deviation $\bar\sigma^{\mathrm{emp}}_{\opt}$ of the broadband estimator $\bar{C}_{\opt}$ to its theoretical value $\bar\sigma^\mathrm{calc}_{\opt}$. 
We calculate 
$\bar\sigma^{\mathrm{emp}}_{\opt}$ by taking the standard deviation of the 2,000 simulated broadband estimators $\bar{C}_{\opt}$.
We calculate $\bar\sigma^{\mathrm{calc}}_{\opt}$ by taking the average of the 2,000 theoretical broadband uncertainties $\bar\sigma_{\opt}$.
These calculations yield $\bar\sigma^{\mathrm{emp}}_{\opt}/\bar\sigma^\mathrm{calc}_{\opt}=0.99$ for $T=192$ s and $\delta\!f=1/32$ Hz, and 0.98 for $T=4$~s and $\delta\!f=1/4$~Hz. 
We highlight that when deriving the theoretical variance of the broadband estimators, cf.~\eqref{eq: C_opt_var_opt_I_non_coase_grained}, we combine \textit{all} frequency bins (from 0~Hz to Nyquist). 
Only by doing this can we obtain the elegant closed form expressions for $\bar\sigma^2_{\mathrm{opt};I}$ and $\bar\sigma^2_\mathrm{opt}$.
In practice, however, one will always pick $f_{\min}\sim\mathcal{O}(10)~\mathrm{Hz}$ and $f_{\max}<f_\mathrm{Nyquist}$, which causes $\bar\sigma^2_\mathrm{opt}$ to deviate from its “full” value. 
But this deviation is at the percent level, and can be reduced by choosing a lower $f_\mathrm{min}$ and higher $f_\mathrm{max}$, or by manually multiplying the resulting $\bar\sigma^2_\mathrm{opt}$ by a small correction factor to make it consistent with the distribution of broadband estimator $\bar C_\mathrm{opt}$. 

\begin{figure}[!htbp]
\centering
\includegraphics[width=0.45\linewidth]{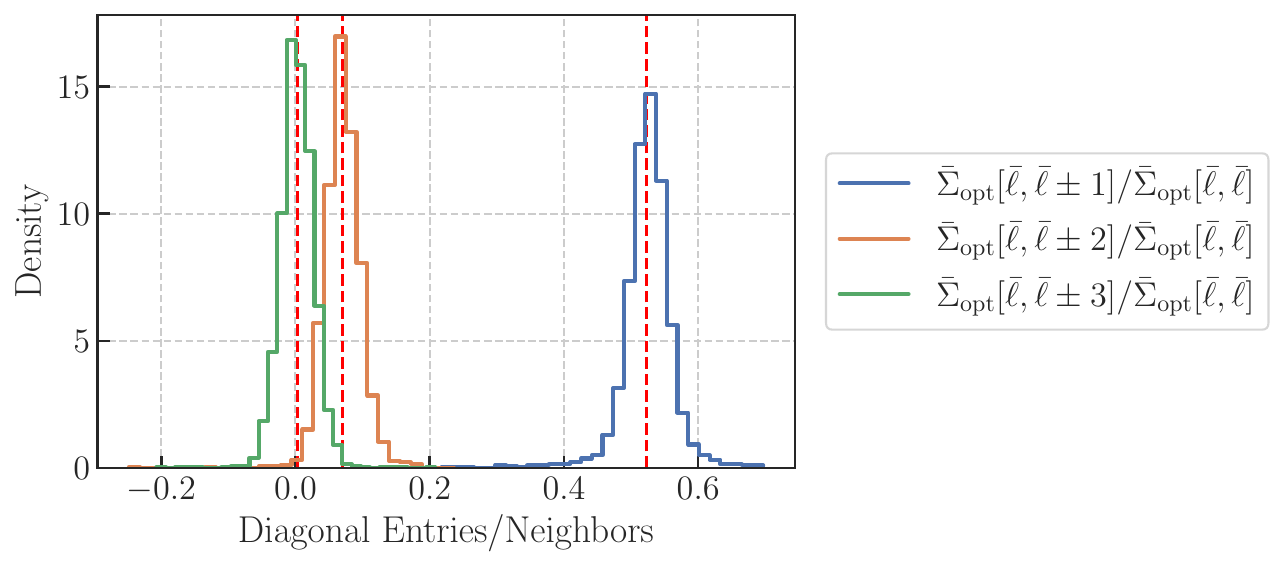}
        \includegraphics[width=0.45\linewidth]{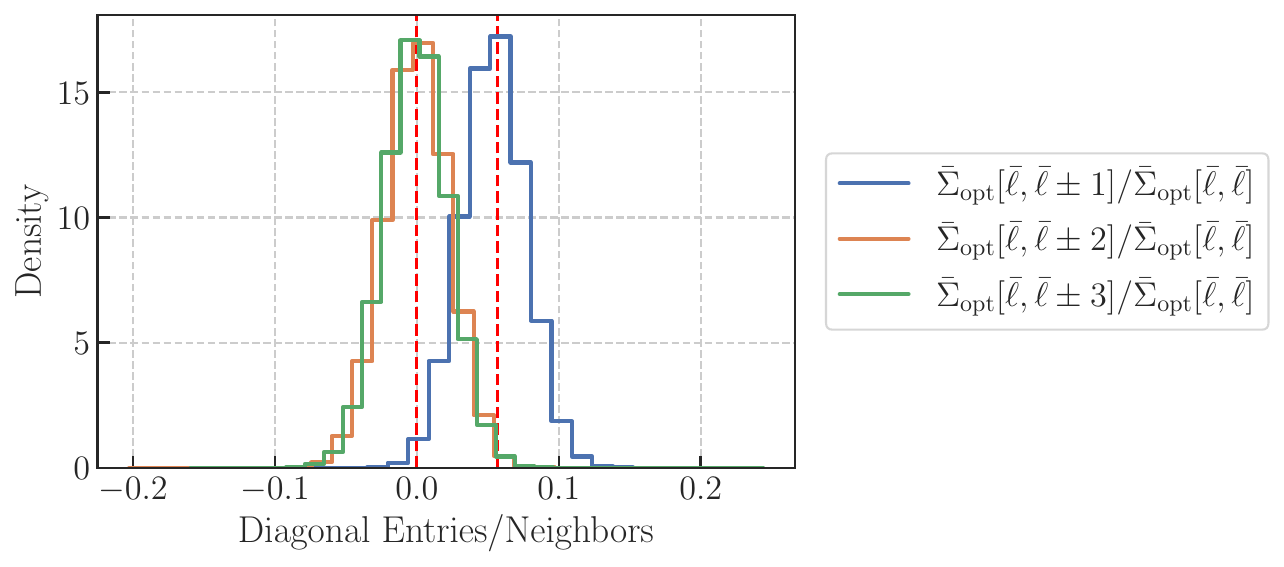}
    \caption{Ratios between the diagonal elements $\bar\Sigma_{\opt;\barell\barell}$ and its first three neighbors. Left panel: $T=4$~s and $\delta\!f=1/4$~Hz. Right panel: $T=192$~s and $\delta\!f=1/32$~Hz. 
    The theoretical values of these ratios are indicated using red dashed vertical lines. }
    \label{fig: covariance}
\end{figure}

Next, we examine the covariance between the narrowband estimators $\bar C_{\opt;\barell}$ and $\bar C_{\opt;\barm}$.
Results are shown in Fig.~\ref{fig: covariance} for $T=4$ s and $\delta\!f=1/4$~Hz (left panel) and $T=192$~s and $\delta\!f=1/32$~Hz (right panel).
Since the covariance between $\bar C_{\opt;\barell}$ and $\bar C_{\opt;\barm}$ is negligible when $|\barell-\barm|>3$ (as one can check numerically via~\eqref{eq: bar_Sigma_opt_barlbarm}), we only consider $|\barell-\barm|\leqslant 3$ in these plots. 
The histograms are obtained by empirically estimating the covariance matrix $\bar\Sigma_{\opt;\barell\barm}$ using the 2,000 simulated frames, and then computing the ratio between the diagonal entries of this matrix with the three neighboring off-diagonal elements. 
The red dashed vertical lines represent the theoretical values of the corresponding ratios.
For both cases of segment duration and frequency resolution, we observe consistency between the the mean of the corresponding histograms and the expected theoretical values.

\begin{figure}[!htbp]
    \centering
    \includegraphics[width=0.45\linewidth]{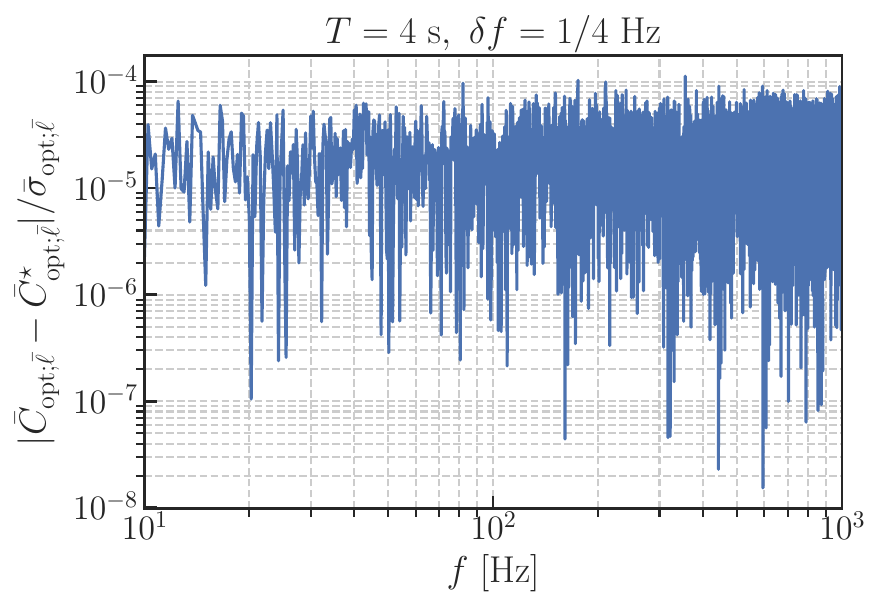}
    \includegraphics[width=0.45\linewidth]{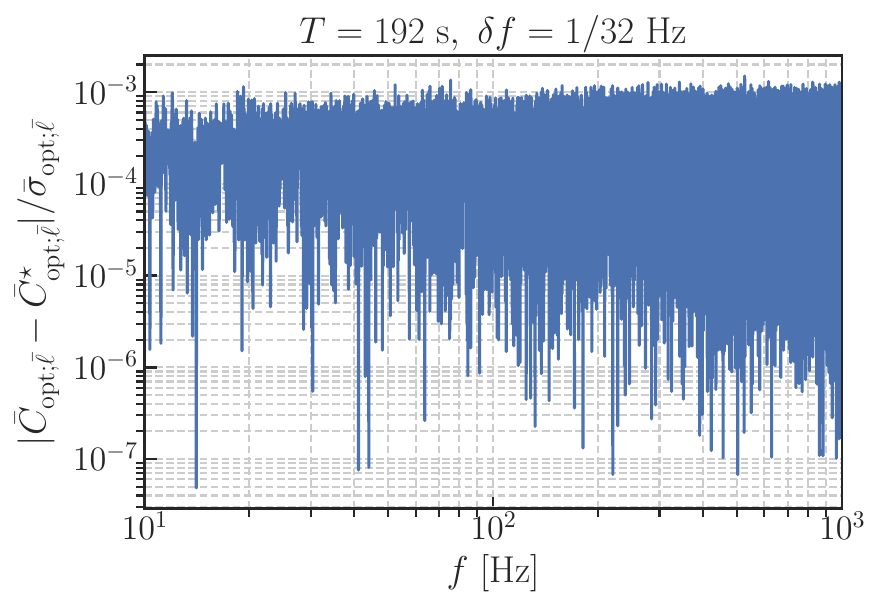}
    \caption{Ratio between the absolute difference of $\bar C_{\opt}(f)$ and $\bar C_{\opt}^\star(f)$ and $\bar\sigma_{\opt}(f)$. 
    The left panel corresponds to $T=4$~s and $\delta\!f= 1/4$~Hz case, while the right panel corresponds to $T=192$~s and $\delta\!f=1/32$~Hz case.}
    \label{fig: narrowband_spetra_comparison}
\end{figure}

\begin{figure}[!htbp]
    \centering
    \includegraphics[width=0.8\linewidth]{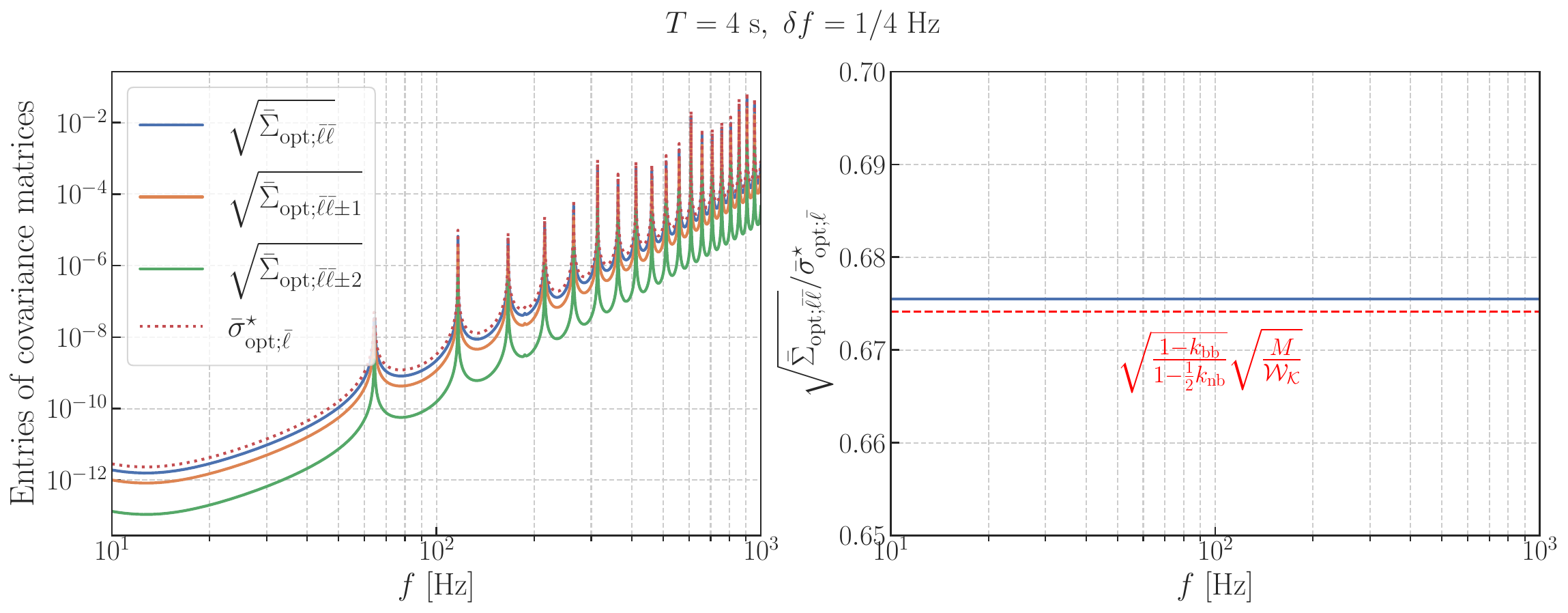}
    \includegraphics[width=0.8\linewidth]{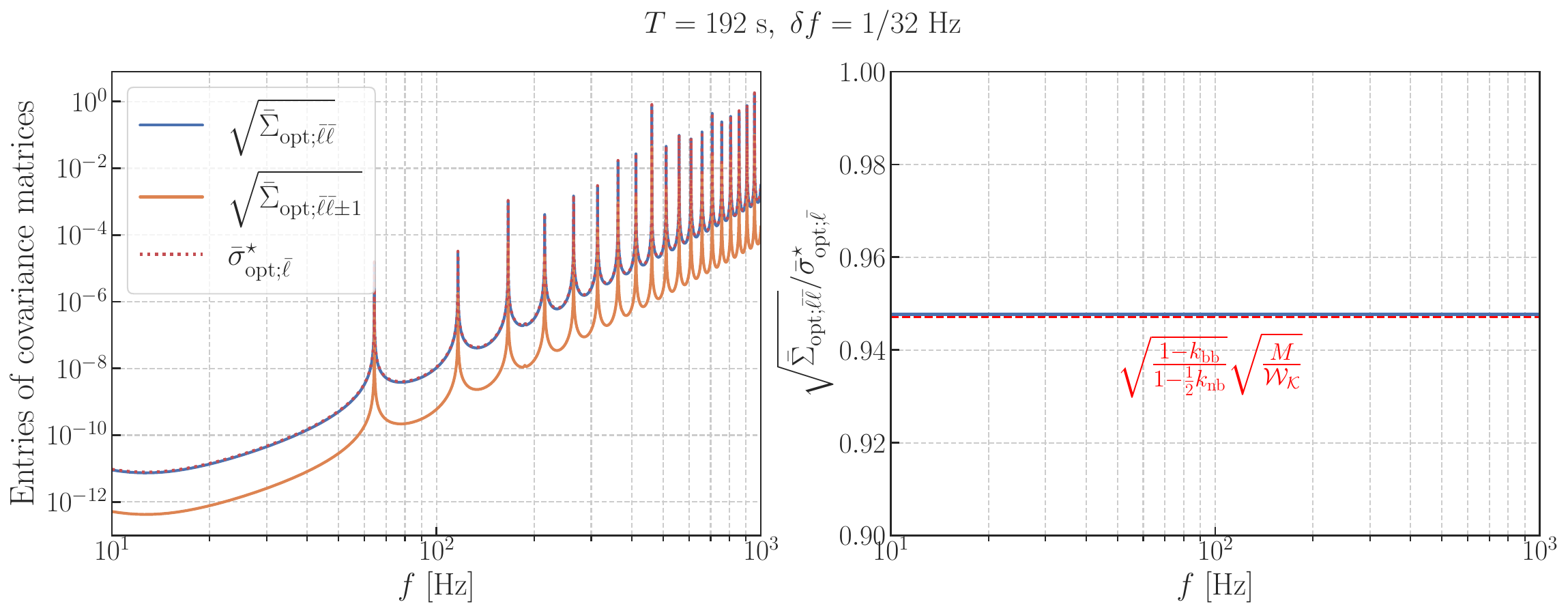}
    \caption{Comparison of $\sqrt{\Sigma_{\opt;\barell\barell}}\,,\sqrt{\Sigma_{\opt;\barell\barell\pm1}}$, and $\sqrt{\Sigma_{\opt;\barell\barell\pm2}}$ computed in this paper with $\bar\sigma^\star_{\opt;\barell}$ computed using \texttt{pygwb}. 
    The left panels show these spectra as functions of frequency, while the right panels show the ratio between $\sqrt{\bar\Sigma_{\opt;\barell\barell}}$ and $\bar\sigma^\star_{\opt;\barell}$, again as functions of frequency. The red dashed lines correspond to the theoretical expectation of this ratio, see discussion around~\eqref{eq: theoretical_ratio} for details. The upper two panels correspond to the case $T=4$~s and $\delta\!f=1/4$~Hz, while the lower two panels correspond to the case $T=192$~s and $\delta\!f=1/32$~Hz.}
    \label{fig: narrowband_variance_comparison}
\end{figure}

Finally, in Figs.~\ref{fig: narrowband_spetra_comparison} and~\ref{fig: narrowband_variance_comparison}, we compare the narrowband estimator and its variance calculated in this paper to the corresponding quantities calculated by \texttt{pygwb}. 
In Fig.~\ref{fig: narrowband_spetra_comparison}, we compare the spectra $\bar C_{{\rm opt};\barell}$ and $\bar C_{{\rm opt};\barell}^\star$ for both $T=4$~s and $\delta\!f=1/4$~Hz (left panel) and $T=192$~s and $\delta\!f=1/32$~Hz (right panel). 
These plots shows that, although the expressions for the narrowband estimators are different, the fractional difference is negligible compared to $\bar\sigma_{\opt;\barell}$.

In Fig.~\ref{fig: narrowband_variance_comparison}, we compare the standard deviation of the narrowband estimators. 
The top two panels of this plot are for $T=4$~s and $\delta\!f=1/4$~Hz, while the bottom two panels are for $T=192$~s and $\delta\!f=1/32$~Hz.
In the left column, we plot the diagonal entries of $\sqrt{\bar\Sigma_{\opt;\barell\barell}}$ and its neighbors, including $\bar\sigma^\star_{{\opt};\barell}$ for comparison. 
In both cases, $\bar\sigma^\star_{\opt;\barell}$ is greater than $\sqrt{\bar\Sigma_{\opt;\barell\barell}}$ as expected, since $\bar\sigma^\star_{{\opt};\ell}$ can be regarded as the diagonal entries of an effective covariance matrix, whose off-diagonal entries are all zero, with the diagonal entries absorbing all the correlation between the different frequency bins. 
(We will discuss this property of $\bar\sigma^\star_{\opt;\barell}$ in more detail at the end of the next section.)

In the right panel of Fig.~\ref{fig: narrowband_variance_comparison}, we plot the ratio between $\sqrt{\bar\Sigma_{\opt;\barell\barell}}$ and $\bar\sigma^\star_{\opt;\barell}$. 
For $T=4$~s and $\delta\!f=1/4$~Hz, this ratio is 0.67, which implies a non-negligible correlation between different frequency bins. 
For $T=192$~s and $\delta\!f=1/32$ Hz, this ratio increases to approximately 0.95, as the frequency resolution is greater than $\Delta\!f=1/T$ due to coarse-graining.
For this case, the correlation between different frequency bins is consequently reduced. 

We also note that the ratio shown in the right panel of of Fig.~\ref{fig: narrowband_variance_comparison} is largely independent of frequency. 
This can be understood by comparing~\eqref{eq: bar_varopt} of this work with (25) of~\cite{pygwb}.
For our analysis, we have $O^2k_{\rm nb}\ll 1$ and $\tilde\sigma_{\barell}^{-2}\simeq(\bar\sigma^{-2}_{o;\barell}+\bar\sigma^{-2}_{e;\barell})$. 
Thus, keeping term up to first-order in $k_{\rm{nb}}$, we find 
\begin{equation}
\frac{1}{\bar\Sigma_{\opt;\barell\barell}}\equiv
\bar\sigma^{-2}_{\opt;\barell}\simeq\left(1-\frac{1}{2}k_{\rm nb}\right)(\bar\sigma_{o;\barell}^{-2}+\bar\sigma^{-2}_{e;\barell})\,.
\end{equation}
A similar expression can also be obtained using the formalism in \cite{pygwb}: 
\begin{equation}
(\bar\sigma^\star_{\opt;\barell})^{-2}\simeq (1-k_{\rm bb})\left((\bar\sigma^\star_{o;\barell})^{-2}+(\bar\sigma^\star_{e;\barell})^{-2}\right)\,.
\end{equation}
Consequently, the ratio between $\sqrt{\bar\Sigma_{\opt;\barell\barell}}$ and $\bar\sigma^\star_{\opt;\barell}$ has the expected form
\begin{equation}
\frac{\sqrt{\bar\Sigma_{\opt;\barell\barell}}}{\bar\sigma^\star_{\opt;\barell}}\simeq\sqrt{\frac{1-k_{\rm bb}}{1-\frac{1}{2}k_{\rm nb}}}\sqrt{\frac{(\bar\sigma^\star_{o;\barell})^{-2}+(\bar\sigma^\star_{e;\barell})^{-2}}{\bar\sigma_{o;\barell}^{-2}+\bar\sigma_{o;\barell}^{-2}}}=\sqrt{\frac{1-k_{\rm bb}}{1-\frac{1}{2}k_{\rm nb}}}\sqrt{\frac{M}{\mathcal{W}_\mathcal{K}}}\,,
    \label{eq: theoretical_ratio}
\end{equation}
explaining the nearly frequency-independent behavior observed in the figure. 
We show the corresponding values of~\eqref{eq: theoretical_ratio} in red dashed lines in Fig.~\ref{fig: narrowband_variance_comparison}, which are consistent with the blue lines as we expect.

\section{Parameter estimation using the optimal narrowband estimator}
\label{sec: PE}
In the last section, we derived the expressions for the optimal narrowband estimator $\Bar{C}_{\opt;\barell}$ and its covariance matrix $\bar{\Sigma}_{\opt;\barell \barm}$. 
We also compared these quantities with the narrowband estimator $\bar{C}^\star_{\opt;\barell}$ and its “variance” $(\bar{\sigma}^\star_{\opt;\barell})^2$ as defined and computed in the \texttt{pygwb} package~\cite{pygwb}. 
In this section, we perform Bayesian parameter estimation (PE) using $\bar{C}_{\opt;\barell}$ and $\bar{\Sigma}_{\opt;\barell \barm}$, and demonstrate that, although $\bar{C}^\star_{\opt;\barell}$ and $(\bar{\sigma}^\star_{\opt;\barell})^2$ do not correspond to the true optimal narrowband estimator and its associated variance, using them for \ac{PE} yields results that are identical to those obtained using $\bar{C}_{\opt;\barell}$ and $\bar{\Sigma}_{\opt;\barell \barm}$. 
For simplicity, we do not include windowing in the following discussion. 
Nevertheless, the conclusions we draw are expected to hold irrespective of whether the data are windowed.

\subsection{Likelihood functions and parameter estimation for simple power-law backgrounds}
\label{s:likelihoods}

We start by considering the simple case where the \ac{SGWB} can be modeled by a single power-law
\begin{equation}    \OGW(f)=\Omega_\mathrm{ref}\left(\frac{f}{f_\mathrm{ref}}\right)^{\alpha},
\label{eq: omega_powerlaw}
\end{equation}
in which case we have two free parameters to fit, i.e., $\bm{\theta}=\{\alpha,\Omega_\mathrm{ref}\}$. 
Since $\bar{C}_{\opt;\barell}$ is computed by combining a large number of estimates from different segments, we can approximate its probability distribution by a Gaussian distribution based on the Central-Limit Theorem. 
Consequently, the likelihood function can be written as
\begin{equation}
    \mathscr{L}(\bar{C}_{\opt;\barell},\bar{\Sigma}_{\opt;\barell \barm}|\bm{\theta})\propto \exp\left(-\frac{1}{2}\sum_{\barell}\sum_{\barm}\left(\bar{C}_{\opt;\barell}-\Omega_{\mathrm{gw};\barell}(\bm\theta)\right)\bar{\Sigma}^{-1}_{\opt;\barell\barm}\left(\bar{C}_{\opt;\barm}-\Omega_{\mathrm{gw};\barm}(\bm\theta)\right)\right).
\end{equation}
One can then rewrite the above expression by multiplying it by an $1=\mathcal{R}_{\barell}(\bm\theta)\mathcal{R}_{\barm}(\bm\theta)\mathcal{R}_{\barell}^{-1}(\bm\theta)\mathcal{R}^{-1}_{\barm}(\bm\theta)$, where $\mathcal{R}_{\barell}(\bm\theta)\equiv(f_{\barell}/f_\mathrm{ref})^{-\alpha}$ for this particular \ac{SGWB} model:
\begin{equation}
\begin{aligned}
 \mathscr{L}(\bar{C}_{\opt;\barell},\bar{\Sigma}_{\opt;\barell \barm}|\bm{\theta})&\propto\exp\left(-\frac{1}{2}\sum_{\barell}\sum_{\barm}\mathcal{R}_{\barell}(\bm{\theta})(\bar{C}_{\opt;\barell}-\Omega_{\mathrm{gw};\barell}(\bm{\theta}))
 \right.\\
 &\hspace{1.5in}\times
 \left.
 (\mathcal{R}^{-1}_{\barell}(\bm{\theta}){\bar{\Sigma}^{-1}_{\opt;\barell\barm}}\mathcal{R}^{-1}_{\barm}(\bm{\theta}))\mathcal{R}_{\barm}(\bm{\theta})(\bar{C}_{\opt;\barm}-\Omega_{\mathrm{gw};\barm}(\bm{\theta}))\right)\\
&=\exp\left(-\frac{1}{2}\sum_{\barell}\sum_{\barm}(\bar{C}^\mathcal{R}_{\opt;\barell}-\Omega_\mathrm{ref})(\bar{\Sigma}^\mathcal{R}_\opt)^{-1}_{\barell\barm}(\bar{C}^\mathcal{R}_{\opt;\barm}-\Omega_\mathrm{ref})\right).
\end{aligned}
\end{equation}
In the last line, we have $\bar{C}^\mathcal{R}_{\opt;\barell}\equiv \mathcal{R}_{\barell}(\bm\theta)\bar{C}_{\opt;\barell}$ (see~\eqref{eq: CR}), and recognize that $\mathcal{R}_{\barell}(\bm\theta)\Omega_{\mathrm{gw};\barell}=\Omega_\mathrm{ref}$. We also define%
\footnote{It is easy to show that $(\bar{\Sigma}_\opt^\mathcal{R})^{-1}$ is indeed the inverse of the covariance matrix $\bar{\Sigma}_\opt^\mathcal{R}\equiv\mathcal{R}_{\barell}(\bm\theta)\bar{\Sigma}_{\opt;\barell\barm}\mathcal{R}_{\barm}(\bm\theta)$.}
$(\bar{\Sigma}^\mathcal{R}_\opt)^{-1}_{\barell\barm}\equiv\mathcal{R}^{-1}_{\barell}(\bm\theta)\bar{\Sigma}_{\opt;\barell\barm}^{-1}\mathcal{R}^{-1}_{\barm}(\bm\theta)$. 

We now break the summation in the above parenthesis into three pieces
\begin{equation}
\begin{aligned}
\ln\mathscr{L}(\bar{C}_{\opt;\barell},\bar{\Sigma}_{\opt;\barell \barm}|\bm{\theta})&=-\frac{1}{2}\sum_{\barell}\sum_{\barm} \bar{C}^\mathcal{R}_{\opt;\barell}(\bar{\Sigma}^\mathcal{R}_{\opt})^{-1}_{\barell \barm}\bar{C}^\mathcal{R}_{\opt;\barm}\\
&+\Omega_\mathrm{ref}\sum_{\barell}\sum_{\barm}\left(\bar{C}^\mathcal{R}_{\opt;\barell}(\bar{\Sigma}^\mathcal{R}_{\opt})^{-1}_{\barell\barm}-\frac{1}{2}\Omega_{\rm ref}(\bar{\Sigma}^\mathcal{R}_{\opt})^{-1}_{\barell \barm}\right)+\mathrm{const}\\
&=-\frac{1}{2}\sum_{\barell}\sum_{\barm} \bar{C}^\mathcal{R}_{\opt;\barell}(\bar{\Sigma}^\mathcal{R}_{\opt})^{-1}_{\barell \barm}\bar{C}^\mathcal{R}_{\opt;\barm}+{\frac{\Omega_\mathrm{ref}\bar{C}^\mathcal{R}_\opt}{(\bar{\sigma}^\mathcal{R}_\opt)^2}-\frac{\Omega_\mathrm{ref}^2}{2(\bar{\sigma}^\mathcal{R}_\opt)^2}}+\mathrm{const}\\
&=-\frac{1}{2}\sum_{\barell}\sum_{\barm} \bar{C}_{\opt;\barell}{\bar{\Sigma}}^{-1}_{\opt; \barell\barm}\bar{C}_{\opt;\barm}+{\frac{\Omega_\mathrm{ref}\bar{C}^\mathcal{R}_\opt}{(\bar{\sigma}_\opt^\mathcal{R})^2}-\frac{\Omega_\mathrm{ref}^2}{2(\bar{\sigma}^{\mathcal{R}}_\opt)^2}}+\mathrm{const}\,,
\end{aligned}
\label{eq: L1}
\end{equation}
where we used~\eqref{eq: lambda_i_closedform}, adapted to this problem as
\begin{equation}
\sum_{\barell}\sum_{\barm}\bar{C}^\mathcal{R}_{\opt;\barell}(\bar{\Sigma}^\mathcal{R}_\opt)^{-1}_{\barell \barm}=\frac{\bar{C}^\mathcal{R}_\opt}{(\bar{\sigma}_\opt^\mathcal{R})^2}\,,\qquad
\sum_{\barell}\sum_{\barm}(\bar{\Sigma}^\mathcal{R}_{\opt})^{-1}_{\barell\barm}=\frac{1}{(\bar{\sigma}_\opt^\mathcal{R})^2}\,,
\end{equation}
to obtain the second line.
(NOTE: In the previous sections, we assumed $\alpha=0$.
So we did not distinguish between $\bar{C}^\mathcal{R}_{\barell}$ and $\bar{C}_{\barell}$ (as discussed below~\eqref{eq: CR}), and we denoted the optimal broadband estimator of $\Omega_\mathrm{ref}$ by $\bar{C}_\opt$.
But, here we are allowing the spectral index $\alpha$ to vary, so we need to explicitly denote the optimal broadband estimator as $\bar{C}^\mathcal{R}_\opt$, and its variance as $(\bar{\sigma}^\mathcal{R}_\opt)^2$.)
To obtain the third line, we transformed $\bar{C}^\mathcal{R}_{\opt;\barell}$ and $(\bar{\Sigma}^\mathcal{R}_{\opt})^{-1}_{\barell\barm}$ back to $\bar{C}_{\opt;\barell}$ and $\bar{\Sigma}^{-1}_{\opt, \barell\barm}$ respectively.  
Note that the first term depends only on the data, while the second and third terms depend on the parameters $\bm{\theta}=\{\alpha, \Omega_\mathrm{ref}\}$.
The $\alpha$ dependence is encoded in the $\mathcal{R}$ factors appearing in $\bar{C}^\mathcal{R}_\opt$ and $(\bar\sigma^\mathcal{R}_\opt)^2$. 

Although we have pointed out that $\bar{C}_{\opt;\barell}^\star$ is not the optimal narrowband estimator and $(\bar{\sigma}_{\opt;\barell}^\star)^2$ is not its variance, we can still consider the following likelihood used in many works~\cite{O3stoch, O4aStoch, pygwb}
\begin{align}
\mathscr{L}^\star(\bar{C}^\star_{\opt;\barell},(\bar{\sigma}^\star_{\opt;\barell})^2|\bm{\theta})=\prod_\ell\frac{1}{\sqrt{2\pi(\bar{\sigma}^\star_{\opt;\barell})^2}} \exp\left(-\frac{1}{2}\frac{(\bar{C}^\star_{\opt;\barell}-\Omega_{\mathrm{gw};\barell}(\bm{\theta}))^2}{(\bar{\sigma}_{\opt;\barell}^\star)^2}\right),
\end{align}
Similar to~\eqref{eq: L1}, we can rewrite the above expression as follows:
\begin{equation}
\begin{aligned}
\ln\mathscr{L}^\star(\bar{C}^\star_{\opt;\barell},(\bar{\sigma}^\star_{\opt;\barell})^2|\bm{\theta})&=-\frac{1}{2}\sum_{\barell}\sum_{\barm} \bar{C}_{\opt;\barell}^{\star,\mathcal{R}}(\bar{\Lambda}^\mathcal{R})_{\barell\barm}^{-1}\bar{C}_{\opt;\barm}^{\star,\mathcal{R}}+{\frac{\Omega_\mathrm{ref}\bar{C}^\mathcal{R}_\opt}{(\bar{\sigma}_\opt^\mathcal{R})^2}-\frac{\Omega_\mathrm{ref}^2}{2(\bar{\sigma}^\mathcal{R}_\opt)^2}}+\mathrm{const}'\\
&=-\frac{1}{2}\sum_{\barell}\sum_{\barm} \bar{C}_{\opt;\barell}^{\star}(\bar{\Lambda})^{-1}_{\barell\barm}\bar{C}^{\star}_{\opt;\barm}+{\frac{\Omega_\mathrm{ref}\bar{C}^\mathcal{R}_\opt}{(\bar{\sigma}_\opt^\mathcal{R})^2}-\frac{\Omega_\mathrm{ref}^2}{2(\bar{\sigma}^\mathcal{R}_\opt)^2}}+\mathrm{const}',
\end{aligned}
\label{eq: L2}
\end{equation}
where we denote $\bar{\Lambda}\equiv\mathrm{diag}\left[(\bar{\sigma}_{\mathrm{opt};0}^\star)^2,(\bar{\sigma}_{\mathrm{opt};1}^\star)^2,...\right]$, and $(\bar{\Lambda}^\mathcal{R})^{-1}_{\barell\barm}\equiv \mathcal{R}^{-1}_{\barell}(\bm\theta)\bar{\Lambda}^{-1}_{\barell\barm}\mathcal{R}^{-1}_{\barm}(\bm\theta)$. 

Now, a direct comparison between \eqref{eq: L1} and \eqref{eq: L2} shows that $\ln\mathscr{L}$ and $\ln\mathscr{L}^\star$ have exactly the same dependence on parameters.
Hence, they yield identical posterior distributions for $\alpha$ and $\Omega_\mathrm{ref}$.
Note that, in this scenario, the optimal broadband estimator and its variance are sufficient statistics for this problem. 
(See App.~\ref{app: sufficient_statistics} for more discussion of sufficient statistics.)
\medskip

\subsection{Beyond power-law backgrounds}

Importantly, the above conclusion regarding the equivalence of the two methods for parameter estimation still holds even if $\OGW(f)$ is not described by a single power-law.
To show this, we consider two scenarios:

\begin{enumerate}

\item  Single-component SGWB:

In this case, we assume that there is only one source of the \ac{SGWB}, so that we can always express $\OGW(f|\bm\theta)$ by the product of a single amplitude $\Omega_s$ and frequency-dependent function $F(f|\bm\theta^-)$:
\begin{equation}
    \Omega_\mathrm{gw}(f)=\Omega_s F(f|\bm{\theta}^-),
\end{equation}
where $\bm\theta\equiv\{\Omega_s\}\oplus\bm{\theta}^-$, with $\bm\theta^-\equiv\{\theta_1,\theta_2,...\}$ encapsulating all the remaining model parameters. 
Given this parametrization, the previous discussion remains valid, as one can always define a reweighting function 
$\mathcal{R}_{\barell}(\bm{\theta}^-)\equiv F^{-1}(f_\barell|\bm\theta^-)$ such that the narrowband estimator $\bar{C}_{\opt;\barell}$ can be transformed to $\bar{C}^\mathcal
R_{\opt;\barell}\equiv \mathcal{R}_{\barell}(\bm\theta^-)\bar{C}_{\opt;\barell}$, with $\langle\bar{C}^\mathcal{R}_{\opt;\barell}\rangle=\Omega_s$. 
A corresponding broadband estimator $\bar{C}^\mathcal{R}_\opt$ and its variance $(\bar{\sigma}_\opt^\mathcal{R})^2$ can then be defined. 
Likewise, one may define $\bar{C}_{\opt;\barell}^{\star,\mathcal{R}}$ and $(\bar{\sigma}^{\star,\mathcal{R}}_{\opt;\barell})^2$, such that combining $\bar{C}^\mathcal{R}_{\opt;\barell}$ with its covariance matrix ${\bar{\Sigma}^\mathcal{R}_{\opt;\barell\barm}}$ yields an identical broadband estimator and variance $\bar{C}^\mathcal{R}_\opt$ and $(\bar{\sigma}_\opt^\mathcal{R})^2$ as those obtained by combining $\bar{C}^\star_{\opt;\barell}$ and $(\bar{\sigma}^{\star,\mathcal{R}}_\opt)^2$.
So, as long as the two sets of estimators produce identical broadband quantities---that is, they give rise to the same sufficient statistics---they lead to identical posterior distributions for the parameters, following the argument established above.

\item 
Multi-component SGWB:

If we assume there are multiple sources contributing to $\Omega_\mathrm{gw}(f)$, then we can parametrize $\Omega_\mathrm{gw}(f)$ as follows:
\begin{equation}
    \Omega_\mathrm{gw}(f)=\sum_{i=1}^{N_s}\Omega_{s;i}F_i(f|\bm{\theta}_i^-),
\end{equation}
where $\Omega_{s;i}$ and $F_i(f_i|\bm{\theta}_i^-)$ describe the amplitude and frequency-dependence  of $\Omega_\mathrm{gw}(f)$ for the $i$th source model, respectively. 
In this case, the definition of $\mathcal{R}_{\barell}$ requires all parameters $\bm\theta=\{\Omega_{s;i}\}_{i=1}^{N_s}\oplus\{\bm\theta_i^-\}_{i=1}^{N_s}$. One may then define the broadband estimator and corresponding $\mathcal{R}_{\barell}(\bm\theta)$ as follows:
\begin{equation}
  \mathcal{R}_{\barell}(\bm\theta)\Omega_\GW(f_\barell)=\sum_{i=1}^{N_s}\Omega_{s;i}\quad\Longrightarrow \quad \mathcal{R}_{\barell}(\bm\theta)\equiv\frac{\sum_{i=1}^{N_s}\Omega_{s;i}}{\sum_{i=1}^{N_s}\Omega_{s;i}F_i(f_\barell|\bm\theta_i^-)}\,,
\end{equation}
for which $\langle\bar{C}^\mathcal{R}_\opt\rangle=\sum_{i=1}^{N_s}\Omega_{s;i}$.
Consequently, the previous discussions remain valid for this case as well. 
\end{enumerate}

We highlight an important subtlety: although we claim that one can rewrite the likelihood using the broadband estimator $\bar{C}^\mathcal{R}_\opt$ and its variance $(\bar{\sigma}^\mathcal{R}_\opt)^2$, both quantities are inherently \textit{parameter-dependent}. 
Indeed, neither $\bar{C}^\mathcal{R}_\opt$ nor $(\bar{\sigma}^\mathcal{R}_\opt)^2$ can be defined without first specifying a specific physical model for the underlying \ac{SGWB}. 
However, once a model is chosen, and the likelihood function is constructed, the model parameters $\bm{\theta}$ naturally induce well-defined broadband quantities $\bar{C}^\mathcal{R}_\opt$ and $(\bar{\sigma}^\mathcal{R}_\opt)^2$. 
This point may be seen more clearly by considering the case of Bayesian inference using MCMC sampling. 
Once the sampler draws a $\bm\theta$ from the chosen prior, this specific $\bm\theta$ immediately gives rise to a specific $R_{\barell}(\bm\theta)$, and therefore the corresponding broadband estimator.

\subsection{Evidence, Bayes factors, and model selection}

In Bayesian inference, one computes the Bayes factor (i.e., evidence ratio) $\mathcal{B}:=\mathcal{Z}_1/\mathcal{Z}_2$ to test whether model $\mathscr{M}_1$ or $\mathscr{M}_2$ is preferred by the data. 
The evidence is defined by
\begin{equation}
    \mathcal{Z}_i\equiv \int\D\bm\theta_i\>\pi(\bm\theta_i|\mathscr{M}_i)\mathscr{L}(\mathrm{data}|\bm{\theta}_i, \mathscr{M}_i)\,,
\end{equation}
where $\pi(\bm\theta_i|\mathscr{M}_i)$ is the prior distribution of parameters $\bm\theta_i$ for model $\mathscr{M}_i$. 

Now, the evidence $\mathcal{Z}$ computed using $\mathrm{data}\equiv\{\bar{C}_{\opt;\barell}, \>\bar{\Sigma}_{\opt,\barell\barm}\}$ will differ, in general, from $\mathcal{Z}^\star$ obtained using $\mathrm{data}^\star\equiv\{\bar{C}^\star_{\opt;\barell}, (\bar{\sigma}^\star_{\opt;\barell})^2\}$, since the evidence itself explicitly depends on the data-dependent terms in the expression of the likelihood (see~\eqref{eq: L1} and~\eqref{eq: L2}). 
However, the data-dependent terms will cancel out when computing the Bayes factor, as both the numerator and denominator have the same contributions from those terms. 
Consequently, using $\bar{C}^\star_{\opt;\barell}$ and $(\bar{\sigma}_{\opt;\barell}^\star)^2$ will still lead to the correct Bayes factor for model selection.

\subsection{Numerical test}

We conclude this section by performing a simple numerical test to verify the discussion in the previous subsections. 
We use \texttt{pygwb} to inject a \ac{SGWB} signal with $\alpha=0$ and $\Omega_\mathrm{ref}=5\times 10^{-12}$ to the noise-only data that we used for previous tests.

To estimate the posterior distribution of the parameters, we make use of Bayes' theorem:
\begin{equation}
p(\bm{\theta}|\mathrm{data})\propto \pi(\bm{\theta})\mathscr{L}(\mathrm{data}|\bm{\theta})\,,
\end{equation}
where the (joint) posterior distribution $p(\bm{\theta}|\mathrm{data})$ is determined by the prior $\pi(\bm{\theta})$ and the likelihood $\mathscr{L}(\mathrm{data}|\bm{\theta})$ of obtaining the data given $\bm{\theta}$. 
For our simulations, we choose a log-uniform distribution from $10^{-13}$ to $10^{-8}$ as the prior for $\Omega_\mathrm{ref}$, and a uniform distribution from $-3$ to $3$ as the prior for $\alpha$. 
We also use the two likelihood functions given in~\eqref{eq: L1} and~\eqref{eq: L2} for comparison. 
The likelihood~\eqref{eq: L1} requires the $\bar{C}_{\mathrm{opt};\barell}$ spectrum and complete covariance matrix $\bar{\Sigma}_{\mathrm{opt};\barell\barm}$, while the likelihood~\eqref{eq: L2} used in previous studies requires $\bar{C}^\star_{\mathrm{opt};\barell}$ and $(\bar{\sigma}^\star_{\mathrm{opt};\barell})^2$. 

\begin{figure}[!htbp]
\centering
\includegraphics[width=0.4\linewidth]{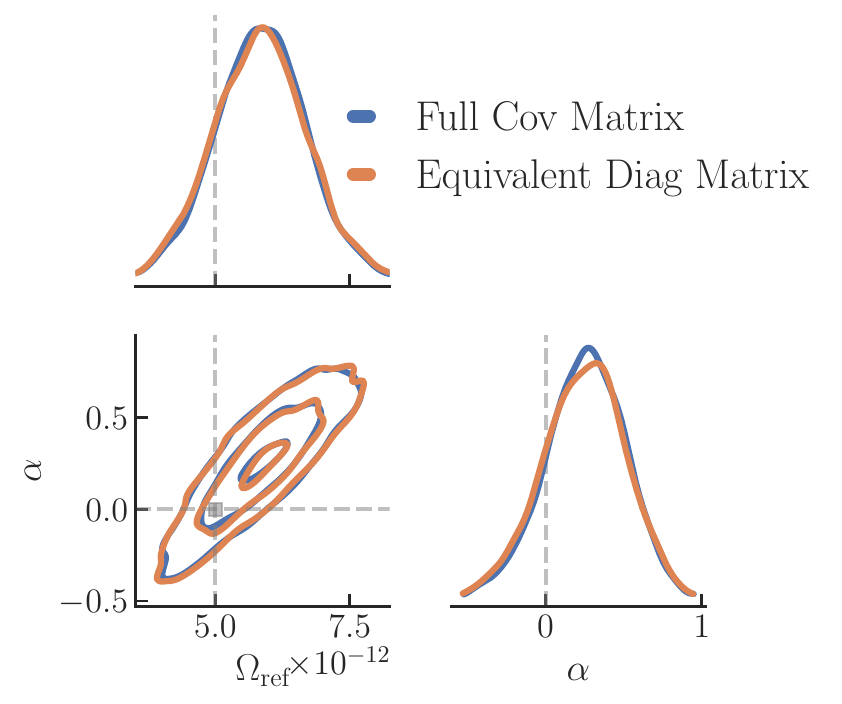}
\includegraphics[width=0.4\linewidth]{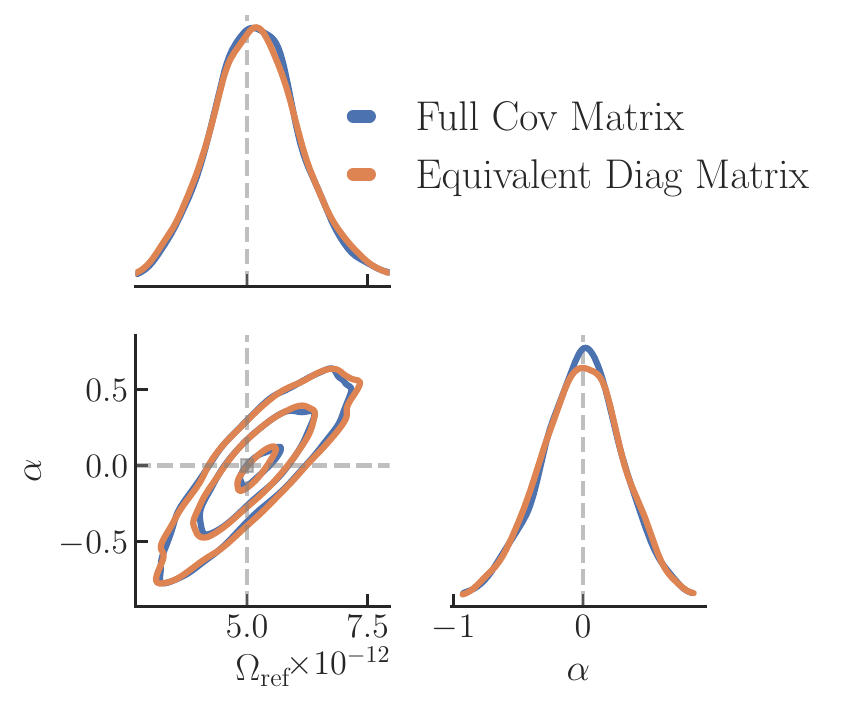}
\caption{The joint posterior distributions of $\Omega_\mathrm{ref}$ and $\alpha$ for $T=4$\>s and $\delta\!f=1/4$\>Hz (left panels) and $T=192$\>s and $\delta\!f=1/32$\>Hz (right panels). 
The blue contours correspond to the case where we use $\bar{C}_{\opt;\barell}$, $\bar{\Sigma}_{\opt;\barell\barm}$, and~\eqref{eq: L1} to compute the likelihood, whereas the orange contours correspond to the case where we use $\bar{C}^\star_{\opt;\barell}$, $(\bar{\sigma}_{\opt;\ell}^\star)^2$, and~\eqref{eq: L2} to compute the likelihood. 
The dashed gray lines show the true values of the parameters. 
Contours for the joint posterior distributions represent 10\%, 60\% and 90\% of the probability.}
\label{fig: Posterior samples}
\end{figure}

For this choice of priors and likelihood functions, we use \texttt{Numpyro}~\cite{phan2019composable,bingham2019pyro} to realize the \ac{NUTS} \ac{MCMC} algorithm to draw posterior samples of $\alpha$ and $\Omega_\mathrm{ref}$. 
We show the resulting joint posteriors of $\{\Omega_\mathrm{ref},\alpha\}$ in Fig.~\ref{fig: Posterior samples}, from which we clearly observe that the blue contours coincide with the orange contours as expected. 
Thus, the quantities $\bar{C}_{\mathrm{opt};\barell}^\star$ and $(\bar{\sigma}_{\mathrm{opt};\barell}^\star)^2$ used in previous studies yield correct \ac{PE} and Bayes factors for the simple Gaussian Likelihood function~\eqref{eq: L2}. 
However, we stress that we must interpret these quantities not as the  \textit{optimal} spectra and associated variances, but rather as \textit{effective} spectra and variances, which can be used for Bayesian analyses.

\section{Biases}
\label{s:bias}
In the previous sections, we derived expressions for the narrowband and broadband estimators of the GW energy density spectrum, along with their corresponding variances. 
Optimal estimators such as these involve weighting the cross-correlated data by its inverse variance, which involves products of the PSDs of the detector output.
But the detector PSDs are not known {\it a~priori}; instead they must also be estimated from the data.
Failure to do so properly can lead to biases in expressions for the variances of the optimal estimators when combined over multiple time segments or frequencies.
As we shall show below, previous analyses have overlooked an additional source of bias, which leads to additional variance on the optimal estimators.

In the following subsections, we  provide an outline / summary of the main results related to bias.
Detailed derivations can be found in App.~\ref{s:biases_details}.
To simplify the calculations, we will restrict attention to non-coarse-grained data and consider only a single frequency bin $\ell$, and we will drop this $\ell$ index from the power spectra to simplify the notation.
We will also restrict attention to combining data over segments.
But the same discussion and conclusions will hold for coarse-grained data and broadband estimators, which are combined over frequency bins as well.

\subsection{Biased estimators of the optimal weights and optimal variance}
Let us suppose that we have $N_{\rm seg}$ independent cross-correlation estimators $\hat C_I$ of the GW energy density spectrum.
As discussed in previous sections, we know that to construct the optimal combined estimator $\Copt$ we should weight the segment-dependent estimators $\hat C_I$ by the inverse of their variances $\sigma^2_I\propto P_{1,I} P_{2,I}$.
If we denote these weights by $w_I$, then we have
\begin{equation}
\Copt\equiv \frac{\sum_I w_I \hat C_I}{\sum_J w_J}\,,
\qquad
\sigma^2_{\rm opt} \equiv {\rm Var}(\Copt)\,,
\qquad
w_I \equiv \frac{\mathcal N}{P_{1,I}P_{2,I}}\,,\qquad \sigma^2_{\rm opt} =\frac{1}{\sum_J w_J}\,.
\label{e:Copt_ideal}
\end{equation}
These equations represent the idealized case, where the PSDs of the detector output are assumed to be known in advance.
But, as mentioned above, these PSDs (and hence the weights $w_I$) must also be estimated from the data, leading to the realistic version of these equations:
\begin{equation}
\Creal\equiv \frac{\sum_I \hat w_I \hat C_I}{\sum_J \hat w_J}\,,
\qquad
\sigma^2_{\rm real} \equiv {\rm Var}(\Creal)\,,\qquad
\hat w_I \equiv \frac{\mathcal{N}}{\hat P_{1,I}\hat P_{2,I}}\,,\qquad
\hat \sigma^2_{\rm real}
\equiv \frac{1}{\sum_J \hat w_J}\,.
\label{e:Copt_real}
\end{equation}
(The normalization factors $\mathcal N$ in \eqref{e:Copt_ideal} and \eqref{e:Copt_real} capture the additional factors that connect $\sigma^2_I$ and $P_{1,I}P_{2,I}$, see e.g., the numerator of the rhs of \eqref{e:what_general}.
But since the inclusion of these factors does not change the final results, we can simplify the following derivation by setting $\mathcal{N}\equiv1$.)
Below, we give results for the mean and variance of the realistic estimator $\Creal$, and compare those to the mean and variance of the idealized estimator $\Copt$, as well as to the expected value of $\hat \sigma^2_{\rm real}$.
Importantly, we will find that $\hat w_I$ and $\hat\sigma^2_{\rm real}$ are biased estimators of $w_I$ and $\sigma^2_{\rm real}$.
The latter is a new result that has previously been overlooked in the literature.

Since we provide details of the bias calculations in App.~\ref{s:biases_details}, we only sketch the main arguments here.
We will assume that $P_{1,I}\equiv P_1$ and $P_{2,I}\equiv P_2$ for all segments, which allows us to easily evaluate several of the sums over segments.
The starting point for our analyses will be segment-dependent estimators $\hat P_{1,I}$, $\hat P_{2,I}$ of the PSDs $P_1$, $P_2$ of the output of the two detectors.
These estimates are typically calculated using Welch's method of PSD estimation, which averages together periodogram estimates from neighboring segments of data.
We will denote the number of effective averages of these estimators by $N_{\rm eff}$.

Assuming that these estimators are unbiased (which they typically are), we can write:
\begin{equation}
\hat{P}_{1,I}=P_{1}+\delta P_{1,I}\,,\qquad
\hat{P}_{2,I}=P_{2}+\delta P_{2,I}\,.
\label{e:Phat-main}
\end{equation}
where
\begin{equation}
    \langle \delta P_{1,I}\rangle
    =\langle \delta P_{2,I}\rangle
    =0\,,\qquad
\langle \delta P^2_{1,I}\rangle\ = \frac{P_{1}^2}{N_{\rm eff}}\,,
\qquad
\langle \delta P^2_{2,I}\rangle\ = \frac{P_{2}^2}{N_{\rm eff}}\,,
\qquad
\langle\delta {P}_{1,I}\,\delta{P}_{2,I}\rangle=\frac{\gamma^2_{12} P_{{\rm gw}}^2}{N_{\rm eff}}\,.
\label{e:evals-main}
\end{equation}
Note that the last result implies
\begin{equation}
    \langle \hat P_{1,I}\hat P_{2,I}\rangle \simeq  P_{1}P_{2}\,,
    \label{e:eval4-main}
\end{equation}
since in the weak-signal limit $\gamma^2_{12} P_{{\rm gw}}^2/P_{1}P_{2}\ll 1$.
The fact that different data are used for PSD estimation and cross-spectrum estimation simplifies several of the expectation value calculations detailed in App.~\ref{s:biases_details}, since these different data segments for estimation can be treated as independent of one another.

Given the above definitions and results, we can show that  (see also App.~B of \cite{Matas:2020roi}):
\begin{equation}
\begin{aligned}
\hat w_I \equiv 
\frac{1}{\hat{P}_{1,I}\hat{P}_{2,I}}
&=\frac{1}{P_1P_2}\left(1-\frac{\delta P_{1,I}}{P_1}-\frac{\delta P_{2,I}}{P_2}+\frac{\delta P_{1,I}^2}{P_1^2}+\frac{\delta P_{2,I}^2}{P_2^2}+\frac{\delta P_{1,I}\delta P_{2,I}}{P_1 P_2}-\cdots\right)\,,
\label{e:w_hat_I_full-main}
\end{aligned}
\end{equation}
where we have used~\eqref{e:Phat-main}, and expanded $\hat w_I$ in a Taylor series expansion in the $\delta P$'s.
Using~\eqref{e:evals-main}, it follows that the expectation value of $\hat w_I$ is given by 
\begin{equation}
\begin{aligned}
\langle \hat w_I\rangle\equiv 
\left\langle\frac{1}{\hat{P}_{1,I}\hat{P}_{2,I}}\right\rangle
&\simeq\frac{1}{P_1P_2}\left(1+\frac{2}{N_{\rm eff}}\right)\,,
\end{aligned}
\end{equation}
where we used the weak-signal limit and ignored third-order terms in $\delta P_{1,I}$ and $\delta P_{2,I}$ (and their products) to obtain the last (approximate) equality.
Thus, $\hat w_I$ is a biased estimator of $w_I\equiv 1/(P_1P_2)$ with bias factor~\cite{Matas:2020roi,pygwb}
\begin{equation}
b \equiv  1 + \frac{2}{N_{\rm eff}}\,.
\label{e:mean_w_hat-main}
\end{equation}
In a similar fashion, we can calculate the variance of $\hat w_I$:
\begin{equation}
{\rm Var}(\hat w_I) 
\equiv\langle \hat w_I^2\rangle - \langle \hat w_I\rangle^2
\simeq\frac{2}{N_{\rm eff}}w_I^2\,,
\label{e:var_w_hat-main}
\end{equation}
and the expected value of the variance estimator $\hat \sigma^2_{\rm real}\equiv {1}/{\sum_I \hat w_I}$ of the realistic estimator:
\begin{equation}
\langle \hat\sigma^2_{\rm real}\rangle=
     \left\langle\frac{1}{{\sum_I\frac{1}{\hat P_{1,I}\hat P_{2,I}}}}\right\rangle\simeq\frac{P_1P_2}{N_{\rm seg}}\left(1-\frac{2}{N_\mathrm{eff}}\right)
     =\sigma^2_{\rm opt}\left(1-\frac{2}{N_{\rm eff}}\right)\,.
\label{e:mean_sigma2_opt_hat-main}
\end{equation}
In these expressions we have used the weak-signal approximation and have ignored higher-order terms in $1/N_{\rm seg}$ and $1/N_{\rm eff}$, since it is usually the case that both $N_{\rm seg}\gg 1$ and $N_{\rm eff} \gg 1$.
See App.~\ref{s:biases_details} for details of these calculations.

\subsection{Relating the means and variances of the optimal and realistic estimators $\Copt$ and $\Creal$}
To calculate the mean and variance of $\Creal$, we begin by expanding $\hat w_I$ and $\hat C_I$ as follows:
\begin{equation}
    \begin{aligned}
&\hat w_I \equiv \langle \hat w_I\rangle + \delta w_I\ = b\,w_I + \delta w_I\,,
\\
&\hat C_I \equiv \langle \hat C_I\rangle + \delta C_I = 
C + \delta C_I\,,
\label{e:dw_dC}
\end{aligned}
\end{equation}
similar to what we did for $\hat P_{1,I}$ and $\hat P_{2,I}$ in the previous subsection.
The first and second-order expectation values of $\delta w_I$ and $\delta C_I$ are given by
\begin{equation}
\langle\delta w_I\rangle =0\,,\quad
\langle\delta C_I\rangle =0\,,\quad
\langle \delta w^2_I\rangle \simeq \frac{2}{N_{\rm eff}}w_I^2\,,
\qquad
\langle \delta C^2_I\rangle \simeq P_1 P_2\,,
\label{e:dw_dC_evals}
\end{equation} 
where we used~\eqref{e:var_w_hat-main} for the third equality.
Similar to what we did above for $\hat w_I$ and $\hat\sigma^2_{\rm est}$, we can expand $\Creal$ in a Taylor series expansion involving $\delta w_I$ and $\delta C_I$, leading to (see App.~\ref{s:biases_details} for details):
\begin{equation}
\Creal = \Copt + \hat\Delta\,,
\label{e:D=C+D}
\end{equation}
where
\begin{equation}
\hat \Delta
\equiv b^{-1}\frac{P_1P_2}{N_{\rm seg}}\left[\sum_I \delta w_I\left(\delta C_I - \frac{1}{N_{\rm seg}}\sum_{J}\delta C_J\right)-\dots\right]\,.
\label{e:Delta_hat}
\end{equation}
Thus, to calculate the mean and variance of $\Creal$, we can make use of two general results:
\begin{equation}
\langle \Creal\rangle = 
\langle \Copt\rangle +
\langle \hat \Delta\rangle\,,
\qquad
{\rm Var}(\Creal) = 
{\rm Var}(\Copt) +
{\rm Var}(\hat \Delta) +
2{\rm Cov}(\Copt, \hat \Delta)
\label{e:mean_var_A+B}
\end{equation}
for the sum of two random variables.
In App.~\ref{s:biases_details}, we calculate $\langle \hat \Delta\rangle$, ${\rm Var}(\hat \Delta)$, and ${\rm Cov}(\Copt, \hat \Delta)$, leading to the final results
\begin{align}
&\langle \Creal\rangle \simeq \langle \Copt\rangle\,,
\label{e:mean_Dopt-main}
\\
&\sigma^2_{\rm real}\equiv{\rm Var}(\Creal)\simeq \frac{P_1 P_2}{N_{\rm seg}} \left(1+\frac{2}{N_{\rm eff}}\right)
=\sigma^2_{\rm opt}\left(1+\frac{2}{N_{\rm eff}}\right)
\simeq \langle \hat \sigma^2_{\rm real}\rangle \left(1+\frac{4}{N_{\rm eff}}\right)\,.
\label{e:var_Dopt-main}
\end{align}
The first result,~\eqref{e:mean_Dopt-main}, for the expectation values was somewhat to be expected, although it wasn't obvious since the weights in the sums for $\Creal$ are estimated from the data.
The second result,~\eqref{e:var_Dopt-main}, for the variance of the realistic combined estimator was not expected.
In retrospect, previous studies had overlooked the fact that, since we have access to only estimates of the PSDs of the detector output, we should be relating the variance of the realistic combined estimator $\Creal$ to the estimated variance $\langle\hat \sigma^2_{\rm real}\rangle$ and not the theoretical variance $\sigma^2_{\rm opt}$.
Hence, there was a missing factor of $b=(1+2/N_{\rm eff})$ in those analyses.

Recall that for the above discussion we:
(i) restricted to a single frequency bin $\ell$, (ii) assumed segment-independent power spectra $P_{1,I}\equiv P_1$ an $P_{2,I}\equiv P_2$, and (iii) combined data only over segments $I$.
But we would obtain the same form for the bias factor if we (i) include the full frequency-dependence of the weights i.e., 
\begin{equation}
    \hat w_{I\ell} = \frac{1}{\hat\sigma_{I\ell}^{2}} = \frac{\gamma_{12;\ell}S_{0;\ell}}{\hat P_{1;I\ell} \hat P_{2;I\ell}},
    \label{e:what_general}
    \end{equation}
(ii) assume segment-dependent PSDs $P_{1;I\ell}$, $P_{2;I\ell}$, and (iii) combine data over frequency bins as well as segments.
One should simply replace the overall factor $P_1P_2/N_{\rm seg}$ by 
\begin{equation}
\frac{P_1 P_2}{N_{\rm seg}} 
\quad\rightarrow\quad
\frac{1}{\sum_{I,\ell}  \frac{\gamma_{12;\ell}S_{0;\ell}}{P_{1;I\ell} P_{2;I\ell}}}\,.
\end{equation}
So the above results are completely general.

\subsection{Visualizing the effect of bias via simulations}
For the simulations performed in Sec.~\ref{s:comparisons}, we showed that when we use the theoretical weights to combine data, the resulting (co)variance is consistent with expectation. 
Here, we show that when we use estimated (biased) weights to combine the data, biases appear.
We do not overlap data in this simulation and consider $T=192$ s and $\delta f=1/32$ Hz case for illustration purpose.

Given 2,000 frames, we obtain 2,000 narrowband estimators and variances, $\bar{C}_{\realistic;\barell}$ and $\bar{\sigma}^2_{\realistic;\barell}$, as well as the corresponding broadband quantities, $\bar{C}_{\realistic}$ and $\bar{\sigma}^2_{\realistic}$. 
Our goal is to determine whether the empirical bias, evaluated directly from the data, is consistent with our expectation $\sqrt{1+4/N_\mathrm{eff}}$. 

To this end, we take the sample standard deviation of $\bar C_{\realistic;\barell}$ as an estimate of $\mathrm{Std}(\bar C_{\realistic;\barell})$, and the sample mean of $\bar\sigma^2_{\realistic;\barell}$ as an estimate of $\langle\bar\sigma^2_{\realistic;\barell} \rangle$. 
We then compute an empirical bias factor $b_{\barell}$ by taking the ratio of $\mathrm{Std}(\bar C_{\realistic;\barell})$ and $\langle \bar\sigma^2_{\realistic;\barell}\rangle^{1/2}$ for each frequency bin $\barell$.
We also apply the same method to the 2,000 broadband estimators and variances to obtain a single empirical bias factor $b$.

In practice, we do not expect the empirical bias factors to be exactly equal to the theoretical prediction due to statistical fluctuations. 
Consequently, it is necessary to quantify the uncertainty associated with these empirical estimates. 
Ideally, one would repeat this entire simulation process $N\gg 1$ times, however, this is computationally prohibitive. 
To circumvent this limitation, we adopt the \textit{bootstrap} method. 

Specifically, given the 2,000 narrow/broadband estimators together with their corresponding variances, we can construct bootstrap samples by \textit{resampling with replacement} from this set of simulated data. 
For each bootstrap sample we recompute the empirical bias factor. 
Repeating this procedure $B\gg 1$ times yields a set of $B$ empirical bias factors, whose distribution can then be used to estimate the uncertainty of the empirical bias.

\begin{figure}[!htbp]
    \centering
    \includegraphics[width=0.7\linewidth]{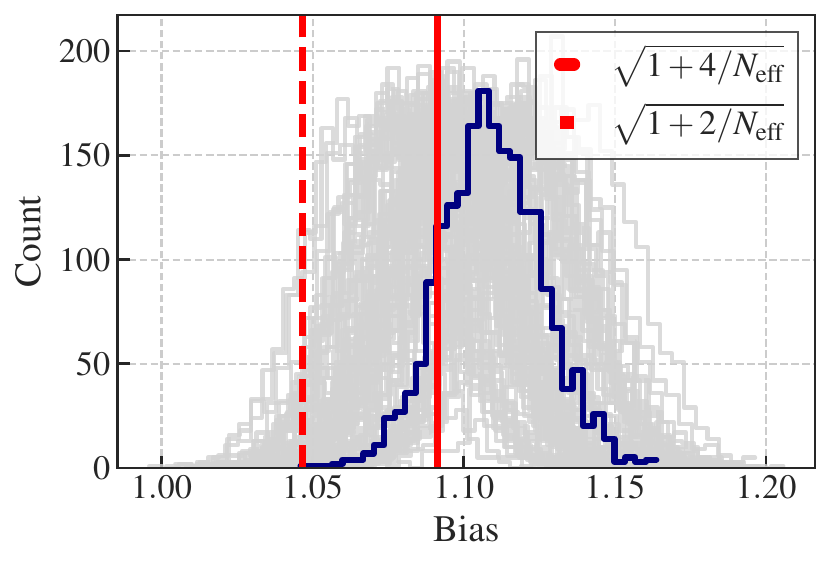}
    \caption{Bootstrap validation of the bias factor. Gray histograms show the bootstrap distributions of the empirical bias factor obtained from narrowband estimators corresponding to 50 randomly selected frequency bins, while the dark-blue histogram corresponds to the result obtained from the the broadband estimator. The solid and dashed red line indicate the theoretical predictions $\sqrt{1+4/N_\mathrm{eff}}$ derived in this work, and the commonly used expression $\sqrt{1+2/N_\mathrm{eff}}$, respectively. The bootstrap results strongly favor the prediction derived in this work.}
    \label{fig: bias_prop}
\end{figure}

We show the results of this bootstrap procedure in Fig.~\ref{fig: bias_prop}. 
We randomly select 50 different $\barell$ values and perform the bootstrap resampling for each of them. 
For each selected $\barell$, the resampling process is repeated $B=2,000$ times, and the resulting distribution of the empirical bias factor is shown as a gray histogram. 
Each gray histogram therefore corresponds to the  bias factor computed from the bootstrap samples based on $\bar C_{\mathrm{real};\barell}$ and $\bar\sigma_{\mathrm{real};\barell}$ for a chosen $\barell$. 
The dark-blue histogram shows the result for the broadband estimator. 

We also indicate two reference values with red vertical lines: the solid line corresponds to the bias factor we derived in this work, while the dashed line corresponds to the bias factor commonly adopted in the literature. 
From these simulations we find that, across the 50 frequency bins, the red solid line falls within the 90\% confidence interval in 91\% of the cases, whereas the red dashed line does so in only 5\% cases. 
This strongly supports the theoretical calculations presented in this section. 
Finally, we also highlight that the dark-blue histogram exhibits no qualitatively different behavior from the gray histograms. This observation indicates that our derivation is not limited to the narrowband estimator in a single frequency bin, but can also be generalized to the broadband estimator.

\section{Conclusions}
\label{sec: conclusions}
In this work, we have carried out a detailed study of the frequency-domain cross-correlation method for the detection and parameter estimation of a persistent, stationary, Gaussian, unpolarized, and isotropic \ac{SGWB} using ground-based GW detectors.
The main focus of our investigations was to define and derive from first principles both optimal and realistic narrowband and broadband estimators for the underlying \ac{SGWB}.
In Secs.~\ref{s:single_segment}, \ref{s:covariance}, \ref{s:multi_segment}, \ref{s:comparisons}, we showed that expressions for our broadband estimators and their variances were consistent with those used in previous searches~\cite{O3stoch, O4_PSD, pygwb}.
However, we found discrepancies for the narrowband estimators and their variances, due to correlations induced by windowing that were neglected in previous studies. 
Nonetheless, we showed in Sec.~\ref{sec: PE} that previous narrowband analyses still yield posterior distributions for SGWB  parameters and Bayes factors that agree with our new results, since both approaches lead to identical sufficient statistics for their likelihood analyses.
Finally, in Sec.~\ref{s:bias}, we discovered an additional source of bias in the variance of the realistic estimators, which comes from needing to estimate the detector noise PSDs when  constructing cross-correlation estimators combined over time segments or frequency bins, or both.
This additional bias had been overlooked in previous analyses.

This comprehensive analysis should serve as a foundational reference for current and future frequency-domain narrowband cross-correlation searches of SGWB signals. 
Nonetheless, it is essential to acknowledge that the methods developed in this study assume the SGWB signal to be persistent, stationary, Gaussian, unpolarized, and isotropic. 
In addition, several of our derivations and all of our simulations assumed that we are in the weak-signal regime, where we can ignore the power in the GWB relative to that from the detector noise.
Detailed discussions of the \textit{strong-signal regime} and \textit{intermittent} SGWB signals will be the subject of future studies.

\begin{acknowledgments}
The authors are grateful for computational resources provided by the LIGO Laboratory and supported by National Science Foundation (NSF) Grants PHY-0757058 and PHY-0823459. H.Z.~and V.M.~are supported in part by NSF grant PHY-2409173.
J.D.R.~acknowledges financial support from NSF grant PHY-2207270 and startup funds from the University of Texas Rio Grande Valley.
\end{acknowledgments}

\bibliography{ref}
\begin{appendix}
\renewcommand{\theequation}{\thesection.\arabic{equation}}
\counterwithin{equation}{section} 

\section{Statistics: standard definitions and results}
\label{app: stats}

In this appendix, we review some standard definitions and results in statistics, which we will use repeatedly in the main text.
We define the mean, variance, and covariance of random variables, and state Isserlis's theorem for zero-mean, Gaussian random variables. 
We also describe how to construct a linear optimal estimator (which, by definition, is unbiased and has minimum variance), and introduce the notion of sufficient statistics.

\subsection{Mean, variance, and covariance}
\label{app: Mean_Var_Cov}

Consider a random variable $x$. Its mean (expected value) and variance are defined as follows:
\begin{align}
     &\mu_x\equiv\langle x\rangle\equiv\int \D x~p(x)x\,,\\
     &\sigma_x^2\equiv\langle x^2\rangle-\langle x\rangle^2\,.
\end{align}
Now, suppose we have $N$ (real) random variables $x_1,~x_2,\cdots,~x_N$.  
Their mutual dependence can be quantified by their covariance, defined as
\begin{equation}
    \Sigma_{ij}\equiv\langle x_ix_j\rangle-\langle x_i\rangle \langle x_j\rangle\,.
\end{equation}
If we normalize the covariance matrix so that all the diagonal elements equal one, we obtain the correlation matrix 
\begin{equation}
    \rho_{ij}\equiv\frac{\Sigma_{ij}}{\sqrt{\Sigma_{ii}\Sigma_{jj}}}\,.
\end{equation}

\subsection{Isserlis's theorem}
\label{app: Isserlis}

Consider $N$ zero-mean, Gaussian random variables $x_1,~x_2,\cdots,x_N$. According to Isserlis's theorem~\cite{isserlis:1918}, the $N$th-order expectation value can be written as
\begin{equation}
    \langle x_1x_2\cdots x_N\rangle=\sum_{p\in P_N^2}\prod_{\{i,j\}\in p}\langle x_ix_j\rangle\,,
    \label{eq: Isserlisis}
\end{equation}
where $p$ denotes all possible distinct ways of partitioning $\{1,2,\cdots,N\}$ into pairs $(i,j)$. The summation above is over all these partitions, and the product is over all pairs in one particular partition. 

If $N$ is odd, no such partioning into pairs exists, so Isserlis's theorem then implies that the expected value of the product of these $N$ zero-mean Gaussian random variables is zero. 
If $N$ is even, then the $N$-th order expectation value can be computed as a product of two-point correlators as shown above. 
A useful example is the case $N=4$:
\begin{equation}
    \langle x_1x_2x_3x_4\rangle = \langle x_1x_2\rangle\langle x_3x_4\rangle + \langle x_1x_3\rangle\langle x_2x_4\rangle+\langle x_1x_4\rangle\langle x_2x_3\rangle\,.
\end{equation}
Note that if we allow for non-zero mean Gaussian random variables, then terms proportional to $\langle x_i\rangle$ or its powers will also show up on the right hand side of~\eqref{eq: Isserlisis}.

\subsection{Optimal estimator}
\label{app: optimal_estimator}

Consider again $N$ random variables $x_1,~x_2,\cdots,x_N$. 
Suppose now that they all have the same mean $\langle x_i\rangle=a,~\forall x_i$, and the covariance $\Sigma_{ij}$ is known.
Then we can construct a linear optimal estimator of $a$ such that the estimator is \textit{unbiased} and has the \textit{minimum variance}. 
We define the optimal estimator $\hat{a}_\opt$ by the linear combination
\begin{equation}
    \hat{a}_\opt\equiv\sum_i\lambda_ix_i\,,
\end{equation}
where $\sum_i$ is shorthand for $\sum_{i=1}^N$.
Since $\hat{a}_\opt$ is to be unbiased, the prefactors $\lambda_i$ must satisfy the following constraint
\begin{equation}
    \sum_i\lambda_i=1\,.
    \label{e:constraint}
\end{equation}
It is also straightforward to show that the variance of this estimator is%
\footnote{Although we use $\Sigma$ as the base symbol for both the covariance matrix $\Sigma_{ij}$ and summation $\sum_i\equiv\sum_{i=1}^N$, it should be clear from the context what $\Sigma$ is being used for.}
\begin{equation}
{\sigma}^2_{\hat{a}_\opt}=\sum_i\sum_j\lambda_i\lambda_j\Sigma_{ij}\,.
\end{equation}
One can adopt the method of Lagrange multipliers to determine the expression of $\lambda_i$'s. 
Consider the following function
\begin{equation}
f(\{\lambda_i\},\Lambda)=\sum_i\sum_j\lambda_i\lambda_j\Sigma_{ij}+\Lambda(1-\sum_i\lambda_i)\,.
\end{equation}
Then minimizing the variance of $\hat a_{\rm opt}$ subject to the constraint~\eqref{e:constraint} is equivalent to requiring that
\begin{equation}
\left\{
\begin{aligned}
&\frac{\partial f}{\partial\lambda_i}=0\,,\qquad\text{for\ } i=1,2,\ldots,N\,,
\\
&\frac{\partial f}{\partial \Lambda}=0\,.
\end{aligned}
\right.
\end{equation}
Then it is straightforward to show that the above equations lead to 
\begin{equation}
    \lambda_i=\frac{\sum_j(\Sigma^{-1})_{ij}}{\sum_j\sum_k(\Sigma^{-1})_{jk}},\qquad{\sigma}^2_{\hat{a}_\opt}=\frac{1}{\sum_j\sum_k(\Sigma^{-1})_{jk}}\,.
    \label{eq: lambda_i_closedform}
\end{equation}
To simplify the notation, we drop the parentheses around $\Sigma^{-1}$, and simply denote $(\Sigma^{-1})_{ij}$ by $\Sigma_{ij}^{-1}$ going forward.

\subsection{Sufficient statistics}
\label{app: sufficient_statistics}

We introduce the idea of sufficient statistics by considering the following toy example, and refer the interested readers to~\cite{Matas:2020roi} for extra discussion.

Consider $N$ Gaussian-distributed random variables,
\begin{equation}
    x_i=a+n_i\,,\qquad i=1,2,\cdots, N,
\end{equation}
with 
\begin{equation}
    \langle n_i\rangle=0,\qquad\langle n_in_j\rangle=\sigma^2\delta_{ij}\,.
\end{equation}
In the above expressions, $a$ denotes the unknown amplitude of a constant signal and $\sigma^2$ is the variance of the data, which we assume to be known.
The likelihood function of observing $\{x_i\}$ given parameter $a$ is 
\begin{equation}
    p(\{x_i\}|a)=\frac{1}{(2\pi)^{N/2}\sigma^N}\exp\left(-\frac{\sum_i(x_i-a)^2}{2\sigma^2}\right)\,.
    \label{e:like}
\end{equation}
It is simple to show that the maximum likelihood estimator of $a$ is given by the sample mean
\begin{equation}
    \hat{a}_\mathrm{ML}=\frac{1}{N}\sum_ix_i\,,
\end{equation}
with variance 
\begin{equation}
{\sigma}^2_{\hat{a}_{\rm ML}}=\frac{\sigma^2}{N}\,.
\end{equation}
Then we can rewrite the likelihood function by noticing the following relation
\begin{equation}
    \sum_i(x_i-a)^2=\sum_ix_i^2-N\hat{a}_{\mathrm{ML}}^2+N(\hat{a}_\mathrm{ML}-a)^2\,.
\end{equation}
Substituting this into~\eqref{e:like} leads to 
\begin{equation}
    p(\{x_i\}|a)=\frac{1}{(2\pi)^{N/2}\sigma^N}\exp\left(-\frac{\sum_ix_i^2}{2\sigma^2}\right)\exp\left(\frac{\hat{a}_\mathrm{ML}^2}{2{\sigma}^2_{\hat{a}_\mathrm{ML}}}\right)\exp\left(-\frac{(\hat{a}_\mathrm{ML}-a)^2}{2{\sigma}^2_{\hat{a}_\mathrm{ML}}}\right).
\end{equation}
Only the last term is related to the parameter $a$, hence the final posterior distribution should be determined only by it. 
Importantly, we highlight that this terms depends only on a special combination of data, $\hat{a}_\mathrm{ML}$, and its variance. 
Therefore, $\hat{a}_\mathrm{ML}$ is a \textit{sufficient statistic} for this particular problem, as only $\hat{a}_\mathrm{ML}$ and its variance appear in the term involving the parameter $a$.
In other words, one obtains the same posterior distribution for $a$ using either the full set of data $\{x_i\}$ or the single data combination $\hat a_\mathrm{ML}$ when performing parameter estimation.
\section{Technical details for single-segment optimal estimator calculations}
\label{s:single_seg_details}

\subsection{Proof of Eq.~\eqref{eq: w2}}
\label{app: proof_w2}

We want to evaluate the summation $\mathcal{S}_\ell\equiv \sum_{j=0}^{N-1}\tilde{w}_{\ell-j}\tilde{w}^*_{\ell-j}$.
Using the definition the \ac{DFT} \eqref{eq: DFT}, we find
\begin{equation}
\begin{aligned}
    \mathcal{S}_\ell&=(\Delta t)^2\sum_{j=0}^{N-1}\sum_{p=0}^{N-1}\sum_{q=0}^{N-1}w_pw_qe^{-2\pi i\frac{p(\ell-j)}{N}}e^{2\pi i\frac{q(\ell-j)}{N}}\\
    &=(\Delta t)^2\sum_{p=0}^{N-1}\sum_{q=0}^{N-1}w_pw_qe^{-2\pi i \frac{\ell(p-q)}{N}}\Big(\underbrace{\sum_{j=0}^{N-1}e^{2\pi i\frac{j(p-q)}{N}}}_{N\delta_{pq}}\Big)\,.
\end{aligned}
\end{equation}
Note that the summation over $j$ yields a Kronecker delta, $\delta_{pq}$, which we can use to  eliminate the summation over e.g., $q$.
We then end up with
\begin{equation}
    \mathcal{S}_\ell=(N\Delta t)(\Delta t)\sum_{p=0}^{N-1}w^2_p
    =(N\Delta t)^2\frac{1}{N}\sum_{p=0}^{N-1}w^2_p
    \equiv T^2\,\overline{w^2}\,.
\end{equation}

\subsection{Proof of Eq.~\eqref{eq: wp_wq}} 
\label{app: proof_wp_wq}

We now want to evaluate
$\mathcal{S}_{pq}\equiv \sum_{j=0}^{N-1}\tilde{w}_{p-j}\tilde{w}_{q-j}^*$.
As before, we use the definition of the \ac{DFT} (Eq. \eqref{eq: DFT}) to obtain:
\begin{equation}
\begin{aligned}
   \mathcal{S}_{pq}&=(\Delta t)^2\sum_{j=0}^{N-1}\sum_{\ell=0}^{N-1}\sum_{m=0}^{N-1}w_\ell w_me^{-2\pi i\frac{\ell(p-j)}{N}}e^{2\pi i\frac{m(q-j)}{N}}\\
    &=(\Delta t)^2\sum_{\ell=0}^{N-1}\sum_{m=0}^{N-1}w_\ell w_me^{-2\pi i \frac{(\ell p-mq)}{N}}\Big(\underbrace{\sum_{j=0}^{N-1}e^{2\pi i\frac{j(\ell-m)}{N}}}_{N\delta_{\ell m}}\Big)\\
    &=\frac{T^2}{N}\sum_{\ell=0}^{N-1}w^2_\ell e^{-2\pi i\frac{\ell(p-q)}{N}}\,.
\end{aligned}
\end{equation}

\subsection{Proof of Eq.~\eqref{eq: bar_Sigma_l_m_final}}
\label{app: proof_bar_Sigma_l_m_final}

To prove~\eqref{eq: bar_Sigma_l_m_final}, we need to evaluate $\mathcal{S}^-_{jk;\barell\barm}$ and $\mathcal{S}^+_{jk;\barell;\barm}$ defined in~\eqref{eq: S_-_S_+}. 
Let us first consider $\mathcal{S}^-_{jk;\barell\barm}$:
\begin{equation}
\begin{aligned}
    \mathcal{S}^-_{jk;\barell\barm}&\equiv \frac{1}{M^2}\sum_{p=M\barell-\frac{M}{2}}^{M\barell+\frac{M}{2}-1}\sum_{q=M\barm-\frac{M}{2}}^{M\barm+\frac{M}{2}-1}e^{-2\pi i(j-k)\frac{(p-q)}{N}}\\
    &=e^{-2\pi i(j-k)\frac{M(\barell-\barm)}{N}}\frac{1}{M^2}\sum_{p,q=0}^{M-1}e^{-2\pi i(j-k)\frac{(p-q)}{N}}\\
    &=e^{-2\pi i(j-k)\frac{M(\barell-\barm)}{N}}\left|\frac{1}{M}\sum_{p=0}^{M-1}e^{-2\pi i(j-k)\frac{p}{N}}\right|^2\,.
\end{aligned}
\end{equation}
To get this expression, we changed the range of the sums over $p$ and $q$ to obtain the second equality.
Then, to evaluate the sum on the rhs of the last line of the above equation, we note that
\begin{equation}
    \sum_{p=0}^{M-1}e^{-2\pi i(j-k)\frac{p}{N}}=\frac{1-e^{-2\pi i(j-k)\frac{M}{N}}}{1-e^{-2\pi i\frac{(j-k)}{N}}}=\frac{e^{-\pi i(j-k)\frac{M}{N}} \left(e^{\pi i(j-k)\frac{M}{N}}-e^{-\pi i(j-k)\frac{M}{N}} \right) }{e^{-\pi i\frac{(j-k)}{N}}\left(e^{\pi i\frac{(j-k)}{N}}-e^{-\pi i\frac{(j-k)}{N}}\right)}\,.
\end{equation}
Thus, 
\begin{equation}
     \sum_{p=0}^{M-1}e^{-2\pi i(j-k)\frac{p}{N}}=e^{-\pi i(j-k)\frac{(M-1)}{N}}\frac{\sin\left(\pi \frac{(j-k)M}{N}\right)}{\sin\left(\pi\frac{(j-k)}{N}\right)}=Me^{-\pi i(j-k)\frac{(M-1)}{N}}\frac{{\rm sinc}\left(\pi \frac{(j-k)M}{N}\right)}{{\rm sinc}\left(\pi\frac{(j-k)}{N}\right)}\,,
\end{equation}
implying that
\begin{equation}
    \mathcal{S}^-_{jk;\barell\barm}=e^{-2\pi i(j-k)\frac{M(\barell-\barm)}{N}}\left(\frac{\sinc\left(\pi \frac{(j-k)M}{N}\right)}{\sinc\left(\pi\frac{(j-k)}{N}\right)}\right)^2\equiv e^{-2\pi i(j-k)\frac{M(\barell-\barm)}{N}}\mathcal{K}_{j,k;M,N}\,,
\end{equation}
where
\begin{equation}
\mathcal{K}_{j,k;M,N}\equiv
\left(\frac{\sinc\left(\pi \frac{(j-k)M}{N}\right)}{\sinc\left(\pi\frac{(j-k)}{N}\right)}\right)^2
=\left|\frac{1}{M}\sum_{p=0}^{M-1}e^{-2\pi i(j-k)\frac{p}{N}}\right|^2\,.
\label{e:K_def_app}
\end{equation}

The calculation of $S^+_{jk;\barell\barm}$ is similar:
\begin{equation}
\begin{aligned}
    \mathcal{S}^+_{jk;\barell\barm}&\equiv \frac{1}{M^2}\sum_{p=M\barell-\frac{M}{2}}^{M\barell+\frac{M}{2}-1}\sum_{q=M\barm-\frac{M}{2}}^{M\barm+\frac{M}{2}-1}e^{-2\pi i(j-k)\frac{(p+q)}{N}}\\
    &=e^{-2\pi i(j-k)\frac{M(\barell+\barm-1)}{N}}\frac{1}{M^2}\sum_{p,q=0}^{M-1}e^{-2\pi i(j-k)\frac{(p+q)}{N}}\\
    &=e^{-2\pi i(j-k)\frac{M(\barell+\barm-1)}{N}}\left(\frac{1}{M}\sum_{p=0}^{M-1}e^{-2\pi i(j-k)\frac{p}{N}}\right)^2\\
    &=e^{-2\pi i(j-k)\frac{M(\barell+\barm)-1}{N}}\mathcal{K}_{j,k;M,N}\,.
\end{aligned}
\end{equation}

\section{Technical details for covariance matrix calculations}
\label{s:covariance_details}

\subsection{Proof of Eq.~\eqref{eq: d1Ip_d1Jq_begin}}
\label{app: proof of d1Ip_d1Jq}
We start by proving~\eqref{eq: d1Ip_d1Jq_begin}, which we can be derived from~\eqref{eq: d_1Id_1J} as follows:
\begin{equation}
\begin{aligned}
 \langle \tilde{d}_{1;I;j}\tilde{d}_{1;I+1;k}^*\rangle&=(\Delta t)^2\sum_{r=0}^{N-1}\sum_{s=0}^{N-1}\langle d_{1;I;r}d_{1;I+1;s}\rangle e^{-2\pi i \frac{rj}{N}}e^{2\pi i\frac{sk}{N}} e^{2\pi i\frac{(1-O)Nk}{N}}\\
 &=(\Delta t)^2\sum_{r=0}^{N-1}\sum_{s=(1-O)N}^{(2-O)N-1}\langle d^{I^+}_{1;r}d^{I^+}_{1;s}\rangle e^{-2\pi i \frac{rj}{N}}e^{2\pi i \frac{[s-(1-O)N]k}{N}}e^{2\pi i \frac{(1-O)Nk}{N}}\\
 &=(\Delta t)^2\sum_{r=0}^{N-1}\sum_{s=(1-O)N}^{(2-O)N-1}\langle d^{I^+}_{1;r}d^{I+}_{1;s}\rangle e^{-2\pi i \frac{(rj-sk)}{N}}.
\end{aligned}
\label{eq: proof_d1Ip_d1Jq_begin}
\end{equation}
where we used~\eqref{e:dI+_def} to obtain the second equality.

\subsection{Proof of Eq.~\eqref{eq: d1Ip_d1Jq_mid}}
\label{app: proof of d1Ip_d1Jq_mid}
To prove~\eqref{eq: d1Ip_d1Jq_mid}, we continue with the last equality of~\eqref{eq: proof_d1Ip_d1Jq_begin}, noting that $\langle d^{I^+}_{1;r}d^{I+}_{1;s}\rangle$ is the definition of the auto-correlation function $\sigma^2_1 R_{1;|r-s|}$ for detector 1.
Then,
\begin{equation}
\begin{aligned}
 \langle \tilde{d}_{1;I;j}\tilde{d}_{1;I+1;k}^*\rangle&=(\Delta t)^2\sum_{r=0}^{N-1}\sum_{s=(1-O)N}^{(2-O)N-1}\sigma_1^2R_{1;|r-s|} e^{-2\pi i \frac{(rj-sk)}{N}}\\
 &=(\Delta t)^2\sum_{r=0}^{N-1}\sum_{s=(1-O)N}^{\min{\big\{r+N-1,(2-O)N-1\big\}}}\sigma_1^2 R_{1;|r-s|}e^{-2\pi i\frac{(rj-sk)}{N}}\\
 &=(\Delta t)^2\sum_{r=0}^{N-1}\sum_{s=(1-O)N}^{\min{\big\{r+N-1,(2-O)N-1\big\}}}\Bigg(\frac{1}{2T}\sum_{n=0}^{N-1}P_{1;n}e^{2\pi i\frac{n(r-s)}{N}}\Bigg)e^{-2\pi i\frac{(rj-sk)}{N}}\\
 &\approx \frac{(\Delta t)^2}{2T}P_{1;j}\sum_{r=0}^{N-1}\sum_{s=(1-O)N}^{\min{\big\{r+N-1,(2-O)N-1\big\}}}\underbrace{\sum_{n=0}^{N-1}e^{2\pi i\frac{n(r-s)}{N}}}_{N\delta_{r,s~(\mathrm{mod}~N)}}e^{-2\pi i\frac{(rj-sk)}{N}}\\
 &=\frac{T}{2N}P_{1;j}\sum_{r=(1-O)N}^{N-1}e^{-2\pi i\frac{r(j-k)}{N}}\,,
\end{aligned}
\label{eq: proof_d1Ip_d1Jq_mid}
\end{equation}
where we changed the upper limit of the second summation over $s$ from $s=(2-O)N-1$ to ${\rm min}\{r+N-1, (2-O)N-1\}$ to ensure $R_{1;|r-s|}=0$ for $|r-s|\geqslant N$. 
To obtain the third line, we rewrote $\sigma^2_1 R_{1;|r-s|}$ in terms of the PSD $P_{1;n}$, using the fact that the auto-correlation function and PSD are a Fourier transform pair. 
To obtain the fourth line, we made the (standard) approximation that the PSD is a slowly varying function of frequency, and that this two-point correlation should be non-negligible only when $j\approx k$.
Thus, we may pull $P_{1;n}$ out of the summation over $n$, replacing the index $n$ with either $j$ or $k$. 
We can also argue this point by rewriting the exponential terms as $\exp(2\pi i(n-j)r/N)\exp(-2\pi i(n-k)s/N)$. 
When summing over $r$ and $s$, the complex exponentials combine constructively only when $n\approx j\approx k$.
Thus, since $P_{1;n}$ is a slowly varying function of frequency compared to these rapidly oscillating terms, we can take it out of the summation, replacing the index $n$ with $j$. 
In this way, the difference between $j$ and $k$ is retained only in the exponential term.
(We will use this “trick” in several other derivations as well.)
Finally, the limits on the summation over $r$ in the final line come from the limits on the sums over $r$ and $s$ in the previous lines restricted by the Kronecker delta $\delta_{r,s~(\mathrm{mod}~N)}$.

\subsection{Proof of Eq.~\eqref{eq: d1Ip_d1Jq_end}}
\label{app: proof of d1Ip_d1Jq_end}
To prove~\eqref{eq: d1Ip_d1Jq_end}, we start with
\begin{equation}
\begin{aligned}
     \Big\langle&\tilde{\bf{d}}_{1;I;p}\tilde{\bf{d}}_{1;J;q}^*\Big\rangle= \frac{T}{2N^3}\sum_{j=0}^{N-1}\sum_{k=0}^{N-1}\sum_{t=(1-O)N}^{N-1}\sum_{r=0}^{N-1}\sum_{s=0}^{N-1}P_{1;j}w_rw_se^{-2\pi i \frac{t(j-k)}{N}}e^{-2\pi i\frac{r(p-j)}{N}}e^{2\pi i\frac{s(q-k)}{N}}e^{2\pi i\frac{(1-O)N(q-k)}{N}}\\
     &\approx\frac{T}{2N^3}P_{1;p}\sum_{j=0}^{N-1}\sum_{t=(1-O)N}^{N-1}\sum_{r=0}^{N-1}\sum_{s=0}^{N-1}w_rw_s\underbrace{\Bigg(\sum_{k=0}^{N-1}e^{2\pi i\frac{k[t-s-(1-O)N]}{N}}\Bigg)}_{N\delta_{t,s+(1-O)N~(\mathrm{mod}~N)}}e^{-2\pi i\frac{tj}{N}}e^{-2\pi i \frac{r(p-j)}{N}}e^{2\pi i \frac{qs}{N}}e^{2\pi i \frac{(1-O)Nq}{N}}\,,
\end{aligned}
\end{equation}
where we used the same argument that we used to derive~\eqref{eq: proof_d1Ip_d1Jq_mid} to pull $P_{1;j}$ out of the summation over $j$, replacing the index $j$ by $p$, to obtain the second line.  
(This is due to the summation of the exponential term $e^{-2\pi i \frac{r(p-j)}{N}}$ times $w_r$, when summed over $r$.)
In addition, since $t$ can only go from $(1-O)N$ to $N-1$, we can simply take $\delta_{t, s+(1-O)N~(\mathrm{mod}~N)}$ as $\delta_{t, s+(1-O)N}$.
Therefore, we have:
\begin{equation}
\begin{aligned}
\Big\langle\tilde{\bf{d}}_{1;I;p}\tilde{\bm{d}}_{1;J;q}^*\Big\rangle&\approx\frac{T}{2N^2}P_{1;p}\sum_{j=0}^{N-1}\sum_{t=(1-O)N}^{N-1}\sum_{r=0}^{N-1}w_rw_{t-(1-O)N}e^{-2\pi i \frac{tj}{N}}e^{-2\pi i\frac{r(p-j)}{N}}e^{2\pi i \frac{q[t-(1-O)N)]}{N}}e^{2\pi i \frac{(1-O)Nq}{N}}\\
&\approx \frac{T}{2N^2}P_{1;p}\sum_{t=(1-O)N}^{N-1}\sum_{r=0}^{N-1}w_rw_{t-(1-O)N}\underbrace{\sum_{j=0}^{N-1}e^{-2\pi i \frac{j(t-r)}{N}}}_{N\delta_{rt}}e^{-2\pi i\frac{rp-qt}{N}}\\
&=\frac{T}{2N}P_{1;p}\sum_{t=(1-O)N}^{N-1}w_t w_{t-(1-O)N}e^{-2\pi i\frac{t(p-q)}{N}}\,.
\end{aligned}
\end{equation}
\section{Technical details for multi-segment optimal estimator calculations}
\label{s:multi_seg_details}

\subsection{Proof of Eq.~\eqref{eq: bar_var}}\label{app: proof_bar_var}
We consider the following summation:
\begin{equation}
    \mathcal{S}_\barell\equiv\sum_{I\in o}\sum_{J=I\pm1}(\bar{\sigma}^{-2}_{I;\barell}+\bar{\sigma}^{-2}_{J;\barell}).
\end{equation}
Expanding the rhs:
\begin{equation}
\begin{aligned}
\mathcal{S}_\barell&=\underbrace{(\bar\sigma^{-2}_{I=1;\barell}+\bar\sigma^{-2}_{J=0;\barell})+(\bar\sigma^{-2}_{I=1;\barell}+\bar\sigma^{-2}_{J=2;\barell})}_{I=1}+\underbrace{(\bar\sigma^{-2}_{I=3;\barell}+\bar\sigma^{-2}_{J=2;\barell})+(\bar\sigma^{-2}_{I=3;\barell}+\bar\sigma^{-2}_{J=4;\barell})}_{I=3}\\
&+\cdots+\underbrace{(\bar\sigma^{-2}_{I=-1;\barell}+\bar\sigma^{-2}_{J=-2;\barell})+0}_{I=N_\mathrm{seg}-1\,(I\equiv -1)}\\
&=2(\bar\sigma^{-2}_{I=1;\barell}+\bar\sigma^{-2}_{I=3;\barell}+\cdots+\bar\sigma^{-2}_{I=-1;\barell})+2(\bar\sigma^{-2}_{I=0;\barell}+\bar\sigma^{-2}_{I=2;\barell}+\cdots+\bar\sigma^{-2}_{I=-2;\barell})-(\bar\sigma^{-2}_{I=0;\barell}+\bar\sigma^{-2}_{I=-1;\barell})\\
&=2\left(\bar\sigma_{o;\barell}^{-2}+\bar\sigma^{-2}_{e;\barell}-\frac{1}{2}\left(\bar\sigma_{0;\barell}^{-2}+\bar\sigma^{-2}_{-1;\barell}\right)\right)\\
&\equiv 2\tilde\sigma^{-2}_{\barell}\,.
\label{e:sigma-tilde-app}
\end{aligned}
\end{equation}

\subsection{Proof of Eq.~\eqref{e:Sigma_oo_ee_oe}}
\label{app: bar_Sigma_oo_oe_ee}
We first consider $\bar\Sigma_{oo;\barell\barm}$:
\begin{equation}
\begin{aligned}
 \bar\Sigma_{oo;\barell\barm}&\equiv\langle \bar C_{o;\barell}\bar C_{o;\barm}\rangle-\langle \bar C_{o;\barell}\rangle\langle\bar C_{o;\barm}\rangle\\
 &=\frac{1}{\sum_{I\in o}\bar\sigma^{-2}_{I;\barell}\sum_{J\in o}\bar\sigma^{-2}_{J;\barm}}\sum_{I\in o}\sum_{J\in o}\bar\sigma^{-2}_{I;\barell}\bar\sigma^{-2}_{J;\barm}\left(\langle\bar{C}_{I;\barell}\bar{C}_{J;\barm}\rangle-\langle\bar{C}_{I;\barell}\rangle\langle\bar{C}_{J;\barm}\rangle\right)\\
 &=\bar\sigma^2_{o;\barell}\bar\sigma^2_{o;\barm}\sum_{I\in o}\bar\sigma_{I;\barell}^{-2}\bar\sigma_{I;\barm}^{-2}\bar\Sigma_{I;\barell\barm}\\
 &\approx\bar\sigma^2_{o;\barell}\bar\sigma^2_{o;\barm}\sum_{I\in o}\bar\sigma_{I;\barell}^{-2}\bar\sigma_{I;\barm}^{-2}\frac{\bar\sigma^2_{I;\barell}}{r_M}\bar{\mathcal{C}}_{w;\barell\barm}\\
 &=\frac{1}{r_M}
\bar\sigma^{2}_{o;\barell}\,\bar{\mathcal{C}}_{w;\barell\barm}\,,
\end{aligned}
\end{equation}
where we used~\eqref{eq: def_of_bar_C_w} to rewrite $\bar\Sigma_{I;\barell\barm}$ in terms of $\bar C_{w;\barell\barm}$ to obtain the fourth line.
Similarly:
\begin{equation}
\bar\Sigma_{ee;\barell\barm}\approx\frac{1}{r_M}\bar\sigma^{2}_{e;\barell}
\,\bar{\mathcal{C}}_{w;\barell\barm}\,.
\end{equation}
Finally,
\begin{equation}
\begin{aligned}
    \bar\Sigma_{oe;\barell\barm}&\equiv\langle \bar C_{o;\barell}\bar C_{e;\barm}\rangle-\langle \bar C_{o;\barell}\rangle\langle\bar C_{e;\barm}\rangle\\
    &=\frac{1}{\sum_{I\in o}\bar\sigma^{-2}_{I;\barell}\sum_{J\in e}\bar\sigma^{-2}_{J;\barm}}\sum_{I\in o}\sum_{J\in e}\bar\sigma^{-2}_{I;\barell}\bar\sigma^{-2}_{J;\barm}\left(\langle\bar{C}_{I;\barell}\bar{C}_{J;\barm}\rangle-\langle\bar{C}_{I;\barell}\rangle\langle\bar{C}_{J;\barm}\rangle\right)\\
&=\bar\sigma^2_{o;\barell}\bar\sigma^2_{e;\barm}\sum_{I\in o}\sum_{J=I\pm1 }\bar\sigma^{-2}_{I;\barell}\bar\sigma^{-2}_{J;\barm}\bar\Sigma_{IJ;\barell\barm}\\
    &\approx\frac{1}{2r_M}\bar\sigma^2_{o;\barell}\bar\sigma^2_{e;\barell}\,\bar{\mathcal{C}}_{w,\ovl;\barell\barm}\sum_{I\in o}\sum_{J=I\pm1}(\bar\sigma^{-2}_{I;\barell}+\bar\sigma^{-2}_{J;\barell})\\
    &=\frac{1}{r_M}\bar\sigma^2_{o;\barell}\bar\sigma^2_{e;\barell}\tilde{\sigma}^{-2}_{\barell}\,\bar {\mathcal{C}}_{w,\ovl;\barell\barm}\,,
\end{aligned}
\end{equation}
where we used~\eqref{eq: def_bar_C_w_ovl} to obtain the fourth line and \eqref{e:sigma-tilde-app} to obtain the last line.

\subsection{Proof of Eq.~\eqref{eq: bar_sigma^2_IJ}}\label{app: proof_bar_Sigma_IJ}
Consider the following summation:
\begin{equation}
    \mathcal{S}_{IJ}\equiv\sum_{\barell=0}^{\frac{N}{2M}-1} \sum_{\barm=0}^{\frac{N}{2M}-1}\bar\sigma^{-2}_{I;\barell}\bar\sigma^{-2}_{J;\barm}\bar\Sigma_{IJ;\barell\barm}\,.
\end{equation}
Substituting~\eqref{eq: bar_Sigma_ovl_lm_complete} for $\bar\Sigma_{IJ;\barell\barm}$ leads to
\begin{equation}
\begin{aligned}
    \mathcal{S}_{IJ}
    &\approx \delta_{I,J=I\pm1}\sum_{\barell,\barm=0}^{\frac{N}{2M}-1}\bar\sigma^{-2}_{I;\barell}\bar\sigma^{-2}_{J;\barm} \sum_{p=M\barell-\frac{M}{2}}^{M\barell+\frac{M}{2}-1}\sum_{q=M\barm-\frac{M}{2}}^{M\barm+\frac{M}{2}-1}\frac{\mathcal{P}_{I;p}+\mathcal{P}_{J;p}}{2M^2N^2}\\
    &\hspace{1in}\times\sum_{j,k=(1-O)N}^{N-1}w_jw_{j-(1-O)N}w_kw_{k-(1-O)N} \left(e^{2\pi i(j-k)\frac{(p-q)}{N}}+e^{2\pi i(j-k)\frac{(p+q)}{N}}\right).
    \end{aligned}
\end{equation}
We then pull $\mathcal{P}_{I;p}$ and $\mathcal{P}_{J;p}$ out of the summation over $p$ (replacing the $p$ indices by $M\barell$), which allows us to do the summations over $p$ and $q$:
\begin{equation}
\begin{aligned}
    \mathcal{S}_{IJ}
    &\approx\delta_{I,J=I\pm1}\frac{1}{2N^2}\sum_{\barell=0}^{\frac{N}{2M}-1}\sum_{\barm=0}^{\frac{N}{2M}-1}\bar\sigma^{-2}_{I;\barell}\bar\sigma^{-2}_{J;\barm}(\mathcal{P}_{I;M\barell}+\mathcal{P}_{J;M\barell})\\
    &\hspace{0.75in}\times\sum_{j,k=(1-O)N}^{N-1}w_jw_{j-(1-O)N}w_kw_{k-(1-O)N}\mathcal{K}_{j,k;M,N}\left(e^{2\pi i(j-k)\frac{M(\barell-\barm)}{N}}+e^{2\pi i(j-k)\frac{M\barell+M\barm-1}{N}}\right)\,.
    \end{aligned}
\end{equation}
To proceed, we first replace $\bar\sigma_{J;\barm}^{-2}$ by $\bar\sigma_{J;\barell}^{-2}$ using the same ``trick" we adopted when deriving~\eqref{eq: proof_d1Ip_d1Jq_mid}.
Then, similar to the calculation of $\bar\sigma^2_{\rm opt}$ in Sec.~\ref{s:single_seg_broad}, we can manipulate the summation over $\barm$ of the second exponential term, combining it with the summation over $\barm$ of the first exponential term.
We are then able to perform the combined summation over $\barm$, giving rise to a $\delta_{jk}$ term, which simplifies the expression for $\mathcal{S}_{IJ}$:
\begin{equation}
\begin{aligned}
    \mathcal{S}_{IJ}&\approx\delta_{I,J=I\pm1}\frac{1}{2MN}\sum_{\barell=0}^{\frac{N}{2M}-1}\bar\sigma^{-2}_{I;\barell}\bar\sigma^{-2}_{J;\barell}(\mathcal{P}_{I;M\barell}+\mathcal{P}_{J;M\barell})\sum_{j=(1-O)N}^{N-1}w^2_jw^2_{j-(1-O)N}\\
    &=\delta_{I,J=I\pm1}\frac{O}{2M}\frac{\overline{w^4_{\ovl}}}{(\overline{w^2_\mathcal{K}})^2}\sum_{\barell=0}^{\frac{N}{2M}-1}\bar\sigma^{-2}_{I;\barell}\bar\sigma^{-2}_{J;\barell}(\bar\sigma^{2}_{I;\barell}+\bar\sigma^{2}_{J;\barell})\\
    &=\delta_{I,J=I\pm1}\frac{O}{2M^2}\frac{\overline{w^4_{\ovl}}}{(\overline{w^2_\mathcal{K}})^2}\mathcal{W}_{\mathcal{K}}(\bar\sigma^{-2}_{\opt;I}+\bar\sigma^{-2}_{\opt;J})\,,
\end{aligned}
\end{equation}
where we defined
\begin{equation}
\overline{w^4_{\ovl}}\equiv \frac{1}{ON}\sum_{j=(1-O)N}^{N-1}w^2_jw^2_{j-(1-O)N}\,,
\label{e:w4ovl}
\end{equation}
and where we used~\eqref{eq: C_opt_var_opt_I_non_coase_grained}, expressed in terms of $\mathcal{W_{\mathcal{K}}}\bar\sigma^2_{\barell}$, to obtain the last line.

\section{Technical details for bias calculations}
\label{s:biases_details}
Here, we provide details for the bias results quoted in the main text.
As described there, to simplify the calculations and notation we will work with non-coarse grained data in what follows.
We will also consider only a single frequency bin $\ell$, but multiple time segments $I$, and we will drop the $\ell$ subscript to simplify the notation.
Finally, we will assume that $P_{1,I}\equiv P_1$ and $P_{2,I}\equiv P_2$ for all segments, so that the theoretical weights $w_I$ are given by $w_I=1/(P_1 P_2)$.

\subsection{Biased estimators of the optimal weights and variances}
Using~\eqref{e:Phat-main}, it follows that
\begin{equation}
\begin{aligned}
\hat w_I \equiv 
\frac{1}{\hat{P}_{1,I}\hat{P}_{2,I}}
&=
\frac{1}{\left(P_1+\delta P_{1,I}\right)
\left(P_2+\delta P_{2,I}\right)}
\\
&=
\frac{1}{P_1P_2}\left(1+\frac{\delta P_{1,I}}{P_1}\right)^{-1}
\left(1+\frac{\delta P_{2,I}}{P_2}\right)^{-1}
\\
&=
\frac{1}{P_1 P_2}\left(1-\frac{\delta P_{1,I}}{P_1} + \frac{\delta P_{1,I}^2}{P_1^2} - \cdots\right)
\left(1-\frac{\delta P_{2,I}}{P_2} + \frac{\delta P_{2,I}^2}{P_2^2} - \cdots\right)
\\
&=\frac{1}{P_1P_2}\left(1-\frac{\delta P_{1,I}}{P_1}-\frac{\delta P_{2,I}}{P_2}+\frac{\delta P_{1,I}^2}{P_1^2}+\frac{\delta P_{2,I}^2}{P_2^2}+\frac{\delta P_{1,I}\delta P_{2,I}}{P_1 P_2}-\cdots\right)\,.
\label{e:w_hat_I_full}
\end{aligned}
\end{equation}
Then using~\eqref{e:evals-main}, it follows that the expectation value of $\hat w_I$ is given by 
\begin{equation}
\begin{aligned}
\left\langle\frac{1}{\hat{P}_{1,I}\hat{P}_{2,I}}\right\rangle
&=\frac{1}{P_1P_2}\left(1-\frac{\langle\delta P_{1,I}\rangle}{P_1}-\frac{\langle\delta P_{2,I}\rangle}{P_2}+\frac{\langle\delta P_{1,I}^2\rangle}{P_1^2}+\frac{\langle\delta P_{2,I}^2\rangle}{P_2^2}+\frac{\langle\delta P_{1,I}\delta P_{2,I}\rangle}{P_1 P_2}-\cdots\right)
\\
&=\frac{1}{P_1P_2}\left(1-0-0+\frac{1}{N_{\rm eff}}+\frac{1}{N_{\rm eff}}+\frac{1}{N_{\rm eff}}\frac{\gamma_{12}^2 P_{\rm gw}^2}{P_1 P_2}-\cdots\right)
\\
&\simeq\frac{1}{P_1P_2}\left(1+\frac{2}{N_{\rm eff}}\right)
\end{aligned}
\end{equation}
where we used the weak-signal limit and ignored third-order terms in $\delta P_{1,I}$ and $\delta P_{2,I}$ (and their products) to obtain the last line.
Thus, we see that $\hat w_I\equiv 1/(\hat P_{1,I}\hat P_{2,I})$ is a biased estimator of $w_I\equiv 1/(P_1P_2)$ with bias factor~\cite{Matas:2020roi,pygwb}:
\begin{equation}
\langle \hat w_I\rangle = b\,w_I\,,
\qquad b \equiv  1 + \frac{2}{N_{\rm eff}}\,.
\label{e:mean_w_hat}
\end{equation}
For the $O=0.5$ case, the exact expression of $N_\mathrm{eff}$ is given by~\cite{pygwb} and~\cite{Welch1967} as $N_\mathrm{eff}=\kappa^{-1}K$, where $K=2T\delta f-1$ and $\kappa$ is given by
\begin{equation}
\kappa=\left[1+2\left(\frac{\displaystyle{\sum_{i=N/2+1}^{N}w_iw_{i-{N/2}}}}{\displaystyle{\sum_{i=1}^Nw_i^2}}\right)^2\frac{K-1}{K}\right].
\end{equation}
Since $N_{\rm eff}$ is typically $\gg 1$, we can approximate the inverse of the bias factor as $b^{-1}\approx (1-2/N_{\rm eff})$, neglecting terms of order $1/N_{\rm eff}^2$ and higher.

Using the same method as above, we can also calculate the variance of $\hat w_I$:
\begin{equation}
{\rm Var}(\hat w_I) 
\equiv\langle \hat w_I^2\rangle - \langle \hat w_I\rangle^2
=\left\langle\frac{1}{\hat{P}^2_{1,I}\hat{P}^2_{2,I}}\right\rangle
-\left\langle\frac{1}{\hat{P}_{1,I}\hat{P}_{2,I}}\right\rangle^2\,.
\label{e:var_w_hat_def}
\end{equation}
To calculate the expectation value of
$1/(\hat P^2_{1,I} \hat P^2_{2,I})$, we first expand $\hat P_{1,I}$ and $\hat P_{2,I}$ as before, yielding:
\begin{equation}
\frac{1}{\hat{P}^2_{1,I}\hat{P}^2_{2,I}}
=\frac{1}{P^2_1P^2_2}\left(1-2\frac{\delta P_{1,I}}{P_1}-2\frac{\delta P_{2,I}}{P_2}+3\frac{\delta P_{1,I}^2}{P_1^2}+3\frac{\delta P_{2,I}^2}{P_2^2}+4\frac{\delta P_{1,I}\delta P_{2,I}}{P_1 P_2}-\cdots\right)\,.
\label{e:1/P^4}
\end{equation}
Then, again using~\eqref{e:evals-main}, we find
\begin{equation}
\left\langle\frac{1}{\hat{P}^2_{1,I}\hat{P}^2_{2,I}}\right\rangle
\simeq\frac{1}{P^2_1P^2_2}\left(1+\frac{6}{N_{\rm eff}}\right)\,.
\label{e:<1/P^4>}
\end{equation}
Thus,
\begin{equation}
{\rm Var}(\hat w_I) 
\simeq 
\frac{1}{P^2_1P^2_2}\left(1+\frac{6}{N_{\rm eff}}\right)
-\frac{1}{P^2_1P^2_2}\left(1+\frac{2}{N_{\rm eff}}\right)^2
\simeq 
\frac{2}{N_{\rm eff}}\frac{1}{P^2_1P^2_2}
=\frac{2}{N_{\rm eff}}w_I^2\,,
\label{e:var_w_hat}
\end{equation}
where we assumed $N_{\rm eff}\gg 1$ to obtain the second-to-last equality.

Finally, we calculate the expected value of the variance estimator $\hat \sigma^2_{\rm real}\equiv 1/\sum_I\hat w_I$ of the realistic estimator $\Creal$ given in~\eqref{e:Copt_real}:
\begin{equation}
\hat \sigma^2_{\rm real}\equiv \frac{1}{\sum_I \hat w_I}
\equiv \frac{1}{\sum_I\frac{1}{\hat{P}_{1,I}\hat{P}_{2,I}}}\,.
\end{equation}
Expanding $\hat P_{1,I}$ and $\hat P_{2,I}$ as before, we find
\begin{equation}
\begin{aligned}
\frac{1}{\sum_I\frac{1}{\hat{P}_{1,I}\hat{P}_{2,I}}}
&=\frac{P_1P_2}{N_{\rm seg}}\bigg(1+\frac{{\sum_I}\delta P_{1,I}}{N_{\rm seg}P_1}+\frac{{\sum_I}\delta P_{2,I}}{N_{\rm seg}P_2}-\frac{{\sum_I}\delta P_{1,I}^2}{N_{\rm seg}P_1^2}-\frac{\sum_I\delta P_{2,I}^2}{N_{\rm seg}P_2^2}-\frac{\sum_I\delta P_{1,I}\delta P_{2,I}}{N_{\rm seg} P_1 P_2}
\\
&\hspace{1in}
+\frac{\sum_{I,J}\delta P_{1,I}\delta P_{1,J}}{N^2_{\rm seg}P^2_1}+\frac{\sum_{I,J}\delta P_{2,I}\delta P_{2,J}}{N^2_{\rm seg}P^2_2}+2\frac{\sum_{I,J}\delta P_{1,I}\delta P_{2,J}}{N^2_{\rm seg}P_1 P_2}
\cdots\bigg)\,.
\label{e:opt_var_hat}
\end{aligned}
\end{equation}
Its expectation value is given by
\begin{equation}
     \left\langle\frac{1}{{\sum_I\frac{1}{\hat P_{1,I}\hat P_{2,I}}}}\right\rangle\simeq\frac{P_1P_2}{N_{\rm seg}}\left(1-\frac{2}{N_\mathrm{eff}}\right).
     \label{e:<opt_hat_var>}
\end{equation}
To obtain this last result, we again used~\eqref{e:evals-main}, remembering that $P_{1,I}=P_1$ and $P_{2,I}=P_2$ are assumed to be independent of the segment value.
This last assumption allows us to cancel the $1/N_{\rm seg}$ factors in the last three terms on the first line of~\eqref{e:opt_var_hat}.
(The two prior terms are zero using the first two equalities in~\eqref{e:evals-main}.)
In addition, we made use of the result
\begin{equation}
\langle \delta P_{1,I}\delta P_{1,J}\rangle \ne 0
\quad\text{if and only if}\quad J=I\ \text{or}\ J=I\pm 2\,,
\label{e:dPIdPJ}
\end{equation}
and similarly for 
$\langle \delta P_{2,I}\delta P_{2,J}\rangle$ and
$\langle \delta P_{1,I}\delta P_{2,J}\rangle$.
This is a consequence of the PSDs of the detector output being estimated from the data segments immediately preceding and following the analysis segment.
Hence, double sums in the last three terms on the last line of~\eqref{e:opt_var_hat} reduce to single sums, with only one factor of $N_{\rm seg}$ in $1/N^2_{{\rm seg}}$ being canceled using the assumption that the expected values of the power spectra are independent of the segment.
Since $N_{\rm seg}\gg 1$, these expectation values are negligible compared to those coming from the first line of~\eqref{e:opt_var_hat}, yielding~\eqref{e:<opt_hat_var>} for the final result.
Thus,
\begin{equation}
\langle \hat\sigma^2_{\rm real}\rangle \simeq \frac{P_1P_2}{N_{\rm seg}}\left(1-\frac{2}{N_{\rm eff}}\right)
=\sigma^2_{\rm opt}\left(1-\frac{2}{N_{\rm eff}}\right)
\label{e:mean_sigma2_opt_hat}
\end{equation}
for the variance estimator $\hat\sigma^2_{\rm real}$.

\subsection{Relating the means and variances of the optimal and realistic estimators}
To calculate the mean and variance of $\Creal$, 
we use~\eqref{e:dw_dC} to perform a Taylor expansion of $\Creal$ around $\hat C_{\rm opt}$, ignoring terms that are third-order or higher in factors of $\delta w_I$ and $\delta C_I$ and their products.
Doing so leads to:
\begin{equation}
\begin{aligned}
\Creal
&\equiv \frac{\sum_I \hat w_I \hat C_I}{\sum_J \hat w_J}
=\frac{\sum_I (b\,w_I + \delta w_I)\hat C_I}{\sum_J (b\, w_J + \delta w_J)}
=\frac{\sum_I (w_I + b^{-1}\delta w_I)\hat C_I}{\sum_J (w_J + b^{-1}\delta w_J)}
\\
&=\frac{\sum_I w_I\hat C_I + b^{-1}\sum_I\delta w_I\hat C_I}{\sum_J w_J\left(1+ b^{-1}\left(\sum_L w_L\right)^{-1}\sum_K\delta w_K\right)}
\\
&= \left(\hat C_{\rm opt} + b^{-1}\Big(\sum_J w_J\Big)^{-1}\sum_I \delta w_I\hat C_I\right)\left( 1 + b^{-1}\Big(\sum_L w_L\Big)^{-1}\sum_K \delta w_K\right)^{-1}
\\
&= \left(\hat C_{\rm opt} + b^{-1}\frac{P_1 P_2}{N_{\rm seg}}\sum_I \delta w_I\hat C_I\right)\left( 1 + b^{-1}\frac{P_1 P_2}{N_{\rm seg}}\sum_K \delta w_K\right)^{-1}
\\
&= \left(\hat C_{\rm opt} + b^{-1}\frac{P_1 P_2}{N_{\rm seg}}\sum_I \delta w_I\hat C_I\right)\left( 1 -b^{-1}\frac{P_1 P_2}{N_{\rm seg}}\sum_K \delta w_K+
b^{-2}\frac{P^2_1 P^2_2}{N^2_{\rm seg}}\sum_K \delta w_K\sum_L \delta w_L-\cdots\right)\,,
\end{aligned}
\end{equation}
where we used the simplification $w_I = 1/(P_1 P_2)$ to get the second-to-last line.
If we further isolate $\hat C_{\rm opt}$ as the first term, then we obtain
\begin{equation}
\Creal= \hat C_{\rm opt} + b^{-1}\frac{P_1 P_2}{N_{\rm seg}}\sum_I \delta w_I(\hat C_I-\hat C_{\rm opt})
- b^{-2}\frac{P^2_1 P^2_2}{N^2_{\rm seg}}\sum_{I,K} \delta w_I \delta w_K (\hat C_I-\hat C_{\rm opt})
-\cdots\,.
\label{e:D_temp}
\end{equation}
But note that the factors $(\hat C_I -\hat C_{\rm opt})$ can be simplified using~\eqref{e:dw_dC}:
\begin{equation}
\hat C_I - \hat C_{\rm opt}= C+\delta C_I - \left(C+\frac{1}{N_{\rm seg}}\sum_J \delta C_J\right)
= \delta C_I - \frac{1}{N_{\rm seg}}\sum_J \delta C_J\,.
\end{equation}
This means that the final term in~\eqref{e:D_temp} is third-order in $\delta$'s and hence can be ignored.
Thus, we obtain the relatively simple final expression
\begin{equation}
\Creal = \hat C_{\rm opt} + \hat\Delta\,,
\label{e:D=C+D}
\end{equation}
where
\begin{equation}
\hat \Delta
\equiv b^{-1}\frac{P_1P_2}{N_{\rm seg}}\left[\sum_I \delta w_I\left(\delta C_I - \frac{1}{N_{\rm seg}}\sum_{J}\delta C_J\right)-\dots\right]\,.
\label{e:Delta_hat}
\end{equation}
Thus, to calculate the mean and variance of $\Creal$, we can make use of
\begin{equation}
\langle \Creal\rangle = 
\langle \hat C_{\rm opt}\rangle +
\langle \hat \Delta\rangle\,,
\qquad
{\rm Var}(\Creal) = 
{\rm Var}(\hat C_{\rm opt}) +
{\rm Var}(\hat \Delta) +
2{\rm Cov}(\hat C_{\rm opt}, \hat \Delta)\,,
\label{e:mean_var_A+B}
\end{equation}
which holds for the sum of any two random variables.
So we are left to evaluate 
$\langle \hat \Delta\rangle$, 
${\rm Var}(\hat \Delta)$, and
${\rm Cov}(\hat C_{\rm opt}, \hat \Delta)$.

From~\eqref{e:Delta_hat}, we see that to evaluate $\langle\hat\Delta\rangle$, we need to calculate $\langle \delta w_I\delta C_I\rangle$ and $\langle\delta w_I \delta C_J\rangle$.
The first of these expectation values is identically zero, since $\delta w_I$ and $\delta C_I$ are uncorrelated (since they are estimated using different segments of data) and $\delta w_I$ and $\delta C_I$ both have zero mean.
Thus,
\begin{equation}
\langle \delta w_I\delta C_I\rangle = \langle \delta w_I\rangle \langle\delta C_I\rangle = 0\,.
\end{equation}
In addition, 
\begin{equation}
\langle \delta w_I\delta C_J\rangle\ne 0
\quad\text{if and only if}\quad J = I \pm 1\,,
\label{e:dwI_dCJ}
\end{equation}
which is a consequence of the PSDs of the detector output being estimated from the data segments immediately preceding and following the analysis segment, similar to what we saw in the previous subsection. 
This means that the double sum in the second term of~\eqref{e:Delta_hat} reduces to a single sum.
So, only one factor of $N_{\rm seg}$ in $1/N^2_{\rm seg}$ will be canceled under the assumption that the expected value of the power spectra are independent of the data segment.
Thus, the second term in the expectation value of~\eqref{e:Delta_hat} scales like $1/N_{\rm seg}$ in comparison to $\langle \hat C_{\rm opt}\rangle$, which is of order unity.
Since $N_{\rm seg}\gg 1$, the expectation value of $\hat\Delta$ is thus negligible compared to the expectation value of $\hat C_{\rm opt}$, allowing us to conclude that
\begin{equation}
\langle \Creal\rangle \simeq \langle \hat C_{\rm opt}\rangle\,.
\end{equation}
This result was somewhat to be expected, since the expectation value of a weighted estimator is usually independent of the weights.
This is trivial to verify for non-data-dependent weights.
But for data-dependent weights, it is not an obvious result.

A similar argument can be used to show that ${\rm Cov}(\hat C_{\rm opt}, \hat\Delta)\simeq 0$.
To prove this, we start by noting that
\begin{equation}
{\rm Cov}(\hat C_{\rm opt}, \hat\Delta)
\equiv \langle \hat C_{\rm opt}\hat\Delta\rangle-\langle\hat C_{\rm opt}\rangle\langle\hat \Delta\rangle \simeq \langle \hat C_{\rm opt}\hat\Delta\rangle\,,
\end{equation}
with the final (approximate) equality following from $\langle\hat\Delta\rangle\simeq 0$.
To evaluate $\langle \hat C_{\rm opt}\hat\Delta\rangle$, we need to calculate
\begin{equation}
\frac{1}{N^2_{\rm seg}}\sum_{I,K} \langle\delta C_K\delta w_I\delta C_I\rangle
\quad\text{and}\quad
\frac{1}{N^3_{\rm seg}}\sum_{I,J,K} \langle\delta C_K\delta w_I\delta C_J\rangle\,.
\label{e:two-terms}
\end{equation}
The first expression is identically zero, since:
\begin{equation}
\begin{aligned}
&\text{if\ } I=K, \quad 
\langle\delta C_K\delta w_I\delta C_I\rangle=\langle\delta C^2_I\delta w_I\rangle=\langle\delta C^2_I\rangle\langle\delta w_I\rangle=0\,,
\\
&\text{if\ } I=K\pm1, \quad 
\langle\delta C_K\delta w_I\delta C_I\rangle=\langle\delta C_{I\mp1}\delta w_I\delta C_I\rangle=\langle\delta C_{I\mp1}\delta w_I\rangle\langle\delta C_I\rangle=0\,,
\\
&\text{if\ } I\neq K\,,K\pm1, \quad 
\langle\delta C_K\delta w_I\delta C_I\rangle=\langle\delta C_K\rangle\langle\delta w_I\delta C_I\rangle=\langle\delta C_K\rangle\langle\delta w_I\rangle\langle\delta C_I\rangle=0\,.
\end{aligned}
\end{equation}
For the second expression,
\begin{equation}
\langle\delta C_K\delta w_I\delta C_J\rangle\
\ne 0\quad\text{if and only if}\quad
J=K=I\pm 1\,,
\end{equation}
which reduces the triple sum in~\eqref{e:two-terms} to a single sum, leading to only one factor $N_{\rm seg}$ being canceled from $1/N_{\rm seg}^3$.  
Thus, $\langle \hat C_{\rm opt}\hat\Delta\rangle$ and hence ${\rm Cov}(\hat C_{\rm opt}, \hat\Delta)$ scale like $1/N_{\rm seg}^2$ compared to ${\rm Var}(\hat C_{\rm opt})$ and ${\rm Var}(\hat \Delta)$, which (as we shall show below) both scale as a $1/N_{\rm seg}$.
Thus, in the limit $N_{\rm seg}\gg 1$, we can ignore ${\rm Cov}(\hat C_{\rm opt}, \hat\Delta)$ from the expression for ${\rm Var}(\Creal)$ given in~\eqref{e:mean_var_A+B}.

We now compute the remaining two terms in the expression for $\sigma^2_{\rm real}\equiv {\rm Var}(\Creal)$.
The first term is easy since we already know that
\begin{equation}
\sigma^2_{\rm opt}
\equiv {\rm Var}(\hat C_{\rm opt}) 
\simeq \frac{P_1 P_2}{N_{\rm seg}}\,,
\end{equation}
which is valid in the weak-signal approximation.
Second, to evaluate ${\rm Var}(\hat \Delta)$, we first split the rhs of~\eqref{e:Delta_hat} into two terms
\begin{equation}
\hat \Delta
\simeq A + B\,,
\end{equation}
where
\begin{equation}
A \equiv b^{-1}\frac{P_1P_2}{N_{\rm seg}}\sum_I \delta w_I\delta C_I 
\,,
\qquad
B \equiv - b^{-1}\frac{P_1P_2}{N^2_{\rm seg}}
\sum_{I,J}\delta w_I\delta C_J\,.
\label{e:Delta_hat-2}
\end{equation}
We will again use
${\rm Var}(A+B) = {\rm Var}(A)+{\rm Var}(B) + 2{\rm Cov}(A,B)$ to calculate the variance of $\hat\Delta$.
First,
\begin{equation}
{\rm Var}(A) \equiv \langle A^2\rangle - \langle A\rangle^2
=b^{-2}\frac{P_1^2 P_2^2}{N^2_{\rm seg}}
\sum_{I,J}\langle \delta w_I\delta C_I\delta w_J\delta C_J\rangle\,,
\label{e:varA_temp}
\end{equation}
where we used $\langle A\rangle=0$.
Now,
\begin{equation}
\langle \delta w_I\delta C_I\delta w_J\delta C_J\rangle
=\delta_{IJ}\langle \delta w^2_I\rangle\langle\delta C^2_I\rangle
+\delta_{I,J=I\pm1}\langle \delta w_I\delta C_{I\mp1}\rangle \langle\delta w_{I\mp1}\delta C_I\rangle
\label{e:two-terms}
\,.
\end{equation}
Note that the first term on the rhs of \eqref{e:two-terms} can be written as
\begin{equation}
\delta_{IJ}\langle \delta w^2_I\rangle\langle\delta C^2_I\rangle\simeq\delta_{IJ}\,\frac{2}{N_{\rm eff}} \frac{1}{P^2_1P^2_2}\, P_1P_2
=\delta_{IJ}\left(\frac{2}{N_{\rm eff}}\right)\frac{1}{P_1 P_2}\,,
\label{e:term1}
\end{equation}
where we used~\eqref{e:dw_dC_evals} to get the first equality.
The second term on the rhs of \eqref{e:two-terms} can be written as
\begin{equation}
 \delta_{I,J=I\pm1}\langle \delta w_I\delta C_{I\mp1}\rangle \langle\delta w_{I\mp1}\delta C_I\rangle   
\propto
\delta_{I,J=I\pm1}
\,P^2_\mathrm{gw}/P^2_1P^2_2\,.
\label{e:term2}
\end{equation}
Since the rhs of \eqref{e:term2} is smaller than the rhs of \eqref{e:term1} by a factor of 
$P^2_{\rm gw}/(P_1 P_2)\ll 1$ (assuming the weak-signal limit), the contribution of this term to ${\rm Var}(A)$ can be neglected.
Thus, substituting \eqref{e:term1} into~\eqref{e:varA_temp}, we have
\begin{equation}
{\rm Var}(A) = b^{-2}\frac{P_1 P_2}{N_{\rm seg}}\left(\frac{2}{N_{\rm eff}}\right)\simeq \frac{P_1 P_2}{N_{\rm seg}}\left(\frac{2}{N_{\rm eff}}\right)\,,
\end{equation}
where we were able to replace the $b^{-2}$ factor by unity, ignoring terms proportional to $1/N_{\rm eff}^2$.

To evaluate ${\rm Var}(B)$, we have
\begin{equation}
{\rm Var}(B)\equiv \langle B^2\rangle - \langle B\rangle^2
=b^{-2}\frac{P_1^2 P_2^2}{N^4_{\rm seg}}
\left(\sum_{I,J,K,L}\langle \delta w_I\delta C_J\delta w_K\delta C_L\rangle
-
\Big(\sum_{I,J}\langle \delta w_I\delta C_J\rangle\Big)^2\right)
\,.
\end{equation}
But using~\eqref{e:dwI_dCJ} and the fact that
\begin{equation}
\langle \delta w_I\delta C_J\delta w_K\delta C_L\rangle\ne 0
\quad\text{if and only if}\quad
J=I\pm 1, L=K\pm1 \text{\ or\ }
K=I, L=J \,,
\end{equation}
we can conclude that ${\rm Var}(B)$ scales like $1/N^2_{\rm seg}$, which is a factor of $1/N_{\rm seg}$ smaller than ${\rm Var}(A)$ and hence can be ignored in the limit $N_{\rm seg}\gg 1$.

Finally, a similar conclusion can be made for 
\begin{equation}
{\rm Cov}(A,B) \equiv \langle A B\rangle - \langle A\rangle\langle B \rangle
= -b^{-2}\frac{P_1^2 P_2^2}{N^3_{\rm seg}}
\sum_{I,J,K}\langle\delta w_I\delta C_I\delta w_J \delta C_K \rangle\,,
\end{equation}
where we again used $\langle A\rangle=0$ to simplify the rhs of the above expression.
Since
\begin{equation}
\langle\delta w_I\delta C_I\delta w_J \delta C_K \rangle\ne 0
\quad\text{if and only if}\quad
J=K=I\quad\text{or}\quad J=K=I\pm1\quad\text{or}\quad J=I\pm1\,,\ K=I\mp1\,,
\end{equation}
it follows that ${\rm Cov}(A,B)$ scales like $1/N^2_{\rm seg}$ and hence can be ignored relative to ${\rm Var}(A)$ in the calculation of the variance of $A+B$.

Putting all these results together gives
\begin{equation}
{\rm Var}(\hat \Delta)=
{\rm Var}(A+B)
\simeq {\rm Var}(A)
\simeq \frac{P_1 P_2}{N_{\rm seg}}\left(\frac{2}{N_{\rm eff}}\right)\,,
\end{equation}
and therefore
\begin{equation}
    {\rm Var}(\Creal)\simeq \frac{P_1 P_2}{N_{\rm seg}} \left(1+\frac{2}{N_{\rm eff}}\right)
    ={\rm Var}(\hat C_{\rm opt})\left(1+\frac{2}{N_{\rm eff}}\right)
    \equiv \sigma^2_{\rm opt}\left(1+\frac{2}{N_{\rm eff}}\right)
    \,.
\label{e:var_Dopt}
\end{equation}
This shows that the variance of the realistic (i.e., non-optimal) estimator $\Creal$ is greater than the variance of the truly optimal estimator by a factor of $1+2/N_\mathrm{eff}$.

If we further compare ${\rm Var}(\Creal)$ given in~\eqref{e:var_Dopt} and the expected value of $\hat \sigma^2_{\rm real}$ given in~\eqref{e:mean_sigma2_opt_hat}, we see that 
\begin{equation}
\sigma^2_{\rm real}\equiv {\rm Var}(\Creal) 
\simeq\sigma^2_{\rm opt} \left(1+\frac{2}{N_{\rm eff}}\right)
\simeq \langle \hat \sigma^2_{\rm real}\rangle \left(1+\frac{4}{N_{\rm eff}}\right)\,.
\label{e:overlooked_var}
\end{equation}
Thus, $\sigma^2_{\rm real}$ is
greater than the expected value of the realistic variance estimator $\hat \sigma^2_{\rm real}$ by a factor of 
$1+4/N_\mathrm{eff}$.
The additional factor of $1+2/N_\mathrm{eff}$ that appears on the rhs of this equation can be regarded as a new source of bias that people overlooked in past studies.
In those earlier studies, people compared the variance of the realistic estimator $\Creal$ to the theoretical variance $\sigma^2_{\rm opt}$, as opposed to $\langle\hat \sigma^2_{\rm real}\rangle$.
This is why there was a missing factor of $b=(1+2/N_{\rm eff})$ in those analyses.
\section{Zero-padding}\label{app: zero_padding}
We discuss the necessity of zero-padding in this appendix. 
We will introduce the wrap-around issue first, which arises when one aims to compute the convolution of two finite-duration, discrete time series using the convolution theorem. 
Then we will review the traditional broadband stochastic cross-correlation analysis, and state the importance of zero-padding in this particular setup. 
Finally, we will discuss the narrowband stochastic cross-correlation analysis, and compare it with the broadband analysis to show that zero-padding is necessary in this case as well.

\subsection{Wrap-around issue}
The \textit{convolution theorem} states that the convolution $(x*y)(t)$ of two time series $x(t)$ and $y(t)$
corresponds to the product of their  Fourier transforms $\tilde x(f)$ and $\tilde y(f)$ in the frequency domain:
\begin{equation}
(x*y)(t) \equiv \int_{-\infty}^\infty \D t' x(t')y(t-t')
\quad\xLeftrightarrow[\mathscr{F}^{-1}]{\mathscr{F}}
\quad \tilde x(f) \tilde y(f)\,,
\end{equation}
where
\begin{align}
&[\mathscr{F}(x)](f)\equiv\tilde x(f) \equiv \int_{-\infty}^\infty \D t\> x(t) e^{-2\pi i ft}\,,
\\
&[\mathscr{F}^{-1}(\tilde x)](t)\equiv x(t) = \int_{-\infty}^\infty \D f\> \tilde x(f) e^{2\pi i ft}\,,
\end{align}
define the Fourier transform of $x(t)$ and the inverse Fourier transform of $\tilde x(f)$, respectively.
Since convolution in the time domain is typically more computationally-expensive than taking Fourier (and inverse Fourier) transforms, one can speed up the calculation by first Fourier transforming $x(t)$ and $y(t)$ to the frequency domain, then taking their product, and finally inverse-Fourier transforming $\tilde x(f)\tilde y(f)$ back to the time domain.
However, as we shall show below, some subtleties arise when working with finite-duration, discrete time-series data. 

So, suppose now that we represent the functions $x(t)$ and $y(t)$ by a finite number of discrete time-series samples $x[i]\equiv x(i\Delta t)$ and $y[i]\equiv y(i\Delta t)$, where $\Delta t$ is the sample period, $i=0,1,\cdots,N-1$, and $T\equiv N\Delta t$ is the total duration of the data.
We define the \textit{linear convolution} of $x[i]$ and $y[i]$ as
\begin{equation}
    (x*y)[n]\equiv\sum_{j=\mathrm{max}(0,n-N+1)}^{\mathrm{min}(n,N-1)}x[j]y[n-j],
\label{e:lin_conv}
\end{equation}
where the lower and upper bound of the summation is to ensure that both $j$ and $n-j$ lie in the range $0,1,\cdots, N-1$.
The array $(x*y)[n]$ will have length $2N-1$, with the index $n$ taking values $0,1,\cdots, 2N-2$.

Geometrically, we can interpret linear convolution as follows:
Flip the order of the $y$-array relative to $x$, initially positioning the arrays so that the  the last element of the flipped $y$-array lines up with the first element of the $x$-array.  
Then multiply those two elements together to get $(x*y)[0]$.
Next, slide the flipped $y$-array one element to the right, multiplying the last element of the flipped $y$-array with the second element of the $x$-array, and the second-to-last element of the flipped-$y$ array with the first element of the $x$-array.
Then add those two products together to get $(x*y)[1]$.
Continue this process of sliding, multiplying, and adding, until the first element of the flipped $y$-array overlaps with the last element of the $x$-array, multiplying them together to get the final value $(x*y)[2N-2]$.
Note that we can also obtain the same answer by evaluating the sum
\begin{equation}
    (x*y)[n]\equiv\sum_{j=0}^{N-1}x[j]y[n-j]\,,
    \label{e:lin_conv}
\end{equation}
with limits 0 and $N-1$
provided we set $y[n-j]=0$ for indices $n-j\notin \{0,1,\cdots, N-1\}$.

As mentioned earlier, to speed up the calculation of the convolution~\eqref{e:lin_conv} of $x[i]$ and $y[i]$, we will use discrete Fourier transforms to do the calculation in the frequency domain.
But note that if we define the length-$N$ discrete Fourier transform (\ac{DFT}) and inverse discrete Fourier transform (\ac{IDFT}) via%
\footnote{Note that this definition differs from that given in  Sec.~\ref{sec: data_analysis} by an overall factor of $\Delta f$ or $\Delta T$.}
\begin{align}
&\mathscr{F}_N(x)[j] \equiv \tilde{x}[j]=\sum_{m=0}^{N-1}x[m]e^{-2\pi i\frac{jm}{N}}\,,
\label{e:DFT_N}
\\
&\mathscr{F}^{-1}_N(\tilde x)[m]\equiv x[m]=\frac{1}{N}\sum_{j=0}^{N-1}\tilde{x}[j]e^{2\pi i\frac{jm}{N}}\,,
\label{e:IDFT_N}
\end{align}
then
\begin{equation}
\begin{aligned}    \mathscr{F}^{-1}_N\left(\mathscr{F}_N(x)\,\mathscr{F}_N(y)\right)&=\frac{1}{N}\sum_{j=0}^{N-1}\sum_{k=0}^{N-1}\sum_{m=0}^{N-1}x[j]y[k]e^{-2\pi i\frac{jm}{N}}e^{-2\pi i\frac{km}{N}}e^{2\pi i\frac{nm}{N}}\\
    &=\frac{1}{N}\sum_{j=0}^{N-1}\sum_{k=0}^{N-1}x[j]y[k]\sum_{m=0}^{N-1}e^{2\pi i\frac{m(n-k-j)}{N}}\\
    &=\sum_{j=0}^{N-1}x[j]y[(n-j)~\mathrm{mod}~N]\\
    &\neq \sum_{j=\max(0,n-N+1)}^{\min{(n,N-1)}} x[j]y[n-j]\,.
\end{aligned}
\label{eq: circular_convolution}
\end{equation}
The final inequality arises because the presence of $\mathrm{mod}~N$ in $y[(n-j) ~\mathrm{mod}~N]$ implies that the $y[k]$ array can be extended \textit{periodically} for indices outside of the range $0,1,\cdots, N-1$.
The resulting ``mod $N$" convolution operation defined by the second-to-last line in~\eqref{eq: circular_convolution} is called \textit{circular convolution}, and the fact that linear convolution does not satisfy a convolution theorem for length-$N$ discrete Fourier transforms is called the \textit{wrap-around issue}~\cite{Press:1992zz}. 

To overcome this problem, we note that the length of the array $(x*y)[n]$ in~\eqref{e:lin_conv} is $2N-1$.
The solution to solving the wrap-around issue for linear convolution is simply to use DFTs of length $2N-1$ (or greater).%
\footnote{This amounts to simply replacing $N$ by $2N-1$ in the argument of the exponentials in~\eqref{e:DFT_N} and~\eqref{e:IDFT_N}.}
This is equivalent to extending the time-series $x[i]$ and $y[i]$ to length $2N-1$ by padding the end of the arrays (for indices $N, N+1, \cdots, 2N-2$) with zeroes~\cite{Press:1992zz}, and then using DFTs  appropriate for arrays of that length.
Since a DFT is faster when the length of the DFT is a power of 2, we will zero-pad the data to length $2N$:
\begin{equation}
x_{\rm zp}[i] \equiv
\begin{cases}
x[i], & i=0,1,\cdots, N-1 \\
0, & i=N, N+1, \cdots 2N-1
\end{cases}
\end{equation}
and similarly for $y_{\rm zp}[i]$.
Then, if we repeat the calculation shown in~\eqref{eq: circular_convolution}, we find
\begin{equation}
\begin{aligned}
\mathscr{F}^{-1}_{2N}\left(\mathscr{F}_{2N}(x_\mathrm{zp})\, \mathscr{F}_{2N}(y_\mathrm{zp}) \right)
&=\sum_{j=0}^{2N-1}x_{\mathrm{zp}}[j]y_\mathrm{{zp}}[(n-j)~\mathrm{mod}~2N]\\
&=\sum_{j=\max(0,n-N+1)}^{\min{(n,N-1)}} x[j]y[n-j]\,,
\end{aligned}
\end{equation}
which is the desired expression for the linear convolution of the original time-series data $x[i]$ and $y[i]$, where $i=0,1,\cdots, N-1$.

So, linear convolution of two length-$N$ time-series satisfies a discrete convolution theorem provided the DFT and IDFT operations are extended to lengths $2N-1$ (or greater).
We achieve this in practice by zero-padding the time-series data from length $N$ to length $2N$.

\subsection{Optimal filtering and zero-padding for the broadband analysis}
We now consider the traditional (broadband) stochastic cross-correlation search, which takes as its fundamental data product a broadband point estimator $\hat{Y}$, with corresponding standard deviation $\hat{\sigma}$.
Given strain data from two detectors $d_1(t)$ and $d_2(t)$, we define $\hat{Y}$ in the time domain by (ignoring windowing in this section)
\begin{equation}
    \hat{Y}=\int_{0}^{T}\D t\>\int_{0}^{T}\D t'\>d_1(t)d_2(t')Q(t-t')\,,
\end{equation}
where $Q(t-t')$ is called the \textit{optimal filter}, since it is chosen to maximize the SNR of $\hat Y$, see e.g.,~\cite{Allen:1997ad} and equations \eqref{e:Y_int} and \eqref{e:Q_opt} below.
The discrete form of $\hat{Y}$ is given by
\begin{equation}
    \hat{Y}=\sum_{j=0}^{N-1} \sum_{k=0}^{N-1}d_1[j]d_2[k]Q[j-k]\,.
    \label{eq: def_Y}
\end{equation}
We can rewrite the above equation as
\begin{equation}
\begin{aligned}
 \hat{Y}=\sum_{j=0}^{N-1}d_1[j]\left(\sum_{k=0}^{N-1}d_2[k]Q[j-k]\right)
 =\sum_{j=0}^{N-1}d_1[j]\left(\sum_{k=0}^{N-1}d_2[k]Q[|j-k|]\right)\,,
 \end{aligned}
\end{equation}
where the last equality makes use of the fact that the filter is an even function, since $Q(t-t')$ is symmetric under interchange of $t$ and $t'$.
Accordingly, the index of $Q[j]$ ranges from $-(N-1)$ to $N-1$, with $Q[j]=Q[-j]$%
\footnote{In Python, this can be easily implemented by defining an array $Q$ with $2N-1$
entries, such that $Q[j]=Q[2N-1-j]=Q[-j]$.}. 
Because we explicitly allow negative indices and assign them the expected symmetric values, we avoid any wrap-around issues associated with $Q[j-k]$ modulo $2N-1$.

Consequently, the convolution in the above equation can be efficiently computed in the frequency domain using length $2N-1$ DFTs and IDFTs. 
We first zero-pad $d_1$ and $d_2$ to length $2N-1$ and then DFT these zero-padded data, obtaining:
\begin{equation}
\begin{aligned}
\hat Y&=\sum_{j=0}^{N-1}d_1[j](d_2*Q)[j]\\
&=\frac{1}{2N-1}\sum_{j=0}^{N-1}d_1[j]\left(\sum_{k=0}^{2N-2}\tilde{d}_{2,\mathrm{zp}}[k]\tilde{Q}[k]e^{2\pi i\frac{kj}{2N-1}}\right)\\
&=\frac{1}{2N-1}\sum_{j,k=0}^{2N-2}d_{1,\mathrm{zp}}[j]\tilde{d}_{2,\mathrm{zp}}[k]\tilde{Q}[k]e^{2\pi i\frac{kj}{2N-1}}\\
&=\frac{1}{(2N-1)^2}\sum_{j,k,\ell=0}^{2N-2}\tilde{d}^*_{1,\mathrm{zp}}[\ell]\tilde{d}_{2,\mathrm{zp}}[k]\tilde{Q}[k]e^{-2\pi i\frac{j\ell}{2N-1}}e^{2\pi i\frac{kj}{2N-1}}\\
&=\frac{1}{2N-1}\sum_{k=0}^{2N-2}\tilde{d}^*_{1,\mathrm{zp}}[k]\tilde{d}_{2,\mathrm{zp}}[k]\tilde{Q}[k]\,,
\label{e:opt_filter_FD}
\end{aligned}
\end{equation}
We note that given our definition of $Q$, if one wants to zero-pad $d_1$, $d_2$, and $Q$ to length longer than $2N-1$, one should zero-pad at the tail of $d_1$ and $d_2$, but zero-pad in the middle of $Q$ to make sure that $Q[j]=Q[-j]$ is still valid.
As mentioned previously, one typically chooses to zero-pad the data and filter to length $2N$ for both simplicity and DFT calculation speed.

Finally, we note that the form of the optimal filter $Q[j-k]$ is not calculated directly in the time domain, but is derived rather from the theoretical expression for the broad-band estimator in the frequency domain, see e.g., \cite{Allen:1997ad}:
\begin{equation}
\hat Y =\int_{-\infty}^\infty \D f\>
\tilde Q(f) \tilde d_1^*(f) \tilde d_2(f)\,.
\label{e:Y_int}
\end{equation}
Maximizing the SNR leads to~\cite{Allen:1997ad, Romano:2016dpx}
\begin{equation}
    \tilde{Q}(f)={\cal N} \frac{\gamma_{12}(|f|)H_\beta(|f|)}{P_1(|f|)P_2(|f|)}\,,
\label{e:Q_opt}    
\end{equation}
where
\begin{equation}
    H_\beta(f)\equiv S_0(f)\left(\frac{f}{f_{\rm ref}}\right)^\beta\,,
    \qquad
    S_0(f)\equiv \frac{3H_0^2}{2\pi^2}\frac{1}{f^3}\,.
    \label{eq:H_beta, S_0}
\end{equation}
Here, $H_\beta(f)$ is the spectral shape appropriate for a GWB having a power-law energy density spectrum
\begin{equation}
    \Omega_{\rm gw}(f) = \Omega_\beta \left(\frac{f}{f_{\rm ref}}\right)^\beta\,.
\end{equation}
The overall normalization ${\cal N}$ of the optimal filter can be chosen so that the expectation value of $\hat Y$ in the presence of a signal having the above energy density spectrum is the dimensionless amplitude $\Omega_\beta$:
\begin{equation}
{\cal N}= \left[\frac{T}{10}\int_{-\infty}^\infty \D f\> \frac{\gamma_{12}^2(|f|)H_\beta^2(|f|)}{P_1(|f|)P_2(|f|)}
\right]^{-1}\,,
\label{e:calN}
\end{equation}
where $T$ is the length of the data (usually, the length of an analysis segment).
Since $T$ is typically much greater than the coherence time of the signal for the ground-based analyses we are considering in this paper, we have
\begin{equation}
    \tilde{Q}[k]\approx\tilde Q(f_k)
\end{equation}
to a very good approximation.
This is the expression that we use for $\tilde Q[k]$ in~\eqref{e:opt_filter_FD}.

\subsection{Zero-padding for the narrowband analysis}
\label{s:zp_for_narrowband}
The starting point for the narrowband cross-correlation search is the narrowband estimator of the energy density spectrum $\Omega_{\rm gw}(f)$.
The optimal narrow-band estimator can be obtained from the optimal broadband estimator $\hat Y$ defined in~\eqref{e:Y_int} by simply restricting attention to the positive and negative frequency bins having frequency $\pm f_k$.
The final result (as we shall show below) is 
\begin{equation}
\hat\Omega_{\rm gw} [k]
=\frac{10}{T}\frac{\mathfrak{R}\left[\tilde{d}^*_{1,\mathrm{zp}}[k]\tilde{d}_{2,\mathrm{zp}}[k]\right]}{\gamma_{12}(|f_k|)S_0(|f_k|)}\,.
\label{e:Omega_hat_estimator}
\end{equation}
The data are necessarily zero-padded for precisely the reasons given above, which led to~\eqref{e:opt_filter_FD}.
Not zero-padding the data would lead to a non-optimal starting point for the narrowband analysis.

To derive~\eqref{e:Omega_hat_estimator}, we first note the normalization factor ${\cal N}$ and optimal filter $\tilde Q(f)$ defined in~\eqref{e:calN} and~\eqref{e:Q_opt} reduce to 
\begin{equation}
{\cal N}\rightarrow\left[\frac{T}{10} \,2\Delta f\, \frac{\gamma_{12}^2(|f_k|)H_\beta^2(|f_k|)}{P_1(|f_k|)P_2(|f_k|)}
\right]^{-1}
=\frac{5}{T \Delta f}\, \frac{P_1(|f_k|)P_2(|f_k|)}{\gamma_{12}^2(|f_k|)H_\beta^2(|f_k|)}
\end{equation}
and
\begin{equation}
\tilde Q(f) \rightarrow
{\cal N}\,\frac{\gamma_{12}(|f_k|)H_\beta(|f_k|)}{P_1(|f_k|)P_2(|f_k|)}
=\frac{5}{T\Delta f} 
\frac{1}{\gamma_{12}(|f_k|) H_\beta(|f_k|)}\,.
\end{equation}
Using these narrow-band expressions implies that the integral in~\eqref{e:Y_int} becomes
\begin{equation}
\hat Y \rightarrow \frac{5}{T} 
\frac{1}{\gamma_{12}(|f_k|) H_\beta(|f_k|)}
\left[\tilde d_1^*(-f_k)\tilde d_2(-f_k) 
+\tilde d_1^*(f_k)\tilde d_2(f_k)\right]
=\frac{10}{T} 
\frac{\mathfrak{R}[\tilde d_1^*(f_k)\tilde d_2(f_k)]}
{\gamma_{12}(|f_k|) H_\beta(|f_k|)}
\equiv\hat\Omega_\beta(f_k) 
\label{e:C.22}
\end{equation}
when summed over the two bins at $\pm f_k$.
The final step is to use~\eqref{eq:H_beta, S_0} to write $H_\beta(|f_k|)$ as $S_0(|f_k|)(f_k/f_{\rm ref})^\beta$ and then move $(f_k/f_{\rm ref})^\beta$ to the rhs of~\eqref{e:C.22} defining 
\begin{equation}
\hat\Omega_{\rm gw}(f_k) \equiv \hat \Omega_\beta(f_k)\left(\frac{f_k}{f_{\rm ref}}\right)^\beta=
\frac{10}{T} 
\frac{\mathfrak{R}[\tilde d_1^*(f_k)\tilde d_2(f_k)]}
{\gamma_{12}(|f_k|) S_0(|f_k|)}\,.
\label{e:Omega_hat_estimator0}
\end{equation}
The discretized form of~ \eqref{e:Omega_hat_estimator0} becomes~\eqref{e:Omega_hat_estimator} when we replace the continuous Fourier transforms $\tilde d_1^*(f_k)$ and $\tilde d_2(f_k)$ by $\tilde d_{1,{\rm zp}}^*[k]$ and $\tilde d_{2,{\rm zp}}[k]$ obtained by taking the DFT of the zero-padded discrete time-series values.
Note that the final result is {\it independent} of the spectral index $\beta$.
\section{Extensions}
\label{s:extenstions}
In this appendix, we briefly discuss several extensions of the analyses presented in the main text: (i) analyzing data corresponding to non-power-law GW backgrounds, (ii) allowing for zero-padded data, and (iii) analyzing cross-correlated data when we have more than two detectors.

\subsection{Beyond power-law backgrounds}
\label{s:ext-bpl}
Since the expected value of the narrowband estimator is proportional
to $\OGW(f)$, one does not need to change any of the calculations for the narrowband estimator when considering energy-density spectra that are not simple power laws.
However, for the broadband estimator,  one needs to modify the calculations to allow for non-power-law backgrounds.

In the analyses presented in the main text, we assumed that the \ac{SGWB} can be modeled by a single power-law, and for simplicity we took the power-law index to be 0. 
For more general spectra, the narrowband estimator must be rescaled by a factor of $\mathcal{R}(f_{\barell};\bm{\theta})$ to ensure that the expected value of the rescaled estimator equals  $\Omega_\mathrm{ref}$ in every frequency bin. 
The definition of $\mathcal{R}(f_{\barell};\bm\theta)$ was given in the main text by $\mathcal{R}(f;\alpha)\equiv(f/f_\mathrm{ref})^{-\alpha}$, where $\bm\theta=\{\alpha\}$ for the single power-law model.
But the parameters $\bm\theta$ can represent \textit{any} physical parameters in general.
Correspondingly, the variance $\bar\sigma^2_{I;\barell}$ should be multiplied by $\mathcal{R}^2(f_{\barell};\bm\theta)$, and the covariance
$\bar\Sigma_{I;\barell\barm}$ should be multiplied by
$\mathcal{R}(f_{\barell};\bm\theta)\mathcal{R}(f_{\barm};\bm\theta)$. With
these rescalings, all other expressions presented in the main text are unchanged.

\subsection{Analyzing zero-padded data}
\label{s:ext-zp-data}
Previous searches for a \ac{SGWB} conducted by the \ac{LVK}
collaboration~\cite{O3stoch, O4aStoch} have zero-padded the data to twice its length before performing the \ac{DFT}s. 
In App.~\ref{app: zero_padding}, we gave arguments explaining why zero-padding is necessary for optimal estimation.
But in the main text, we ignored zero-padding to simplify the presentation.
Here, we show that it is a simple matter to modify the analyses to allow for zero padding.

We assume below that the data is zero-padded to twice its length, which is the standard procedure for past stochastic analyses.
We use $N$ to denote the original number of data points per data segment, prior to zero-padding.

\begin{enumerate}
\item Once the data are zero-padded, coarse-graining is unavoidable, and the coarsening factor becomes to $M=2T\delta\!f$.

\item  One needs to change the definition of the kernel function $\mathcal{K}_{j,k;M,N}$ in~\eqref{eq: K} to the following:
\begin{equation}
\mathcal{K}_{j,k;M,N}\equiv\left(\frac{\sinc\left(\pi\frac{(j-k)M}{2N}\right)}{\sinc\left(\pi\frac{(j-k)}{2N}\right)}\right)^2.
\end{equation}

\item  One needs to change how we compute the variance of the broadband estimator of a single segment:
\begin{equation}
    \bar\sigma^2_{\opt;I}=\frac{\mathcal{W}_\mathcal{K}}{\displaystyle{M\sum_{\barell=0}^{\frac{N}{2M}-1}\bar\sigma^{-2}_{I;\barell}}}\quad\rightarrow\quad \bar\sigma^2_{\opt;I}=\frac{2\mathcal{W}_\mathcal{K}}{\displaystyle{M\sum_{\barell=0}^{\frac{N}{M}-1}\bar\sigma^{-2}_{I;\barell}}}.
\end{equation}

 The need for the factor of “2” in the numerator can be understood intuitively, as zero-padding \textit{artificially} increases the frequency resolution of the data without actually contributing any additional information.
\end{enumerate}

\subsection{Multiple detectors}
\label{s:ext-mult-det}

In the main text, we restricted attention to cross-correlation measurements from a single pair of ground-based detectors.
Here, we show how to extend those calculations to a network containing more than two detectors.

For concreteness, we will consider a network consisting of four detectors, e.g., LIGO Hanford (H), LIGO
Livingston (L), Virgo (V), and KAGRA (K).
We will assume that we have already computed the optimal narrowband estimator for each of them.
The goal now is to construct an optimal narrowband estimator that combines data from all $6=4\times 3/2$ distinct detector pairs (i.e., baselines). 

For this detector network, one should consider the following $6\times6$ covariance matrix:
\begin{equation}
    \Sigma_\mathrm{net;\ell}\equiv
    \begin{pmatrix}
        \hat\sigma^2_{\mathrm{HL-HL;\ell}} & \hat\sigma^2_{\mathrm{HL-HV;\ell}}& \hat\sigma^2_{\mathrm{HL-HK;\ell}}& \hat\sigma^2_{\mathrm{HL-LV;\ell}}&
        \hat\sigma^2_{\mathrm{HL-LK;\ell}}& \hat\sigma^2_{\mathrm{HL-VK;\ell}}\\
        
        \hat\sigma^2_{\mathrm{HV-HL;\ell}} & \hat\sigma^2_{\mathrm{HV-HV;\ell}}& \hat\sigma^2_{\mathrm{HV-HK;\ell}}& \hat\sigma^2_{\mathrm{HV-LV;\ell}}&
        \hat\sigma^2_{\mathrm{HV-LK;\ell}}& \hat\sigma^2_{\mathrm{HV-VK;\ell}}\\
        
        \hat\sigma^2_{\mathrm{HK-HL;\ell}} & \hat\sigma^2_{\mathrm{HK-HV;\ell}}& \hat\sigma^2_{\mathrm{HK-HK;\ell}}& \hat\sigma^2_{\mathrm{HK-LV;\ell}}&
        \hat\sigma^2_{\mathrm{HK-LK;\ell}}& \hat\sigma^2_{\mathrm{HK-VK;\ell}}\\
        
        \hat\sigma^2_{\mathrm{LV-HL;\ell}} & \hat\sigma^2_{\mathrm{LV-HV;\ell}}& \hat\sigma^2_{\mathrm{LV-HK;\ell}}& \hat\sigma^2_{\mathrm{LV-LV;\ell}}&
        \hat\sigma^2_{\mathrm{LV-LK;\ell}}& \hat\sigma^2_{\mathrm{LV-VK;\ell}}\\
        
        \hat\sigma^2_{\mathrm{LK-HL;\ell}} & \hat\sigma^2_{\mathrm{LK-HV;\ell}}& \hat\sigma^2_{\mathrm{LK-HK;\ell}}& \hat\sigma^2_{\mathrm{LK-LV;\ell}}&
        \hat\sigma^2_{\mathrm{LK-LK;\ell}}& \hat\sigma^2_{\mathrm{LK-VK;\ell}}\\
        
        \hat\sigma^2_{\mathrm{VK-HL;\ell}} & \hat\sigma^2_{\mathrm{VK-HV;\ell}}& \hat\sigma^2_{\mathrm{VK-HK;\ell}}& \hat\sigma^2_{\mathrm{VK-LV;\ell}}&
        \hat\sigma^2_{\mathrm{VK-LK;\ell}}& \hat\sigma^2_{\mathrm{VK-VK;\ell}}\\
    \end{pmatrix}.
\end{equation}
To evaluate each of the terms in the covariance matrix, we can repeatedly apply Isserlis's theorem. 
Doing so shows that all diagonal entries (which share two detectors in common) are proportional to the product of two detector noise \ac{PSD} terms, $P_{n_a}P_{n_b}$. 
Terms that share a single detector are proportional to $P_{n_a}P_\mathrm{gw}$, and terms that do not have any detector in common, such as $\hat\sigma^2_{\mathrm{HL-VK};\ell}$ and $\hat\sigma^2_{\mathrm{VK-HL};\ell}$, are proportional to $P_\mathrm{gw}^2$. 

Since we work in the weak-signal limit, the off-diagonal terms are much smaller
than the diagonal ones. 
This implies that correlations between different baselines are negligible.
Consequently, the standard inverse-variance-weighted average can still be used to combine narrowband or broadband estimators from different baselines.

\section{Simulation details}\label{app: simulation}
In this appendix, we provide details of our simulations. 

Since the results derived in the main text are valid for arbitrary detector noise \ac{PSD}s, we consider, without loss of generality, a network of two next-generation detectors.
These are two 40-km-long Cosmic Explorer (\ac{CE}) detectors located at LIGO Hanford and the LIGO Livingston~\cite{Harry:2010zz,LIGOScientific:2014pky}, respectively. 
For the \ac{CE} detectors, we adopt the noise \ac{PSD} corresponding to the second-stage of development~\cite{CE2_PSD}. 
Furthermore, the expressions for the estimators and their covariance are independent of the underlying \ac{SGWB} signal, provided that we are in the weak-signal regime. 
Consequently, for simplicity, we do not simulate any \ac{SGWB} data. 

We simulated 2,000 noise-only frames for each detector using the \texttt{Bilby} package~\cite{Ashton:2018jfp,Bilby}. 
Each frame has a duration of 2048 s and a sampling frequency of 2048 Hz. We then used \texttt{pygwb}~\cite{pygwb} to perform the cross-correlation calculations. 
Since the current implementation of \texttt{pygwb} does not support all of the calculations we have proposed in the main text, we had to modify the  \texttt{pygwb} code accordingly. 

As discussed in Sec.~\ref{s:bias}, using suboptimal weights to combine estimators leads to a bias in the estimated variance. 
To avoid this issue, we do not use the estimated \acp{PSD} when calculating $\bar{\sigma}^2_{\mathrm{opt};\bar{\ell}}$; rather, we use the true (injected) noise \acp{PSD}. 
With this approach, the variance estimates are free from bias.

In practice, analyses are typically performed with either $T=192$~s segments and $\delta f=1/32$~Hz, or $T=4$~s segments and $\delta f=1/4$~Hz. 
We therefore consider both configurations here. 
For both analyses, we choose $f_\mathrm{min}=10$ Hz and $f_\mathrm{max}=1000$ Hz, and adopt 50\%-overlapping Hann windows.

\end{appendix}
\end{document}